\def\lam{$\lambda$}

\documentclass[traditabstract]{aa}
\usepackage{graphicx}
\usepackage{txfonts}
\usepackage{natbib}
\usepackage[figuresright]{rotating}
\usepackage{subfigure}
\usepackage{multirow}
\usepackage{psfig}
\usepackage{rotating}
\usepackage{lscape}
\usepackage{color}
\usepackage{afterpage}
\usepackage{natbib}

\begin{document} 
   \title{Gemini GMOS spectroscopy of HeII nebulae in M33}

%   \subtitle{HeII nebulae in M33}

   \author{C. Kehrig \inst{1,2}
          \and M.S. Oey \inst{2}\fnmsep\thanks{Visiting Astronomer, Kitt Peak National Observatory, National Optical Astronomy Observatory, which is operated by the Association of Universities for Research in Astronomy (AURA) under cooperative agreement with the National Science Foundation.}
          \and P.A. Crowther\inst{3}
          \and J. Fogel\inst{2}
          \and E. Pellegrini\inst{2}
          \and O. Schnurr\inst{1,3}
          \and \\ D. Schaerer\inst{4}
          \and P. Massey\inst{5,}$^\star$
          \and K. Roth\inst{6}}

\offprints{C. Kehrig}

   \institute{Astrophysikalisches Institut Potsdam (AIP), innoFSPEC Potsdam, An der Sternwarte 16, 14482 Potsdam, Germany \\ 
              \email{ckehrig@aip.de, oschnurr@aip.de}
         \and
             Department of Astronomy, University of Michigan, 500 Church Street, Ann Arbor, MI 48109 \\
             \email{msoey@umich.edu, fogel@umich.edu, pelleger@umich.edu}
         \and 
            Department of Physics and Astronomy, University of Sheffield, Hounsfield Road, Sheffield, S3 7RH, UK \\
            \email{Paul.Crowther@sheffield.ac.uk}
         \and 
             Geneva Observatory, University of Geneva, 51, Ch. des Maillettes, 1290 Versoix, Switzerland \\
            \email{daniel.schaerer@unige.ch} 
         \and L
             owell Observatory, 1400 W. Mars Hill Road, Flagstaff, AZ 86001, USA \\
            \email{massey@lowell.edu}
         \and 
             Gemini Observatory, Hilo, HI 96720, USA \\
            \email{kroth@gemini.edu}
             }

   \date{} 

 
%%%%%%%%%%%%%%%%%%%%%%%%%%%%%%%%%%%%%%%%%%%%%%%%%%%%%%%%%%%%%%%%%%%%%%%%%%%%% 
%%%%    Abstract                                                         %%%% 
%%%%%%%%%%%%%%%%%%%%%%%%%%%%%%%%%%%%%%%%%%%%%%%%%%%%%%%%%%%%%%%%%%%%%%%%%%%%% 
 
\abstract{We have carried out a narrow-band survey of the Local Group galaxy, 
M33, in the HeII $\lambda$4686 emission line, to identify 
HeII nebulae in this galaxy.  
With spectroscopic follow-up 
observations, we confirm three of seven candidate objects, including 
identification of two new HeII nebulae, BCLMP651, HBW673.  We also 
obtain spectra of associated ionizing stars for all the HII regions, 
identifying two new WN stars.  
We demonstrate that the ionizing source  
for the known HeII nebula, MA 1, is consistent with being 
the early-type WN star MC8 (M33-WR14), by carrying out a combined  
stellar and nebular analysis of MC8 and MA1. We were unable to identify the
helium ionizing sources for HBW 673 and BCLMP 651, which do not appear to be Wolf-Rayet stars.   
According to the [OIII]$\lambda$5007/H$\beta$ vs 
[NII]$\lambda$6584/H$\alpha$ diagnostic diagram, excitation mechanisms 
apart from hot stellar continuum are needed to account for the nebular 
emission in HBW 673, which appears to have no stellar source at all.}

\maketitle 
 
%%%%%%%%%%%%%%%%%%%%%%%%%%%%%%%%%%%%%%%%%%%%%%%%%%%%%%%%%%%%%%%%%%%%%%%%%%%%% 
%%%%    Introduction                                                     %%%% 
%%%%%%%%%%%%%%%%%%%%%%%%%%%%%%%%%%%%%%%%%%%%%%%%%%%%%%%%%%%%%%%%%%%%%%%%%%%%% 
\section{Introduction}\label{intro} 
 
The nebular HeII $\lambda$4686 emission line is often
observed in the integrated spectra of blue compact dwarf galaxies [BCDs; e.g.,
 \cite{ks81}; \cite{dm98}; \cite{i06a}; \cite{k08}], together with other
signatures of high ionization (e.g.,[Ne V] and [Fe V] emission)
(\citealt{ti05}).  It is often seen in Wolf-Rayet (WR) galaxies, especially in
low-metallicity systems (e.g., \citealt{s99}; \citealt{le10}).  Locally, however,
HeII emission is extremely rare in ordinary HII regions (e.g.,
\citealt{g91}; \citealt{n03a}), excluding planetary nebulae. What is the
origin of this line ?  Only certain classes of WR stars, and possibly also certain O stars, are predicted to be hot enough
to ionize nebular HeII (e.g.,\citealt{k02}; \citealt{c07}), whose ionization potential
$h\nu = 54.4$ eV.

In the few, known examples of HeII nebulae in the Local Group, 
nebular HeII emission appears to be mostly associated with individual 
stars (e.g., \citealt{g91}).  For example, 
two nebulae (G2.4+1.4 in the Milky Way and S3 in IC 
1613) are ionized by WO stars (\citealt{j75}; \citealt{dk82}); and
four nebulae, N79, N206 in the LMC, and N76, NGC249 in the SMC, are associated with 
early WN stars (\citealt{p91}; \citealt{n91}; \citealt{tp89}; \citealt{p09}).  
The N159~F nebula in the LMC is ionized by a massive X-ray binary (LMC 
X-1;\citealt{pa86}), and the LMC N44~C nebula appears to be 
associated with a normal O7 star (\citealt{g00}), although it 
has also been suggested to be ionized by a ``fossil'' X-ray source 
(\citealt{pm89}) that is no longer present.

The WR phase occurs after 3-5 Myr for high mass stars, and has relatively
short durations (few 10$^{5}$ yr), making WR stars vital, age-sensitive probes
of high-mass star formation. Their
HeII-ionizing luminosities are among the highest of any stars, reaching up to
$10^{38}erg s^{-1}$ (e.g., \citealt{snc02}).
The nebular HeII emission therefore offers a powerful probe to 
infer properties such as initial mass function, metallicities, and 
ages for distant star-forming regions.
To properly understand and interpret HeII emission as a diagnostic, it is
essential to establish which WR stars are responsible for nebular HeII.
WR spectral energy distributions (SEDs) are notoriously difficult to model, involving complex radiation
transfer in non-LTE, supersonic expanding atmospheres with important
ionization stratification, inhomogeneities, and line blanketing
(e.g., \citealt{c95}, 2002; \citealt{hk98}).  As a result, model atmospheres
have varied greatly; for example,
\cite{sd99} point out very different predictions for the
core effective temperature $T_\star$ of the
prototype WN4b star, HD~50896: \cite{s93}
predicted $T_\star = $ 45,000 K, while \cite{hk98}
predicted $T_\star = $ 100,000 K.  There are especially 
dramatic differences in the model SEDs for the HeII edge at 228 \AA.  
Fifteen years ago, early WC and WN stars often had been
predicted to ionize He$^+$ (e.g., \citealt{s96a}), whereas more recent
models now show softer SEDs and only occasional He$^+$-ionizing emission (e.g., 
\citealt{c02a}; \citealt{snc02}).  The elusiveness of HeII nebulae 
is consistent with this latter trend.  However, 
it contrasts with the observed large number of
nebular HeII-emitting WR galaxies (\citealt{s99}).  

A possible explanation for
nebular HeII emission in metal-deficient objects is 
that in such low metallicity environments, WR stars possess 
sufficiently weak winds that are optically thin in the He$^{+}$ continuum 
(e.g., \citealt{s92}), thereby allowing the He$^{+}$-ionizing
radiation to escape from the stellar atmospheres.  Thus, empirical determination
of the stellar parameter space associated with nebular HeII photoionization
will provide much-needed constraints for WR model 
atmospheres (e.g., \citealt{s96a}; \citealt{c99}; \citealt{s02}).

Outside the Local Group, \cite{b08} analysed spectra of 570 WR galaxies from SDSS6 and conclude that at 
$12 + \log {\rm O/H}< 8.0$, the main sources of HeII ionizing photons 
appears to be O stars, arguing for a less dense stellar wind at these 
metallicities, while at higher abundances WN stars might increasingly 
dominate the ionization budget. \Citet{dm98} showed that the observed equivalent widths of nebular
HeII$\lambda$4686 emission in the BCD IZw18 can be reproduced from an 
instantaneous burst model with the present WC/WO stars. They also demonstrate that there is a 
spatial correlation of nebular HeII$\lambda$4686 with the position of 
WR stars. A similar result was reported by \cite{k08} who 
found WR stars in the BCD IIZw70 by means of integral field 
spectroscopy. They find that the region occupied by the likely 
ionizing cluster is associated to the nebular emission in {\mbox 4686 \AA~} 
indicating that the hard ionizing radiation responsible for the 
nebular HeII$\lambda$4686 emission can be related to the WR stars. 
However, although WR and O stars could
be responsible for most of the HeII emission in BCDs and WR galaxies
(e.g.,\citealt{s96a}), other processes are apparently needed to
explain this high ionization in a significant fraction of them
(e.g., \citealt{guseva00}; \citealt{ti05})

Models 
of HeII $\lambda$4686 as a diagnostic have already been computed, 
both without (e.g., \citealt{s96a}, 2002; \citealt{snc02}) and 
with supernovae remnants [SNRs; \cite{cmk02}].  The metal-poor sensitivity is a primary 
motivation to develop diagnostics for relatively unevolved starbursts, and 
strong nebular HeII$\lambda$1640,4686 emission is expected to be one of the 
best signatures of massive Population III stars (e.g., \citealt{s02}; 
\citealt{s03} and references therein).  Several searches for this 
spectral signature in high-redshift galaxies have been carried out 
(e.g., \citealt{d04}; \citealt{n08}; \citealt{o08} and review in 
\citealt{s08}).   

To better understand the actual source of HeII nebular emission, we have carried out a systematic search for HeII nebulae in the Local Group 
galaxy, M33.  The proximity of M33, together with its relatively low 
inclination ($i = 53^\circ$) and high star-formation rate, makes this an ideal 
target for such a study.  At its distance of 840 kpc (\citealt{f01}), 1$\arcsec$ 
$\sim$ 4 pc.  Moreover, M33 has a well-determined metallicity gradient 
(\citealt{rs08}; hereafter RS08), allowing us to search for objects over a 
range of metallicities.  For example, the case of G2.4+1.4, in the direction 
of the Galactic Center, suggests that HeII nebulae can also form in metal-rich 
environments. 

The paper is organized as follows. In \S \ref{obs_datared}, we report 
narrow-band imaging to identify candidate HeII nebulae, and 
follow-up spectroscopic observations. The nebular properties of 
our HII regions are discussed in \S \ref{el_np}. In \S \ref{confirmed_heii_neb}, we 
describe our sample of confirmed HeII nebulae, including a detailed analysis of the object MA 1 and its resposible 
ionizing star, MC8. In \S \ref{discuss} we discuss our results. Finally, 
\S \ref{summary} summarizes the main conclusions derived from this work. 
 
%%%%%%%%%%%%%%%%%%%%%%%%%%%%%%%%%%%%%%%%%%%%%%%%%%%%%%%%%%%%%%%%%%%%%%%%%%%%% 
%%%%    Observations    and Data Reduction                                                 
%%%%%%%%%%%%%%%%%%%%%%%%%%%%%%%%%%%%%%%%%%%%%%%%%%%%%%%%%%%%%%%%%%%%%%%%%%%%% 
 
\section{Observations and data reduction}\label{obs_datared} 
 
\subsection{Narrow-band imaging}\label{image} 
 
\begin{figure*} 
\centering 
\includegraphics[width=11cm]{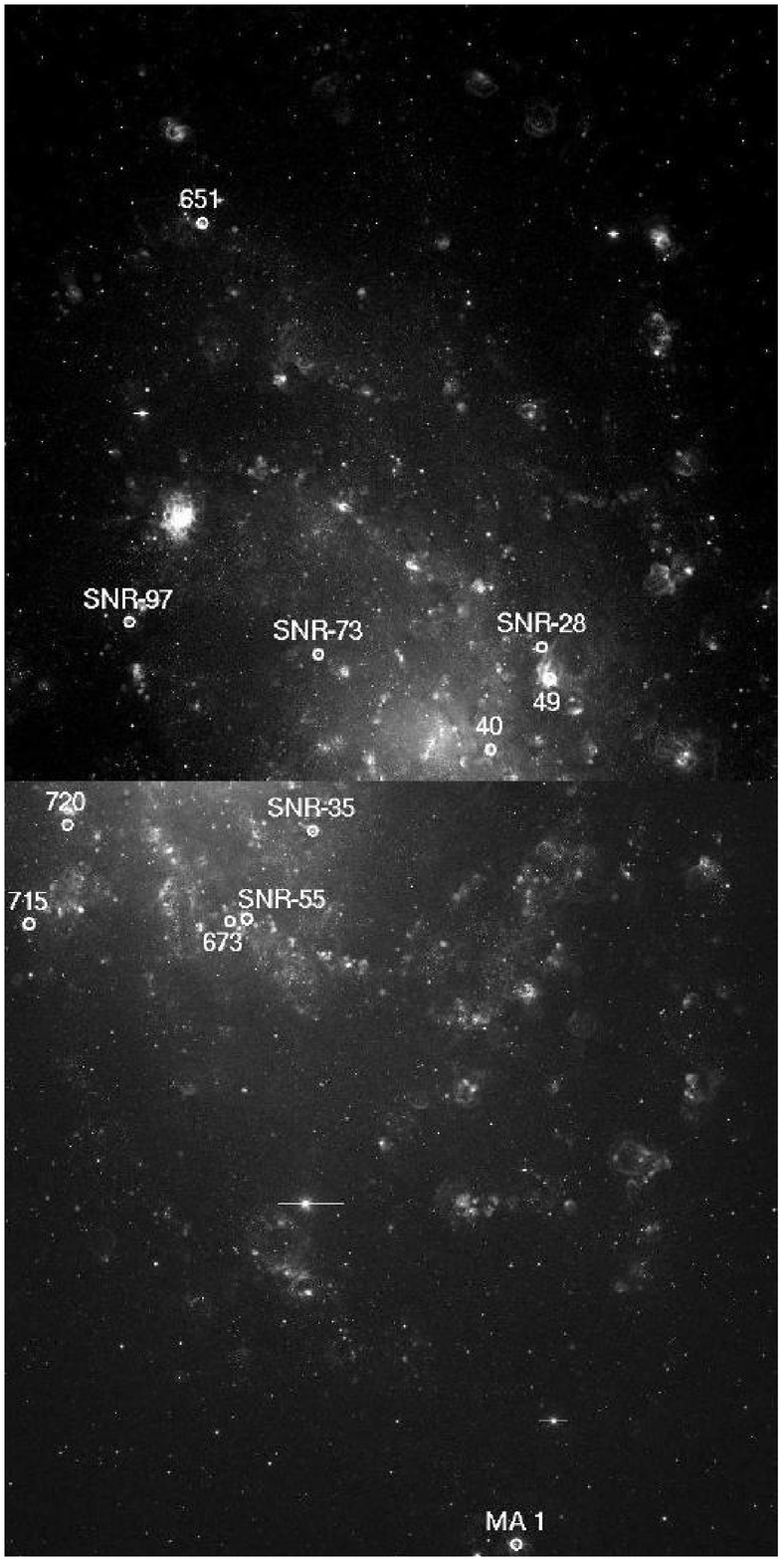}\\
\caption{Locations of the 12 cadidate HeII nebulae (seven HII regions and five SNRs)
superimposed on the H$\alpha$ image 
  from the NOAO Survey of Resolved Stellar Content of Local Group Galaxies
\mbox{(PI: P. Massey)}. The labeled ID of the
  seven HII regions correspond to the first column of Table~\ref{sample}. North and South fields of 
  M33 are shown by the top and bottom figures, respectively. Angular size is
  0.5$^\circ$ x 0.5$^\circ$. The spatial resolution is
  0.26$\arcsec$/pix. North is up and east to the left.} 
\label{m33} 
\end{figure*}

We obtained HeII narrow-band imaging of M33 with the MOSAIC imaging camera on 
the 4-m Mayall Telescope at Kitt Peak National Observatory, on November 3, 
2003.  We used the ``WR HeII'' filter centered at 4690 \AA\ for  
emission-line imaging, and the ``WR475'' filter centered at 4750 \AA\ for 
continuum observations.  Both filters have a transmission full width at 
half-maximum (FWHM) of about 50 \AA.  The 36$\arcmin$ MOSAIC field of view 
allowed us to image essentially the entire star-forming disk of M33 in two 
fields, North and South (Figure~\ref{m33}). These fields are centered at the
same positions as the H$\alpha$ narrow-band images obtained by one of us (PM) as part of the
NOAO Survey of Resolved Stellar Content of Local Group Galaxies, which were also obtained with the MOSAIC camera.  
We observed in 20-minute exposures for 
a total of 100 minutes on the South field, and 140 minutes on the North field.   
 
The data were reduced using the MSCRED routines in \textsc{IRAF}\footnote{IRAF
  is distributed by the National Optical Astronomical Observatories, which are
  operated by the Association of Universities for Research in Astronomy, Inc.,
  under cooperative agreement with the National Science Foundation.},
including bias subtraction, flat-fielding, and bad pixel correction.  The
MOSAIC camera also generates a low-level reflection of the pupil, which was
removed with the standard tools for this purpose in the \textsc{MSCRED}
software package.  The final, combined continuum image was then registered
with the corresponding line image, again using the \textsc{MSCRED} tools.
 
From the continuum-subtracted images, we visually identified 12 candidate HeII
nebulae in M33.  The HeII emission is faint and apparently diffuse in all
cases.  Seven of the objects are diffuse HII regions, mostly associated with
WR stars, while five are confirmed young SNRs.  

Figure~\ref{m33} displays a
portion (0.5$^\circ$ x 0.5$^\circ$) of the H$\alpha$ images obtained
previously, as described above. North and South fields of M33 are shown, with
the location of the 12 observed HeII nebulae identified. The HeII
emission-line, continuum and continuum-subtracted images of our seven HII regions are shown
in Figure~\ref{heii}.

\begin{table*}%[ht] 
\caption{The sample of our candidates for HeII-emitting HII regions} 
\label{sample} 
\centering 
\begin{minipage}{14.9cm} 
\centering 
\begin{tabular}{lcccccc} 
\hline\hline 
Name\tablefootmark{a} & Other designation & R.A.  & DEC & 
Galactocentric & Slit PA   &  Observing date  \\ 
 & & (J2000.0) & (J2000.0) & distance (Kpc) & (degrees) & \\ 
\hline 
MA 1 & HII-anon &01 33 02.2   &+30 11 18.8  & 7.3 & 246.6 &25 Sep 2008\\ 
BCLMP 49  & NGC 595   &01 33 33.5  &+30 41 33.6  & 1.4 & 90.0 &24 Sep 2008\\ 
BCLMP 40  &           &01 33 43.1  &+30 39 06.2  & 0.6  & 100.0 &24 Sep 2008\\ 
HBW 673   & HII-z29  &01 33 49.5   &+30 33 19.5 & 1.7 & 216.0 & 29 Aug 2006\\ 
BCLMP 720 &          &01 34 16.0   &+30 36 42.8 & 2.0 & 213.4 & 24 Sep 2008\\ 
BCLMP 715 &          &01 34 22.3   &+30 33 13.8 & 2.6 & 215.3 & 24 Sep 2008\\ 
BCLMP 651 &          &01 34  29.5  &+30 57 16.6 & 4.7 & 340.4  &26 Sep 2008\\ 
\hline 
\end{tabular} 
\end{minipage}
\tablefoot{
\tablefoottext{a}{Catalog abbreviations are the following: BCLMP, \cite{b74}; MA, \cite{ma42}; HBW, \cite{h99}}
}
\end{table*}

\subsection{Spectrophotometric data}\label{spectra} 
We obtained follow-up spectroscopic data of our sample objects with the 
Gemini Multi-Object Spectrograph (GMOS; \citealt{h04}) at the Gemini-North 
observatory.  We were awarded two different runs, one  
in service mode during semester 2006B (program GN-2006B-Q-83), and 
another in classical mode during the nights 2008 September 24 - 26 
(program GN-2008B-C-5).  The seven target HII regions are listed in 
Table~\ref{sample}.  Columns (1) and (2) show the ID for each 
object from the catalogs of \cite{b74}, \cite{h99}, and \cite{ma42}. 
Columns (3) and (4) give the object coordinates.  Column (5) 
quotes the deprojected galactocentric distance of each nebula for our assumed
distance of 840 kpc to M33 and an 
inclination angle of 56 degrees (\citealt{z89}). Column (6) 
gives the long-slit position angle (PA) used for each object.  The last column 
shows the corresponding spectroscopic observing date. 
Figure~\ref{finding_charts} shows our finding charts for the sample, 
using the H$\alpha$ emission-line images in Figure~\ref{m33}.  The 
associated hot stars (see Table~\ref{tab4}) are identified, together with 
the slit positions for our observations (green boxes). 
%The red box indicates the size of the HeII image as shown in 
%Figure~\ref{heii}.  
Yellow boxes show where the nebular HeII 
emission falls within the slit, and represent the apertures for our 
extracted spectra. 
 
The GMOS detector 
array consists of three $2048 \times 4608$ EEV CCDs in a row.  We used the B600 grating in the blue, covering a spectral 
range $\sim$ 3600 -- 6400 \AA~ (centered at 5000 \AA). On the red side, the R600 
(centered at 7730 \AA) was utilized, providing a spectral range from $\sim$ 6300 to 
9110 \AA. The data were binned by a factor of 4 in the spatial dimension and by 2 in the 
spectral dimension, yielding $\sim$ 0.29$\arcsec$/pix and a spectral resolution of $\sim$ 3.6 \AA~FWHM 
sampled at \mbox{$\sim$ 0.9 \AA/pixel.} The complete spectroscopic 
setup information is given in Table~\ref{obs}.  We obtained 
spectra of both the nebulae and their associated ionizing stars, using 
a slit width of 0.75$\arcsec$.  Observations of the spectrophotometric 
standard stars BD+284211 and BD+254655 were obtained during the observing 
nights for flux calibration.  Bias frames, dome flat-fields and copper-argon 
arc exposures were taken as part of the Gemini baseline calibrations.

%%%%%%%%%%%%%%%%%%%%%%%%%%%%%%%%%%%%%%%%%%%%%%%%%%%%%%%%%%%%%%%%%%%%%%%%%%% 
%            Table - Journal of Observations 
%%%%%%%%%%%%%%%%%%%%%%%%%%%%%%%%%%%%%%%%%%%%%%%%%%%%%%%%%%%%%%%%%%%%%%%%%%% 
\begin{table*}[ht] 
\caption{GMOS Spectroscopic setup information} 
\label{obs} 
\centering 
\begin{tabular}{cccccc} 
\hline\hline 
Spectral range &Dispersion &Resolving power &Spatial resolution  & Exposure time  \\ 
(\AA) &(\AA~ pix$^{-1}$) & &('' pix$^{-1}$) &(sec) & \\ 
\hline 
3600 - 6400 & 0.90 & 1688 & 0.29 & $3\times1200$  \\ 
 6300 - 9110 & 0.94 & 3744 &  0.29 & $3\times600$   \\ 
\hline 
\end{tabular} 
\end{table*} 
 
The raw images were processed using the Gemini GMOS routines within 
\textsc{IRAF}. Biases and flat-fields were combined with the tasks 
\textsc{gbias} and \textsc{gsflat}, respectively. Science targets and standard 
stars were reduced with task \textsc{gsreduce}, which applies the overscan 
correction, subtracts off the bias, mosaics the three detectors of GMOS 
interpolating across the chip gaps for science data, and applies the flat field 
correction. The wavelength calibration was established from the arcs with 
the task \textsc{gswavelength} giving residuals $<$ 0.1 \AA. The science 
spectra were further wavelength calibrated by applying the task 
\textsc{gstransform}. We use the task \textsc{gsextract} to extract the one 
dimension spectra performing sky subtraction. The sensitivity function was 
created by using the task \textsc{gsstandard}, and finally we flux calibrated 
the science images making use of the task \textsc{gscalibrate}. The 
uncertainty in the instrument response fitting across the spectrum for blue and red sides 
is $<$ 2 $\%$.

 %%%%%%%%%%%%%%%%%%%%%%%%%%%%%%%%%%%%%%%%%%%%%%%%%%%%%%%%%%%%%%%%%%%%%%%%%%% 
%            The sample of nebulae 
%%%%%%%%%%%%%%%%%%%%%%%%%%%%%%%%%%%%%%%%%%%%%%%%%%%%%%%%%%%%%%%%%%%%%%%%%%% 
 
\begin{table*} 
\caption{HeII nebulae and dominant ionizing stars} 
\label{tab4} 
\centering 
\begin{minipage}{14.9cm} 
\centering 
\begin{tabular}{l|clcc} 
\hline\hline 
HII region  & \multicolumn{4}{c}{Associated ionizing star(s)}    
\\ & Name\tablefootmark{a} & Other designation\tablefootmark{b}  &  
\multicolumn{2}{c}{Spectral Type}   
\\ & & & Previous work\tablefootmark{c} & This work     
\\ \hline 
\multirow{3}{*}{MA 1}  & 	 J013303.19+301124.2  & MC8 = M33-WR14 & WN  & WNE    \\ 
                       & 	 J013303.19+301124.2  & MC9 = M33-WR15 & WN & WNE \\ 
                             &  J013302.66+301120.4  &  K10-01         & ---  & WN3-4\\  \hline  
\multirow{5}{*}{BCLMP 49}    &  J013333.81+304129.6  &  NGC595-WR9         & WN7 &  WN7  \\ 
                              & J013333.66+304129.8  &  K10-02         & ---  & O \\  
                           &    J013333.46+304129.8  & &  & featureless\tablefootmark{d}\\  
                              & J013333.31+304129.7  & AM5 = M33-WR45 = NGC595-WR3 & WC6 & WC6   \\  
                              & J013334.38+304130.2  & AM7 = M33-WR51 = NGC595-WR7 & WNL & WN8  \\ \hline  
\multirow{2}{*}{BCLMP 40}  &    J013343.19+303906.2 & ``W91 129'' (M33-WR72)  &  WN7+abs & WN7-8  \\  
& & & & \\ \hline 
\multirow{2}{*}{HBW 673}   & no obvious star associated   & & &   \\  
                   &  &  & &   \\ \hline 
\multirow{2}{*}{BCLMP 720}   & J013416.07+303642.1 & H108               & P Cyg LBV candidate\tablefootmark{e} & P Cyg LBV candidate  \\ 
                  & 	       J013416.28+303646.4 &  MC70 = M33-WR129 & WC5 & WC5  \\ \hline 
\multirow{2}{*}{BCLMP 715}   & J013422.54+303317.1 &  MC72 = M33-WR133  &  WN  & WN4 \\ 
                    & J013422.37+303313.7 &  K10-03                       & --- & WN4 \\ \hline   
\multirow{2}{*}{BCLMP 651}   & J013429.44+305719.2 & &  & featureless$^{d}$  \\ 
                             & J013429.65+305711.2 & &  & featureless$^{d}$ \\  
\hline 
\end{tabular} 
\end{minipage}
 \tablefoot{
\\
\tablefoottext{a}{LGGS designation (JHHMMSS.ss+DDMMSS.s) from \cite{m06}.}\\ 
\tablefoottext{b}{Star identifications are MC = \cite{mc83}; M33-WR = Massey \& Johnson (1998; hereafter MJ98); NGC595-WR = \cite{d93}; AM = \citet{am85}; W91 = see \cite{m95} and MJ98; Hnnn = \cite{c96}; K10-nn = this work.}\\
\tablefoottext{c}{Spectral types from MJ98, \citet{A04}, \citet{D08} and references therein.}\\
\tablefoottext{d}{We are not able to provide a classification for the star since its spectrum shows no obvious stellar features.}\\
\tablefoottext{e}{P Cyg LBV candidate classification comes from \cite{m07}.}%\end{landscape} 
}
\end{table*} 
 
%\footnote{Spectral classification from this work} 
 
In Table~\ref{tab4} we list the dominant ionizing stars for 
our HII regions, and their rectified, blue spectra are plotted in 
Figure~\ref{star_spectra}.  These stellar spectra were used to estimate spectral 
classifications.  For three of the stars in Table~\ref{tab4}, 
designated as K10-$nn$, spectral types were not known before.  We also 
refine spectral types provided previously by others. We base our 
classification of WN stars on the line ratios proposed by \cite{ssm96} 
(e.g., \mbox{HeII$\lambda$5411/HeI$\lambda$5875}; 
\mbox{NV$\lambda$4604/NIII$\lambda$4640}; 
\mbox{NIV$\lambda$4057/NV-III$\lambda$4604-40}).  The 
two WC stars in our sample were classified using the 
CIII$\lambda$5696/CIV$\lambda$5808 line ratio (\citealt{c98}).  We also 
confirm the luminous blue variable (LBV)-like spectrum of the M33 star J013416.07+303642.1 classified 
previously as a P Cyg LBV candidate by \cite{m07}.  For three stars in our sample we are not 
able to give any spectral type since their spectra have low S/N and show no obvious stellar 
features.

Blue and red spectra for the HII regions are displayed in 
Figures~\ref{neb_blue_spectra} and \ref{neb_red_spectra}, respectively. 
From these spectra, we are able to confirm nebular HeII emission in three of the 
seven HII regions: MA 1, BCLMP651 and HBW673, the last two being new 
HeII-emitting nebulae identified for the first time. 
Portions of our two dimensional spectra of MA 1, BCLMP651 and HBW673 are shown 
in Figure~\ref{2dimage} from which we can see spatially extended 
HeII$\lambda$4686 emission, thus confirming its nebular 
nature. Figure~\ref{zoom} presents the wavelength range 4255 - 4775 
\AA\ for these three nebulae, from which we can see the faint,  
temperature sensitive emission line [OIII]\lam4363  
and the nebular HeII$\lambda$4686.   
 
%%%%%%%%%% DESCRIBING THE SAMPLE %%%%%%%%%%%%%%%%%%%5 
%%%%% HII-anon 
 
%MA 1, one of the first HeII-emitting HII regions to be discovered in a spiral 
%galxy other than the Milky Way, displays the HeII zone centered on WR stars. 

Emission-line fluxes were measured in the 1D spectra (Figures~\ref{neb_blue_spectra} and \ref{neb_red_spectra}) using the \textsc{IRAF} task 
\textsc{SPLOT}, which integrates the line intensity from a locally fitted
continuum. The derived line fluxes were computed via gaussian fitting. 
We carried out an iterative procedure to derive both the logarithmic
reddening, C(H$\beta$), and the  equivalent width of the absorption in the
hydrogen lines, EW$_{abs}$, which we use to correct the measured line intensities for both effects.
We used the Galactic reddening law of \cite{c89}  and assumed that EW$_{abs}$
is the same for all Balmer lines (e.g., \citealt{ks96}, \citealt{k99}). 
We computed a linear fit to the ratio of the measured-to-theoretical 
Balmer decrements
as a function of the aforementioned reddening law, while simultaneously varying the
values for the EW$_{abs}$.  This fit provides a value of C(H$\beta$)
for each value of EW$_{abs}$. We then adopted the EW$_{abs}$  and C(H$\beta$) values from the best fit (i.e. with minimum chi-square). The
uncertainty of the fit is adopted as the error in C(H$\beta$).  
In order to estimate how
the EW$_{abs}$ correction affects our results, we compare the values of
C(H$\beta$) with and without the aforementioned EW$_{abs}$ correction, and we
find that most are the same within the uncertainties. This indicates that, in general, the
contribution of the EW$_{abs}$ in the final flux measurements is not
significant. Moreover, in all cases, our estimations for the $EW_{abs}$  do not 
exceed the typical value of $\sim$ 2\AA\ from \cite{m82}. 

Since the blue and red spectra have no emission lines in common, the two 
sets were matched the following way: in the blue wavelength range, 
the strength of the emission lines was normalized to H$\beta$, while in the red, 
the line fluxes are normalized to H$\alpha$.  After reddening 
correction, the two sides are scaled assuming an intensity ratio of  
H$\alpha$/H$\beta$ = 2.86 (Case B, $T_{\rm e}$ = 10$^{4}$ K, $n_{\rm e}$=100 
cm$^{-3}$). The blue side of our spectra is the one whose flux calibration is
taken as reference. 
 
Reddening-corrected emission-line intensities, normalized to H$\beta$, are
shown in Table~\ref{lines1}, together with the values of C(H$\beta$) and
EW$_{abs}$ that provide the best match between the corrected and the
theoretical Balmer line ratios.  For four objects the uncertainties in
C(H$\beta$) are larger than the measured values, and therefore consistent with
no reddening. Since [OII]$\lambda$3727 is the line most affected by extinction
in our spectra, for these objects, we checked how much
[OII]$\lambda$3727/H$\beta$ changes with and without the reddening
correction. We found that the changes are within the quoted uncertainties.

The line-flux errors are calculated using the expression by \cite{g94}, 

\begin{equation}  
\sigma_{line}=\sigma_{cont}N^{1/2}\left(1 + \frac{\rm EW}{N\Delta\lambda}\right)^{1/2} 
\end{equation} 
where $\sigma_{cont}$ is the standard deviation of the continuum near the 
emission line, $N$ is the width of the region used to measure the line in 
pixels, $\Delta\lambda$ is the spectral dispersion in \AA/pixel, and EW 
represents the equivalent width of the line. The final errors quoted 
in Table~\ref{lines1}  account for the uncertainty in the extinction 
coefficient C(H$\beta$) and  
include contributions of $<$ 2$\%$ due to the uncertainties in the instrument response fitting.
Systematic errors are not included.

\begin{table*} 
%\begin{sidewaystable} 
%\caption{Observed HII region line fluxes\protect\footnotemark[1]}
\caption{Observed HII region line fluxes}
\label{lines1} 
\centering 
\begin{minipage}{14.9cm} 
\begin{tabular}{lccccccc} 
\hline\hline 
Wavelength & MA 1 inner\tablefootmark{a}  &BCLMP49 &BCLMP40 &HBW673 &BCLMP720 &BCLMP715 &BCLMP651   \\  
\hline 
3727 [O~II]  &  54.0 $\pm$ 7.0     & 172 $\pm$ 7         & 195 $\pm$ 53      & 337 $\pm$ 182      & 280 $\pm$ 28    & 97.0 $\pm$ 19.0  & 215 $\pm$ 17       \\ 
3868 [Ne~III] & 55.0 $\pm$ 10.0     & 4.00 $\pm$ 1.00    & ---               & 181 $\pm$ 59       & ---                & 31.0 $\pm$ 7.0  & 43.0 $\pm$ 4.0    \\ 
3889 H8+He~I  & 18.0 $\pm$ 1.0     & 18.0 $\pm$ 1.0      & ---               & ---                & 23.0 $\pm$ 9.0    & 14.0 $\pm$ 6.0  & 21.0 $\pm$ 3.0   \\ 
3968H7+[Ne~III] & 34.0 $\pm$ 6.0  & 14.0 $\pm$ 1.0       & ---               & 77.0 $\pm$ 43.0    & 18.0 $\pm$ 8.0    & ---              & 34.0 $\pm$ 3.0   \\ 
4100 H$\delta$  & 25.0 $\pm$ 4.0     & 26.0 $\pm$ 2.0    & ---               & 24.0 $\pm$ 2.0     & 26.0 $\pm$ 9.0     & 26.0 $\pm$ 5.0  & 29.0 $\pm$ 3.0   \\ 
4340 H$\gamma$  & 49.0 $\pm$ 2.0     & 47.0 $\pm$ 1.0    & 47.0 $\pm$ 12.0   & 48.0 $\pm$ 10.0    & 47.0 $\pm$ 2.0     & 47.0 $\pm$ 3.0  & 46.0 $\pm$ 2.0   \\ 
4363 [O~III]    & 9.00 $\pm$ 3.00    & ---                 & ---               & 18.0 $\pm$ 5.0    & ---                & ---              & 4.70 $\pm$ 0.50  \\ 
4471 He~I       &  ---               & 4.00 $\pm$ 0.40    & ---               & ---               & 8.00 $\pm$ 1.00     & ---              & 4.60 $\pm$ 0.50   \\ 
{\bf 4686 He~II} & {\bf 19.0 $\pm$ 2.0}    & ---          & ---               & {\bf 55.0 $\pm$ 9.0}   & ---                & ---    & {\bf 11.0 $\pm$ 1.0}     \\  
4861 H$\beta$  & 100 $\pm$ 1  & 100 $\pm$ 1               & 100 $\pm$ 6       & 100 $\pm$ 6.0      & 100 $\pm$ 2     & 100 $\pm$ 2      & 100 $\pm$ 1  \\ 
4959 [O~III]  & 207 $\pm$ 1  & 47.0 $\pm$ 1.0             & 12.0 $\pm$ 5.0   & 554 $\pm$ 39        & 25.0 $\pm$ 1.0     & 160 $\pm$ 5     & 147 $\pm$ 2         \\ 
5007 [O~III] & 602 $\pm$ 30     & 139 $\pm$ 2              & 36.0 $\pm$ 6.0   & 1663 $\pm$ 119      & 73.0 $\pm$ 2.0   & 484 $\pm$ 14      & 440 $\pm$ 7         \\ 
5876 He~I   & 9.00 $\pm$ 2.00   & 11.8 $\pm$ 0.3          & ---               & 10.0 $\pm$ 2.0     & 15.0 $\pm$ 1.0     & 12.0 $\pm$ 1.0  & 12.0 $\pm$ 1.0            \\ 
6300 [O~I]  & ---       & 0.70 $\pm$ 0.10  & ---           & ---               & 1.70 $\pm$ 0.60   & 4.00 $\pm$ 1.00  & 3.20 $\pm$ 0.20            \\ 
6548 [N~II] & 3.00 $\pm$ 1.00   & 8.70 $\pm$ 0.30          & 20.0 $\pm$ 8.0   & 26.0 $\pm$ 7.0     & 17.0 $\pm$ 1.00     & 8.00 $\pm$ 1.00  & 8.50 $\pm$ 0.50          \\ 
6563 H$\alpha$  & 286 $\pm$ 6     & 286 $\pm$ 7            & 286 $\pm$ 113     & 286 $\pm$ 79      & 286 $\pm$ 20      & 286 $\pm$ 25  & 286 $\pm$ 16  \\ 
6584 [N~II]     & 6.00 $\pm$ 1.00   & 26.0 $\pm$ 1.0       & 58.0 $\pm$ 23.0   & 86.0 $\pm$ 24.0   & 53.0 $\pm$ 4.0     & 23.0 $\pm$ 3.0  & 24.0 $\pm$ 1.0  \\ 
6678 He~I       & 4.00 $\pm$ 1.00   & 3.50 $\pm$ 0.20      & ---               & ---               & 4.80 $\pm$ 0.40    & 3.30 $\pm$ 0.60 & 2.80  $\pm$ 0.20 \\ 
6717 [S~II]     & 8.00 $\pm$ 1.00  & 5.10 $\pm$ 0.30       & 30.0 $\pm$ 9.0    & 55.0 $\pm$ 16.0   & 43.0 $\pm$ 3.0    & 29.0 $\pm$ 3.0  & 25.0 $\pm$ 2.0 \\ 
6731 [S~II]     & 5.00 $\pm$ 1.00   & 4.00 $\pm$ 0.20      & 22.0 $\pm$ 8.0    & 37.0 $\pm$ 11.0   & 30.0 $\pm$ 2.0    & 20.0 $\pm$ 2.0  & 17.0 $\pm$ 2.0 \\ 
7136 [Ar~III]   & 8.00 $\pm$ 1.00    & 9.30 $\pm$ 0.40     & 4.00 $\pm$ 1.00   & 23.0 $\pm$ 8.0    & 6.70 $\pm$ 0.70    & 11.0 $\pm$ 1.0  & 9.00 $\pm$ 1.00 \\ 
9069 [S~III]  & 12.0 $\pm$ 4.0   & 7.40 $\pm$ 0.50         & 5.00 $\pm$ 2.00   & 8.00 $\pm$ 4.00   & 6.00 $\pm$ 1.00   & 7.00 $\pm$ 3.00  & 5.00 $\pm$ 1.00   \\ \hline 
$C_{H\beta}$     & 0.09 $\pm$ 0.08     & 0.37 $\pm$ 0.03    & 0.22 $\pm$ 0.57   & 0.09 $\pm$ 0.39   & 0.12 $\pm$ 0.10    & 0.05 $\pm$ 0.12  & 0.04 $\pm$ 0.08  \\  
$EW_{abs}$ (\AA)&0.3                  &1.7                 &2.0               & 0.1                 &0.3                 &1.2                &0.1              \\  
\hline        
\end{tabular} 
%\footnotetext[1]{nnnnnn}  
\end{minipage}
\tablefoot{
Reddening-corrected line fluxes, normalized to F(\mbox{H$\beta$})= 100, for apertures defined by the HeII-emitting region (Figure~\ref{finding_charts}).  The reddening coefficients, $C(\mbox{H$\beta$})$, and the value of the absorption correction, $EW_{\rm abs}$ are listed in the last two rows for each HII region. The quoted uncertainties include measurement and reddening errors.  \\
\tablefoottext{a}{Emission-line fluxes from the inner, HeII-emitting part of the nebula (see Section \ref{wr14_cloudy} for details).}
}
%\end{sidewaystable} 
\end{table*}  
 
%%%%%%%%%%%%%%%%%%%%%%%%%%%%%%%%%%%%%%%%%%%%%%%%%%%%%%%%%%%%%%%%%%%%%%%%%%%%% 
%%%%      Analysis                                                       %%%% 
%%%%%%%%%%%%%%%%%%%%%%%%%%%%%%%%%%%%%%%%%%%%%%%%%%%%%%%%%%%%%%%%%%%%%%%%%%%%% 
 
\section{Nebular properties}\label{el_np} 
 
\begin{table*}%{ht} 
%\begin{landscape} 
\caption{Nebular properties} 
\label{np} 
\centering 
\begin{minipage}{14.9cm} 
\centering 
\begin{tabular}{lccccccc} 
\hline\hline 
Parameter &MA 1\tablefootmark{a} &BCLMP 49 &BCLMP 40 &HBW673 &BCLMP 720 &BCLMP 715  &BCLMP651 \\  
\hline 
$n_{\rm e}$([S II])(cm$^{-3}$) 	                          & $<$ 100 &159 & $<$ 100  & $<$ 100  & $<$ 100 & $<$ 100  & $<$ 100   \\ 
%$N_{\rm e}$(S II)(cm$^{-3}$) 	                          & 33 & 2           & & & $\pm$            & &  &                \\ 
{\bf $t_{\rm e}$}([O III])(10$^{4}$ K) 	                          & 1.35 $\pm$ 0.20  & --- & --- & 1.19 $\pm$ 0.11  & --- & ---  & 1.18 $\pm$ 0.04   \\ 
{\bf $t_{\rm e}$}([S III])(10$^{4}$ K)                             & ---   & --- & --- & 1.17 $\pm$ 0.12  & --- & --- & 1.17 $\pm$ 0.04 \\ 
{\bf $t_{\rm e}$}([O II])(10$^{4}$ K)                              & ---   & --- & ---  & 1.14 $\pm$0.08  & --- & ---   & 1.12 $\pm$ 0.03  \\ 
12 + log (O$^{+}$/H$^{+}$) 	                          & 6.80 $\pm$ 0.18  & --- & --- & 7.82 $\pm$ 0.26  & --- & ---  & 7.65 $\pm$ 0.05  \\ 
12 + log (O$^{2+}$/H$^{+}$) 	                          & 7.92 $\pm$ 0.15  & --- & --- & 8.50 $\pm$ 0.10  & --- & ---  & 7.94 $\pm$ 0.04   \\ 
12 + log (O/H)\tablefootmark{b}  & 8.00 $\pm$ 0.05  & 8.35  & 8.65  & 8.66  $\pm$ 0.09   & 8.54 & 8.20  & 8.12  $\pm$ 0.03   \\ 
12 + log (N$^{+}$/H$^{+}$)  	                & 5.77  $\pm$ 0.12  & --- & --- & 7.08 $\pm$ 0.11  & --- & ---  & 6.55 $\pm$ 0.03   \\  
log (N/O) 	                                & -1.03 $\pm$ 0.10  & --- & --- & -0.74 $\pm$ 0.28 & --- & ---  & -1.10 $\pm$ 0.06   \\  
12 + log (Ne$^{2+}$/H$^{+}$)  	                & 7.30 $\pm$ 0.20   & --- & --- & 8.01 $\pm$ 0.19  & --- & ---  & 7.40 $\pm$ 0.06   \\  
log (Ne/O) 	                                & -0.62 $\pm$ 0.09  & --- & --- & -0.49 $\pm$ 0.22 & --- & ---  & -0.54 $\pm$ 0.07   \\ 
12 + log (S$^{+}$/H$^{+}$) 	                & 5.19 $\pm$ 0.10  & --- & --- & 6.19 $\pm$ 0.11  & --- & ---  & 5.86 $\pm$ 0.03  \\ 
12 + log (S$^{2+}$/H$^{+}$) 	                & 6.26 $\pm$ 0.09  & --- & --- & 6.18 $\pm$ 0.08  & --- & ---  & 5.98 $\pm$ 0.03  \\   
12 + log (S/H)                                  & 6.47 $\pm$ 0.10  & --- & --- & 6.67 $\pm$ 0.09  & --- & ---  & 6.26 $\pm$ 0.03  \\
12 + log (Ar$^{2+}$/H$^{+}$) 	                & 5.58 $\pm$ 0.11  & --- & --- & 6.17 $\pm$ 0.08  & --- & ---  & 5.76 $\pm$ 0.03  \\ 
12 + log (Ar/H)                                 & 5.73 $\pm$ 0.11  & --- & --- & 6.21 $\pm$ 0.08  & --- & ---  & 5.79 $\pm$ 0.03  \\
log ([OII]/[OIII])\tablefootmark{c} & -1.18 $\pm$ 0.06 & -0.03 $\pm$ 0.02 & 0.61 $\pm$ 0.24 & -0.82 $\pm$ 0.24 & 0.45 $\pm$ 0.04 &  -0.82 $\pm$ 0.09 & -0.44 $\pm$ 0.03  \\  
log ([SII]/[SIII])\tablefootmark{d} & 0.03 $\pm$ 0.07  & 0.09 $\pm$ 0.03 & 1.02 $\pm$ 0.14 & 1.06 $\pm$ 0.11 &1.09 $\pm$ 0.04  & 0.85 $\pm$ 0.21 & 0.92 $\pm$ 0.03 \\ \hline 
 
\end{tabular} 
\end{minipage} 
\tablefoot{
\\
\tablefoottext{a}{Nebular properties of MA 1 were derived using the emission-line fluxes from the inner, HeII-emitting part of the nebula (see Section \ref{wr14_cloudy} for details).}\\
\tablefoottext{b}{MA 1, HBW673, BCPLM651: O/H abundance derived using $t_{\rm e}$([OIII]); BCLMP49, BCLMP40, BCLMP720,BCLMP715: O/H derived from Pettini \& Pagel (2004) - O3N2 method.}\\
\tablefoottext{c}{[OII]$\lambda $3727/[OIII]$\lambda$$\lambda$4959,5007.}\\
\tablefoottext{d}{[SII]$\lambda$$\lambda$6717,6731/[SIII]$\lambda$9069}%\end{landscape} 
}
\end{table*} 
%\footnote{O/H abundance derived using $T_{\rm e}$[OIII]} 
 
For our seven HII regions, we calculated the physical properties of 
the ionized gas in the HeII-emitting zone (see Figure~\ref{finding_charts}).   
Physical conditions and chemical abundances for the HII regions are given in 
Table~\ref{np}. The quoted uncertainties account for measurement and reddening 
coefficient errors. Electron densities, electron temperatures and 
ionic abundances of the ionized gas were derived using the IRAF nebular package 
(\citealt{sd94}).  
We obtained the electron densities, $n_{\rm e}$, from the [SII]$\lambda 
$6717/$\lambda$6731 line ratio.  The derived estimates and upper limits for 
$n_{\rm e}$ place all of the HII regions in the low-density regime ($n_{\rm 
  e}$ $\lesssim$ 100 cm$^{-3}$). 
 
We were able to measure the faint auroral line [OIII]$\lambda$4363 from the
spectra of the three confirmed HeII nebulae: MA 1, BCLMP 651, HBW 673.  For
these three objects, we derived the electron temperatures $t_{\rm e}$([OIII])
($t_{\rm e}$ = 10$^{-4}$ x $T_{\rm e}$) 
using the [OIII]$\lambda$4363/[OIII]$\lambda\lambda$4959,5007 line ratio. For
MA 1, since there is a gradient in $T_{\rm e}$ between the innermost,
HeII-emitting region (13,500 K) and the outer region (10,000 K), we measure
the physical properties and chemical abundances from the innermost region.
Also, we find MA 1 to be density bounded, and so the ionic species are more
co-spatial (see Section \ref{wr14_cloudy} for more details on the MA 1
nebula); we therefore derive the chemical abundances of the different
elements assuming a single, observed temperature.  

For BCLMP 651 and HBW673,
$t_{\rm e}$([OII]) was calculated from the relation between [OII] and [OIII]
electron temperatures given by \cite{p06}:
\mbox{$t_{\rm e}$([OII]) = 0.72$t_{\rm 
    e}$([OIII])+0.26}. Since no auroral line could be measured in the 
low excitation zone, we assumed the 
approximation $t_{\rm e}$([SII]) $\approx$ $t_{\rm e}$([NII]) $\approx$ 
$t_{\rm e}$([OII]) to be valid. As we did not detect the line 
[SIII]$\lambda$6312 in any of our objects, a theoretical relation between 
[OIII] and [SIII] electron temperatures was used for BCLMP651 and HBW673: \mbox{$t_{\rm e}$([SIII]) = 
  1.05$t_{\rm e}$([OIII]) - 0.08} (\citealt{pd05}).  This 
relation differs slightly from the empirical relation found by \cite{g92}, 
mostly due to the introduction of the new atomic coefficients for S$^{2+}$ 
from \cite{tg99}.     
 
In order to derive the chemical abundances we followed as summarized below.

{\it Oxygen.} A small fraction of oxygen is 
expected to be in the form of the O$^{3+}$ ion in HII regions which show 
high-ionization emission lines like HeII$\lambda$4686. According to the 
photoionization models from \cite{i06b}, the O$^{3+}$/O is $\geq$ 
1$\%$ only in the highest-excitation HII regions for which 
\mbox{O$^{+}$/(O$^{+}$ + O$^{2+}$) $\lesssim $ 10$\%$}, as in the case of MA 1 
and HBW673. Thus, in order to derive the total oxygen abundances for these two 
objects, we proceed as following. For MA 1 we assumed O$^{3+}$/(O$^{+}$ + O$^{2+}$) = 4.43$\%$, as obtained 
from a detailed photoionization model of this object (see
Section\ref{wr14_cloudy}). 

In the case of HBW673, we added the fraction of the
unseen O$^{3+}$ ion in its oxygen abundance following the prescription
by \cite{i06b}, that gives \mbox{O$^{3+}$/(O$^{+}$ + O$^{2+}$) $\sim$ 0.19.} 
We also checked that using the oxygen ionization correction factor (ICF) from the MA1 photoionization
model shifts the total oxygen abundance of
HBW673 by 0.06 dex, that is within the quoted uncertainty. For BCLMP651 we set 
O$^{3+}$/H$^{+}$ to zero, since it shows \mbox{O$^{+}$/(O$^{+}$ + O$^{2+}$) 
  $>$ 10$\%$}, and assume \mbox{O/H = (O$^{+}$/H$^{+}$) +
  (O$^{++}$/H$^{+}$)}. 

%{\bf
%\begin{displaymath}
%\frac{\rm O^{3+}}{\rm H^+} = 0.5\times \frac{\rm He^{2+}}{\rm He^{+} + He^{2+}}\left(\frac{\rm O^{+}}{\rm H^+} + \frac{\rm O^{2+}}{\rm H^+}\right)   
%\end{displaymath} 
%}	

%{\bf that gives O$^{3+}$/(O$^{+}$ + O$^{2+}$) $\sim$ 0.19.}

%Helium abundances were calculated with the formulations of \cite{b99}.

To 
obtain the oxygen abundances in those HII regions in which [OIII]$\lambda$4363 
could not be measured, we applied a commonly used, strong-line method from 
\cite{pp04}: the O3N2 index 
[log{([OIII]$\lambda$5007/H$\beta$)/([NII]$\lambda$6584/H$\alpha$)}], whose 
associated uncertainty is $\sim$ 0.25 dex. Nevertheless one should bear in
mind that calibrations of different strong-line indices provide different
metallicities (see \citealt{b09} and references therein for a more detailed discussion).

{\it Sulfur and argon.} We derive S$^{+}$/H$^{+}$ abundances from the fluxes of 
the [SII]$\lambda$$\lambda$6717,6731 emission lines.  Abundances of S$^{2+}$/H$^{+}$ were obtained from the 
flux of the [SIII]$\lambda$9069 line. We calculate the abundance of Ar$^{2+}$/H$^{+}$ from the emission of 
[ArIII]$\lambda$7136, assuming $t_{\rm e}$([ArIII]) $\approx$ $t_{\rm 
  e}$([SIII]) (\citealt{g92}). To derive the total argon and sulfur abundances
  for BCLMP651 and HBW673, we adopted the
metallicity-dependent ICFs of \cite{i06b}. In the case of MA 1, we assume
\mbox{Ar$^{+}$/Ar$^{2+}$ = 0.016}, \mbox{Ar$^{3+}$/Ar$^{2+}$ = 0.378}, and
\mbox{S$^{3+}$/(S$^{+}$ + S$^{2+}$) = 0.512}, taken from our photoionization
modeling of the MA 1 nebula (see Section \ref{wr14_cloudy}).

{\it Nitrogen and neon.} We calculated the N/O and Ne/O abundance ratios assuming that 
N/O=N$^{+}$/O$^{+}$ and Ne/O = Ne$^{++}$/O$^{++}$, based on the 
similarity of the ionization potentials of the ions involved
(\citealt{pc69}). This ICF scheme has been used in previous work
(e.g., \citealt{ls07}; \citealt{h08}; \citealt{b09}) and
is a good approximation for high ionization degree objects.    
In our spectra, neon is seen via the [NeIII] emission 
line at 3869 \AA. For this ion, we also took the [OIII] electron temperature as 
representative of the high-excitation zone, in BCLMP 651 and HBW673.

Figure~\ref{plots_2} shows the standard [OIII]$\lambda$5007/H$\beta$ versus
[NII]$\lambda$6584/H$\alpha$ diagnostic diagram (\citealt{bpt81}; hereafter BPT)
for our sample of seven candidates HeII-emitting nebulae.  The solid curve,
from \cite{k03}, separates objects photoionized by purely thermal (stellar)
sources from regions that are excited by additional processes.  The majority
of our objects are found in the zone of objects photoionized by thermal sources.  However,
\mbox{HBW 673}, one of our HeII nebulae, is the only object located outside of
this zone (see Section \ref{hbw673} for a more detailed discussion of HBW 673).
  
\section{Description of the individual HeII nebulae}\label{confirmed_heii_neb} 
 
\subsection{MA 1}\label{ma1}
 
This object is Number 1 in the list of emission-line nebulae in M33 by 
\cite{ma42}. MA 1 is one of the first HeII nebulae identified in a spiral 
galaxy beyond the Milky Way (\citealt{gos92}) and is one of McCall's four 
``Rosetta Stone'' extragalactic HII regions (\citealt{m82}).  It shows a 
complex morphology of arcs, loops, and filaments having a diameter of $\sim$ 
50$\arcsec$ (200 pc at our assumed distance of 840 kpc) as seen from the H$\alpha$ image 
(see Figure~\ref{finding_charts}). MA 1 is one of the outermost known HII regions in 
M33, located about 30$\arcmin$, or 7.3 kpc, from the nucleus. 
 
We obtain 12 + log(O/H) $\sim$ 8.00 for MA 1, consistent with previous 
work, which demonstrated the low O/H of this nebula (e.g., \citealt{m85}; 
\citealt{gos92}; \citealt{c06}).  The line intensities 
in this HII region are similar to values seen in metal-poor HII galaxies (see 
\citealt{k04}, Figure 14).  The presence of HeII$\lambda$4686 
emission in MA 1 was reported by \cite{gos92} and \cite{c06}. The latter 
showed that the HeII emission is nebular in nature but they did not identify 
the source of He$^{+}$-ionizing photons in MA 1.  The high 
log([OIII]$\lambda5007$/H$\beta$) of $\sim$ 0.78 (see Table~\ref{lines1}), 
confirms the unsually high excitation and ionization state of this nebula
(e.g., RS08), which is unusual for HII regions with similar O/H. 
Our spectrum of MA 1 confirms that the nebular HeII emission is 
spatially extended (Figure~\ref{2dimage}); its observed
HeII$\lambda$4686/H$\beta$ = 0.19 is consistent with that found for some other
metal-poor HII regions excited by WNE stars (e.g., N76 in 
the SMC; see Table~\ref{heii_lg}).  Below, we investigate whether the 
early-type WN star MC8 [\cite{mc83} 
alias M33-WR14 (MJ98)], 
is capable of supplying the high energy photons required for He II nebular emission 
from MA 1. 
 
\subsubsection{Stellar atmosphere models for the WN source MC8}\label{wr14_atmosphere} 
 
Here, we investigate whether the early-type WN star MC8 [\cite{mc83}
alias M33-WR14 (MJ98)], is capable of supplying the high energy photons
required for He II nebular emission from MA 1. Stellar atmosphere models
are constructed using the code CMFGEN (\citealt{hm98}), which solves the radiative transfer 
equation in the co-moving frame, under the additional constraints of 
statistical and radiative equilibrium.

From Figure~\ref{finding_charts}, the early type WN source MC8 
is by far the brightest source within 
the MA 1 nebula. Indeed, ground-based photometry from \citet{m06} indicates 
V = 16.14 mag, from which E(B-V)=0.06 mag (via $C_{H\beta}$ = 0.09) implies  
$M_{\rm V}$ = --8.7 mag for our assumed distance of 840 kpc. In contrast, absolute magnitudes of 
WN-type stars in the Large Magellanic Cloud range from --2 to --7.5 mag 
(\citealt{crm07}), with $M_{\rm V}$ = --4 mag typical for single
early-type WN stars.  In addition, the stellar He II  $\lambda$4686
equivalent width is \mbox{$W_{\lambda}$ $\sim$ 3 \AA}, versus \mbox{30--500 \AA} for LMC single, early-type
WN stars (\citealt{ch06}). The properties of MC8 are therefore consistent
with a WN star located within a compact, young star cluster such as Brey 65 in the LMC
(\citealt{W99}), in which the He II $\lambda$4686 emission strength 
is severely diluted by other cluster members (mostly O, B and A stars). Alternatively,
could MC8 be an Of supergiant, since they possess much weaker He II $\lambda$4686 emission 
than WN stars plus brighter absolute magnitudes
of $M_{\rm V}$ = --7 mag. We are able to discriminate between these two possibilies since
Of supergiants possess narrow $\lambda$4686 emission, with \mbox{FWHM$\sim$2--7\AA} (\citealt{c02b}) versus \mbox{FWHM $\sim$= 25-60 \AA} for early-type WN stars (\citealt{ch06}).
From our MC8 Gemini GMOS spectrum, He II $\lambda$4686 emission is broad {\mbox (FWHM $\sim$ 30 \AA)}, 
allowing us to conclude that MC8 is a probable cluster hosting an early-type WN star.

Therefore, in our analysis of MC8 we are unable to fix the absolute magnitude of the
early-type WN star, and instead focus upon reproducing the line luminosity of He II 
$\lambda$4686. 
We have convolved the GMOS spectrum of MC8 with standard filters, from which 
synthetic photometry of V = 17.35 mag (and B--V = 0.09 mag) is obtained, indicating that
\mbox{$\sim$ 50 $\%$} of the continuum light from the MC8 source is included within the slit 
(recall Figure~\ref{finding_charts}). Fortunately, since the WN star is included, we are able to estimate 
its He II $\lambda$4686 line luminosity. To assess the extinction, we have compared the GMOS 
spectrum to a 3 Myr old, instantaneous  burst, LMC metallicity starburst99 (\citealt{L99}) 
evolutionary  synthesis model. An extinction of E(B-V) = 0.30 mag is required to reproduce the 
spectral energy distribution of the GMOS dataset (Figure~\ref{figure_paul}), from which
we obtain a $\lambda$4686 line luminosity of  4.5 $\times$ 10$^{35}$ erg\,s$^{-1}$ for the WN star, versus
an average of 8.4 $\pm$ 7.1  $\times$ 10$^{35}$ erg\,s$^{-1}$ for LMC
early-type WN stars (\citealt{ch06}).

In addition, we are unable to derive a stellar temperature for the WN
star since $\lambda$4686 is the sole diagnostic available to us. In view of the 
limitations imposed  by the sparse continuum and line constraints, 
we have calculated generic stellar atmosphere models for the WN star hosted by MC8 
using CMFGEN. Our approach mimics the study of HD 9974 (WN3) by \cite{m04}
except for more extensive model atoms: H I, He I-II, C IV, N 
III-V, O III-VI, Ne IV-V, Si IV, S IV-V, Ar V-VII, Ca V-VII, Fe IV-VII, Ni 
IV-VII, totalling 822 super-levels (23,098 lines). Other than H, He, CNO 
we adopt solar abundances (\citealt{a09}), scaled to 0.3 Z$_{\odot}$ 
(see Table~\ref{np}). In the absence of CNO features, 
we adopt equilibrium WN abundances, namely X(N)=0.2\%, X(C)=X(O)=0.03\%.  
With regard to wind 
clumping (\citealt{h91}), this is incorporated using a radially-dependent 
volume filling factor, $f$, with  $f_{\infty}$ = 0.1 at terminal velocity $v_{\infty}$, 
resulting in a reduction in mass-loss rate by a factor of $\sqrt{(1/f)}$ $\sim$ 3.

In the absence of Wolf-Rayet temperature diagnostics from our GMOS dataset of MC8, 
\citet{R08} has undertaken spectral analyses of 
all WN stars in the LMC including 50 WN3--4 stars. From this comprehensive study,
an average of $T_{\ast}$ = 90 $\pm$ 30 kK was obtained that we shall adopt here for the
WN star in MA8. As discussed previously (e.g., \citealt{s92}; \citealt{s02}), the
wind density of WN atmospheres can dictate the hardness of their ionizing fluxes more
significantly than stellar temperature. We demonstrate this effect by selecting three 
stellar luminosities of the WN component, namely log {\it L}/$L_{\odot}$ = 5.3, 5.6 and 5.9, which 
correspond to stellar masses of 10, 16 and 25 $M_{\odot}$ (\citealt{sm92}), spanning 
a representative range for WN stars (\citealt{c07}). 
For these three cases, the (clumped) mass-loss rate required to 
reproduce the He II $\lambda$4686 feature was relatively uniform, 0.8, 1.1, 1.2 $\times 
10^{-5}$ $M_{\odot}$ yr$^{-1}$, respectively, with $v_{\infty} \sim$ 1750 
km\,s$^{-1}$ for each case. However, since the stellar luminosity varies by a 
factor of four, the wind becomes progressively more transparent at higher 
luminosity, hardening the ionizing spectrum, as revealed by the number of 
He$^{+}$ ionizing photons, $Q_{2}$, shown in  Table~\ref{wn} for 
each model atmosphere. In Figure~\ref{figure_paul} we compare 
the de-reddened GMOS spectrum of MC8 with both the cluster spectral energy distribution
and high luminosity WN model.
 
\begin{table}
\caption{Physical properties of WN atmosphere models for MC8.}
\label{wn}
\centering
\begin{tabular}{lccc}
\hline\hline
Parameter & log $L$/$L_{\odot}$ = 5.3  & log $L$/$L_{\odot}$= 5.6 & log $L$/$L_{\odot}$ = 5.9
\\ \hline
%log L/L$_{\odot}$   & 5.3  &   5.6 &    5.9 \\
$T_{\ast}$ (kK)$^{a}$        & 90   &   90  &    90 \\
$T_{eff}$ (kK)$^{a}$      & 78   &   80  &    85 \\
dM/dt (10$^{-5}$ M$_{\odot}$ yr$^{-1}$)       & 0.8 & 1.1 &  1.2 \\
M$_{\rm V}$ (mag)         & -3.4 &   -4.0 &   -4.4  \\
log Q$_{0}$      & 49.16 &   49.47 &   49.76 \\
log Q$_{1}$      & 48.92 &   49.24 &   49.55 \\
log Q$_{2}$      & 38.40 &  46.46 &   48.32 \\ \hline
\end{tabular}
\tablefoot{\tablefoottext{a}{The stellar temperature, $T_{\ast}$, is derived from the Stefan-Boltzmann relation, with the
stellar radius defined at a Rosseland optical depth of $\tau$=20. The
effective temperature, $T_{eff}$, is similarly derived, using the effective radius defined
at $\tau$=2/3.}}

%The stellar (effective) radius is defined
%at a Rosseland optical depth of 20 (2/3), with the stellar (effective)
%temperature, $T_{\ast}$ ($T_{eff}$) resulting from the Stefan-Boltzmann
%relation.

\end{table}
 
\subsubsection{Photoionization modeling of the MA 1 nebula}\label{wr14_cloudy} 
 
We now model the MA1 HII region with the photoionization code CLOUDY,
v. C08.00 (\citealt{f98}), using these stellar atmosphere models for MC8 to
define the ionizing spectral energy distribution.  While we do not know which,
if any, of the three atmosphere models is representative of MC8, our purpose
here is to investigate feasibility.  We adopt the measured chemical abundances
in the central region listed in Table~\ref{np}, which are consistent with the
reported abundance gradient in M33 (e.g., RS08). He II emission is emitted
fairly uniformly within the inner nebula, 22 pc in radius. In addition, the
outer region of the nebula has a complex morphology extending up to a radius
of 60 pc [see Figure~\ref{heii} and Figure~\ref{2dimage} (top panel)].  To
account for these distinct nebular components we construct a two-zone
photoionization model, composed of an inner and outer region, each with a
constant density listed in Table~\ref{bestmodel}.  A constant cosmic ray
energy density of $2.6\times10^{-9} cm^{-3}$ (\citealt{w98}) is included, but
is negligible.

Of the three atmosphere models, solely the model illuminated by the highest
luminosity \mbox{(log $L/L_{\odot}$ = 5.9)} atmosphere is capable of
reproducing the observed HeII emission, for the observed ionization parameters
log $U$ $\geq -2.0$, where $U =Q^0/4\pi r^2 cn_e$, $Q^0$ is the H-ionizing
photon emission rate, and $r$ is the Str\"omgren radius.  The parameters for
the two parts of this model nebula are listed in Table~\ref{bestmodel}.We self
consistently use the emergent radiation transmitted through the highly ionized
gas of the inner region as the ionization source of the outer region beyond
the He II emission, accounting for an observed difference in $T_e$ between the
inner region (13,500 K) and outer region (10,000 K).  The two lower-luminosity
WN atmospheres predict the nebular intensity of HeII$\lambda$4686/H$\beta$ to
be orders of magnitude lower than observed, and they are not capable of
producing nebular He II $\lambda$4686 emission with realistic choices of
metallicity and $U$.

Table~\ref{narrowbeamfluxes} shows good agreement between the observed nebular 
line strengths and the photoionization model.  The standard deviation between 
the observed and predicted emission is 17$\%$, excluding [SIII]$\lambda9069$, 
which is affected by telluric absorption. These models demonstrate that the 
SED of the log $L/L_{\odot}$ = 5.9 atmosphere is a plausible ionization source for the 
surrounding nebula, without a need to invoke additional ionization sources. 
We note that the central line of sight is optically thin: the 2-D morphology 
revealed by the images and the weakness of [OII] relative to [OIII] indicates 
that an $H^+$ ionization front is not present in the line of sight toward the 
He II nebula, consistent with our model. Early spectral observations 
by \cite{m82} also suggested parts of the nebula to be density bounded 
based on the [OII] / [OIII] ratio. 
 
While neither our model stellar atmosphere nor model nebula is necessarily 
meant be an accurate physical representation of this region, the reasonable agreement 
demonstrates that the observed star is indeed capable of reproducing the observed 
nebula.   
 
\begin{table} 
\caption{Nebular properties of MA 1 zones} 
\label{bestmodel} 
\centering 
\begin{tabular}{lcc} 
\hline\hline 
Parameter                 & Inner region & Outer region\\ 
\hline 
Log $n(H)_0$ (cm$^3$) & 1.85 & 0.95 \\ 
Log $r_0$ (cm)        & 19.50 & 19.91 \\ 
Log $r_{outer}$ (cm)      & 19.78 & 20.09 \\ 
$v_{turb}$ (km/s)        & 4.0   & 4.0 \\ 
12 + log (He/H)       & 11.0  & 11.0 \\ 
12 + log (O/H)        & 8.00  & 8.00 \\ 
12 + log (N/H)        & 6.97  & 6.97 \\ 
12 + log (Ne/H)       & 7.38  & 7.38 \\ 
12 + log (S/H)        & 6.47  & 6.47 \\ 
12 + log (Ar/H)       & 5.73  & 5.73 \\ 
\hline 
\end{tabular} 
\end{table} 
 
\begin{table*} 
\caption{Dereddened line fluxes relative to F(H$\beta$)=100 for central zone of MA 1} 
\label{narrowbeamfluxes} 
\centering 
\begin{tabular}{lcccc} 
\hline\hline 
Wavelength &MA 1 Inner & Inner model & MA 1 Outer& Outer model \\ 
\hline 
3727 [O~II]     & 54.0 $\pm$ 7.0    & 56.0   &  78.6 $\pm$ 1.7 & 81.7 \\ 
3868 [Ne~III]   & 55.0 $\pm$ 10.0   & 70.0   &  47.3 $\pm$ 1.0 & 64.0 \\ 
4363 [O~III]    & 9.00 $\pm$ 3.00   & 10.00   &  3.50 $\pm$ 0.50 & 8.00 \\ 
4686 He~II      & 19.0 $\pm$ 2.0    & 17.0   &  0.10 $\pm$ 0.20 & 0.30 \\ 
4861 H$\beta$   & 100  $\pm$ 1      & 100    &  100  $\pm$ 1 & 100 \\ 
5007 [O~III]    & 602  $\pm$ 30     & 612    &  556  $\pm$ 28 & 569 \\ 
5876 He~I       & 9.00  $\pm$ 2.00  & 10.70   & 13.0 $\pm$ 2.6 & 12.2 \\ 
6584 [N~II]     & 6.00  $\pm$ 1.00  & 6.40   & 12.1 $\pm$ 1.2 & 10.1 \\ 
6678 He~I       & 4.00  $\pm$ 1.00  & 3.00   & 2.60 $\pm$ 0.40 & 3.40 \\ 
6717,6731 [S~II] & 13.0  $\pm$ 1.0  & 13.4   & 26.5 $\pm$ 2.0 & 19.0 \\ 
7136 [Ar~III]   & 8.00  $\pm$ 1.00  & 8.40   & 8.90 $\pm$ 0.40 & 9.30 \\ 
9069 [S~III]    & 12.0 $\pm$ 4.0    & 21.1   & 29.1 $\pm$ 1.2 & 23.6 \\ 
%$C_{H\beta}$     & 0.10 $\pm$ 0.04    &       & 0.000 $\pm$ 0.002 &      \\ 
\hline 
\end{tabular} 
\end{table*} 
 
%%% HII-Z29 %%%%% 
 
\subsection{HBW673}\label{hbw673} 
 
HBW673 is a relatively faint and small HII region about 2 kpc from the center 
of the galaxy.  Its diameter is roughly 20$\arcsec$ ($\sim$ 80 pc at our assumed 
distance of 840 kpc), as measured from the $H\alpha$ image, and in contrast to MA 
1, HBW673 does not show a complex morphology (see 
Figure~\ref{finding_charts}). From Figure~\ref{2dimage} (bottom) we can see 
that there is no obvious star associated with HBW673. The spectrum of HBW673 
has very strong emission at $\lambda$4686 showing one of the highest values of 
HeII$\lambda$4686/H$\beta$=0.55 observed among the local HeII nebulae 
(see Table~\ref{heii_lg}). In addition, HBW673 exhibits extremely high 
excitation in [OIII]$\lambda5007$ and in [NeIII]$\lambda3868$ (see 
Table~\ref{lines1}).  
 
The spectrum of HBW673 is not that of a SNR, lacking the variety of different
ions that are shock-excited in classical SNRs (e.g., \citealt{b81}). A PN-type
nature of HBW673 is also ruled out. The typical size of a Galactic PN is some
tenths of a parsec (e.g., \citealt{p90}), while evolved PNe have been found
with sizes of up to 4 pc (e.g., \citealt{c97}). At our assumed distance of 840 kpc to M33, 1$\arcsec$ corresponds to 4 pc and therefore any extended
emission region in our images is not considered to be a PN. Such high relative
fluxes of the HeII$\lambda$4686 emission line indicate that radiation with
photon energies greater than 4 Rydberg is intense, and difficult to explain by
ionizing stellar radiation (e.g., \citealt{ti05}; \citealt{i04}). In addition,
the BPT diagnostic diagram (Figure~\ref{plots_2}) for our sample of HII
regions shows that HBW 673 is outside the zone of normal HII regions. This
suggests that the main excitation mechanism in HBW673 is not a hot stellar
continuum, but another ionization source. HBW673 is not coincident with
any known X-ray sources.

%%%%%% HII651 %%%%%%% 

\subsection{BCLMP651} 
 
The third HeII nebula we identified is BCLMP651 with a diameter of $\sim$ 
30$\arcsec$ (120 pc at our assumed distance of 840 kpc). The nebular morphology consists 
of a bright knot on the southeast side, and a second, fainter one on the 
northwest side of the nebula. For BCLMP651 there does not appear 
to be any WR star responsible for the extended HeII emission.  
Figure~\ref{2dimage} (middle) shows that the nebular HeII emission is placed 
between two continuum sources.  We were not able to detect any obvious stellar 
features in our spectra of these objects, which only have S/N $\sim 10$.   
Like the other HeII nebulae, BCLMP651 is also a high-excitation 
object, with e.g., high [NeIII]$\lambda3868$/H$\beta$=0.40, in comparison 
to the other five HII regions for which we do not see nebular HeII emission 
(see Table~\ref{lines1}).  This object also is not coincident with any 
known X-ray source. 
   
%%%%%%%%%%%%%%%%%%%%%%%%%%%%%%%%%%%%%%%%%%%%%%%%%%%%%%%%%%%%%%%%%%%%%%%%%%%%% 
%%%%      Discussion                                                     %%%% 
%%%%%%%%%%%%%%%%%%%%%%%%%%%%%%%%%%%%%%%%%%%%%%%%%%%%%%%%%%%%%%%%%%%%%%%%%%%%% 
\section{Discussion}\label{discuss} 
 
From our sample of seven candidate HeII nebulae in 
M33, we confirm the presence of nebular HeII$\lambda$4686 emission 
line in the spectra of three objects: MA 1, HBW673 and BCLMP651.  
In addition to MA 1, there are four more 
confirmed He II nebulae ionized by WN stars in the Local Group: LMC N79 (\citealt{p91}), LMC 
N206 (\citealt{n91}), SMC N76 (\citealt{tp89}; \citealt{g91}) and 
NGC 249 (\citealt{p09}) in the SMC.  
%This supports the idea that a WN star 
%can be hot enough to radiate photons more energetic than 4 Rydbergs (54 eV).  
However, WN stars are not the only sources capable 
to ionize HeII. The Galactic nebula G2.4+1.4 and the S3 nebula in 
IC 1613 exhibit nebular HeII emission associated with WO stars 
(\citealt{j75}; \citealt{d90}; \citealt{dk82}; \citealt{g91}), while the nebular He II 
emission in LMC N44C is puzzling since emission 
is centered on an O7 star (\citealt{g00}) which is  
not expected to be hot enough to produce He$^{+}$-ionizing 
photons. \cite{pm89} have proposed N44C to be a fossil X-ray ionized nebula, but a 
clearer view of this object is still needed (see \citealt{g00})

Among our sample, the HeII nebulae HBW673 and BCLMP651 are also intriguing 
objects.  The former does not seem to be associated with any hot 
star.  Yet, except for the presence of strong nebular HeII emission, it 
shows a fairly normal, highly excited, HII region spectrum. 
Moreover, electron temperatures in these nebulae are not unusally 
high \mbox{($\sim$ 12,000 K)}. 
 
RS08 have analysed Keck spectra of 61 HII regions in M33, and detected
HeII$\lambda$4686 emission from five of these objects (BCLMP38b, C001Ab, BCLMP711a,
BCLMP90, BCLMP208f). They claim the HeII$\lambda$4686 emission is almost certainly
nebular as it is spatially resolved in these five HII regions. Their strongest
source, BCLMP38b, was also detected in our continuum-subtracted He II image,
but it was compact and unresolved, and so we were not able to identify it as
an extended He II nebula.  \cite{mv07} also report the existence of HeII
emission in three M33 HII regions (BCLMP258b, BCLMP220, BCLMP45). Doing a coordinate cross-reference with the catalog of WR stars in
M33 (\citealt{m06}), we find that with the exception of BCLMP38b, no HeII nebula
presented by RS08 and \cite{mv07} appears to be associated with a WR star.
The object BCLMP38b  appears to be associated with the WC star MC45 (M33-WR76),
which might be responsible for the ionization in the nebula, although no WC
stars are currently known to ionize He II nebulae. A more detailed study is
necessary to assess whether MC45 is the He II-ionizing source.

%They do not discuss whether the HeII$\lambda$4686 emission line is nebular or
%stellar in nature.
 
\begin{table*}%{ht} 
%\begin{landscape} 
%\caption{HII Regions with nebular He II Emission in the Local Group\protect\footnotemark[1]}
\caption{HII regions with nebular He II emission in the Local Group} 
\label{heii_lg} 
\centering 
\begin{minipage}{20.0cm} 
\centering 
\begin{tabular}{lcccccc} 
\hline\hline 
Galaxy & Region & 12+log(O/H)  & Ionizing source & Spectral type & HeII$\lambda$4686/H$\beta$ [F(H$\beta$)=100] &References\tablefootmark{b} \\ \hline 
MW  & G2.4+1.4 & 8.45  & WR102 & WO2   & 40-120  & 7, 3, 4, 17      \\ 
LMC & N44C  & 8.32 &X-5? & X-ray Neb?   & 2-14  & 14, 6  \\ 
LMC & N159  & 8.36\tablefootmark{a} & X-1 & HMXB  & 5  & 13, 5          \\ 
LMC & N79   & 8.17-8.27  &BAT99-2 & WN2b(h)  & 21-56 & 15, 11    \\ 
LMC & N206  & 8.36 & BAT99-49 & WN4:b+O8V  & 8-10 & 12, 10  \\ 
SMC & N76   & 7.93  & AB7 & WN4+O6I(f)& 16-24  & 19, 5, 10   \\ 
SMC & NGC249 & 8.11\tablefootmark{a}& SMC-WR10 & WN3ha & & 16 \\ 
IC1613 & S3  & 7.70 & &   WO3  &  23 & 2, 5, 9 \\ 
M33 & BCLMP38b & 8.39  & MC45 & WC4         & 17 & 18, 1  \\ 
M33 & BCLMP90  & 8.50  & &no obvious hot star associated & 8  & 18 \\
M33 & C001Ab & 8.61 & &no obvious hot star associated & 2  & 18 \\
M33 & BCLMP208f & 8.07  & &no obvious hot star associated & 3  & 18 \\
M33 & BCLMP711a & 8.28 & &no obvious hot star associated & 4  & 18 \\ 
M33 & MA 1  & 8.00  & MC8 & WNE  & 19 & 8 \\ 
M33 & HBW673 & 8.66 & & no obvious hot star associated & 55 & 8  \\ 
M33 & BCLMP651 & 8.12 & & no obvious hot star associated & 11 & 8 \\ 
\hline 
\end{tabular}
\end{minipage}
\tablefoot{
\\
\tablefoottext{a}{For LMC N159 and SMC NGC249, since no metallicity measurement has been reported in
the literature,  we show the average values of the LMC and SMC oxygen abundance from \cite{rd90}}\\
\tablefoottext{b}{(1) \cite{A04}; (2) \cite{dk82}; (3) \cite{d90}; (4) \cite{e92}; (5) \cite{g91}; (6) \cite{g00}; (7) \cite{j75}; (8) Kehrig et al. (this paper); (9) \cite{kb95}; (10) \cite{n03a}; (11) \cite{n03b}; (12) \cite{n91}; (13) \cite{pa86}; (14) \cite{pm89}; (15) \cite{p91}; (16) \cite{p09}; (17) \cite{p95}; (18) \cite{rs08}; (19) \cite{tp89}}}

%\footnotetext[2]{(1) \cite{j75}; (2) \cite{dk82}; (3) \cite{pa86}; (4) \cite{pm89}; (5) \cite{tp89}; (6) \cite{d90}; (7) \cite{g91};  (8) \cite{n91}; (9) \cite{p91}; (10) \cite{kb95}; (11) \cite{p95}; (12) \cite{g00}; (13) \cite{n03a}; (14) \cite{n03b}; (15) \cite{A04}; (16) \cite{rs08}; (17) \cite{p09}; (18) this paper} 
% \medskip Notes: The HII regions with HeII emission detected by \cite{mv07} are not listed since the nature (stellar or nebular) of the HeII$\lambda$4686 line in these objects is not clear (see Section \ref{discuss} for details).
%\end{landscape} 
\end{table*} 
   
Table~\ref{heii_lg} summarizes the HII regions with {\it confirmed} nebular 
He II $\lambda$4686 emission detected so far in the Local Group.   
The ionizing sources are apparently quite heterogenous. 
As discussed previously, the HeII-ionizing sources of M33 HBW673, M33 BCLMP651 
and LMC N44C are still unknown. In several studies of nebular HeII emission in BCDs,  Izotov and collaborators
(e.g., \citealt{ti05}; \citealt{i06a}) find the 
presence of nebular HeII$\lambda$4686 emission even at large distances 
from young clusters, indicating that mechanisms other than stellar 
photoionization (e.g., shocks or high-mass X-ray binaries) seem to be 
present, at least in some objects.  A similar result was found
by \cite{m10}. These authors studied the ionized gas  
and massive stellar population in the HII galaxy NGC 5253, and found that 
the nebular HeII$\lambda$4686 emissions to be, in general, not 
coincident with WR stars.  One can easily 
envision a mixture of different contributions, including possibly from 
SNRs, to the observed HeII emission beyond the Local
Group. We should also stress that from our survey for 
HeII nebulae in M33, we identified five SNRs with HeII emission. 
Thus, SNRs or shock-excited objects (e.g., \citealt{p10}) can 
also be responsible for some HeII nebulae.  
 
Even among the WR HeII nebulae, the ionizing stars show different spectral 
types (Table~\ref{heii_lg}) and at present, it is not possible to look for 
systematic properties that differentiate the HeII-ionizing WR stars from the 
remainder of them due to the relatively small number of local WR HeII nebulae 
detected so far. However, an association of nebular HeII emission with WR 
stars (e.g., MA 1) in low-metallicity environments seems to exist. Model 
atmospheres of \cite{snc02} predict that low wind densities favour transparent 
winds such that only weak winds are expected to produce a significant He+ 
continuum below 228 \AA. Since mass-loss rates of WR stars are predicted 
(\citealt{vk05}) and observed to scale with metallicity (\citealt{c02a}), it 
is expected that HeII nebulae are preferentially associated with WR stars in 
metal-poor regions. Indeed, seven of the eight WR stars which exhibit nebular 
HeII are in nebulae with 12 + log(O/H) $<$ 8.4 (see Table~\ref{heii_lg}).

\section{Summary}\label{summary} 
 
To investigate the properties of HeII nebulae and the origin of 
the nebular HeII$\lambda$4686 emission, we have carried out a survey 
for HeII nebulae in the Local Group galaxy M33. Due to its proximity, M33 allows HII 
regions excited by few/single stars to be resolved.  Since M33 shows a well 
defined metallicity gradient, this galaxy also offers the opportunity to probe 
the metallicity dependence of the nebular HeII line formation and strength. 
The rarity of HII regions which show nebular HeII among Local Group galaxies 
illustrates the significance of such regions within M33. 
 
From our HeII narrow-band imaging of M33, 
we identified seven candidates for HeII nebulae, for 
which we obtained follow-up optical spectra. 
We detect nebular HeII emission in the spectra of three of the seven 
candidate HII 
regions: MA 1, BCLMP651, and HBW673, the latter two being newly confirmed HeII 
nebulae.  All three are high-excitation HII regions. 
 
We have assessed whether the observed early-type WN star MC8 is a plausible source 
for the hard ionizing radiation of MA 1 through a combined stellar and nebular 
analysis. A high luminosity, relatively transparent WN wind is capable of 
reproducing the nebular properties, although MC8 is almost certainly an unresolved 
star cluster (cf. Brey 65 and Brey 73 in LMC; \citealt{W99}). 
 
The absence of a hot WR star in BCLMP651 and HBW673 suggests possible 
similarities with N44C and N159 in the LMC, where the high excitation is 
linked to possible X-ray sources within the nebulae. To date, however, no 
X-ray counterpart has been linked to either  HBW673 or BCLMP651.  In HBW673, 
the highest-excitation object in our sample, the presence of a hard radiation 
source other than a hot stellar continuum is indicated by the 
BPT diagnostic diagram (Figure~\ref{plots_2}).  At present we are unable 
to identify the likely source powering the emission in BCLMP 651 and HBW 
673. Further study of these two objects, which contribute to the 
heterogeneity of ionizing sources, is required.  
Additional study of local HII regions with He II emission are needed to lead us 
to a clearer understanding of the sources of nebular HeII emission  
before their applicability to starbursts can be properly assessed.

\bigskip 
 
\noindent \textbf{Acknowledgements}\\ 
 
Our gratitude to the NOAO Gemini Science Center and Gemini Observatory staff,
including Kevin Volk, Dara Norman and Jennifer Holt. We wish to thank the
anonymous referee for his/her useful comments and suggestions. MSO and CK acknowledge
support from the National Science Foundation, grants AST-0448893 and
AST-0806476.  CK, as a Humboldt Fellow, acknowledges support from the
Alexander von Humboldt Foundation, Germany.  OS acknowledges support from the
Science and Technology Facilities Council, UK. DS is supported by the Swiss
National Science Foundation. We thank E. Rosolowsky and
J.D. Simon for providing us with their Keck/LRIS datasets. This work is based
on observations obtained in programs GN-2006B-Q83 and GN-2008B-C5, at the
Gemini Observatory, which is operated by AURA under a cooperative agreement
with the NSF on behalf of the Gemini partnership: the National Science
Foundation (United States), the Science and Technology Facilities Council
(United Kingdom), the National Research Council (Canada), CONICYT (Chile), the
Australian Research Council (Australia), CNPq (Brazil) and CONICET
(Argentina).

%%%%%%%%%%%%%%%%%%%%%%%%%%%%%%%%%%%%%%%%%%%%%%%%%%%%%%%%%%%%%%%%%%%%%%%%%%%%% 
%%%%    References                                                       %%%% 
%%%%%%%%%%%%%%%%%%%%%%%%%%%%%%%%%%%%%%%%%%%%%%%%%%%%%%%%%%%%%%%%%%%%%%%%%%%%% 

%%%%%%%%%%%%%%%%%%%%%%%%%%%%%%%%%%%%%%%%%%%%%%%%%%%%%%%%%%%%%%%%%%%%%%%%%%%%% 
%%%%%%%%%%%%%%%%%%%%%%%%%%%%%%%%%%%%%%%%%%%%%%%%%%%%%%%%%%%%%%%%%%%%%%%%%%%%% 
%\vspace{1cm} 
%\footnotesize{This paper was typeset using a \LaTeX\ file prepared by the author} 
 
%%%%%%%%%%%%%%%%%%%%%%%%%%%%%%%%%%%%%%%%%%%%%%%%%%%%%%%%%%%%%%%%%%%%%%%%%%%%% 
%%%%%%%%%%%%%%%%%%%%%%%%%%%%%%%%%%%%%%%%%%%%%%%%%%%%%%%%%%%%%%%%%%%%%%%%%%%%% 
 
%\clearpage 
 
\newpage
 
\begin{figure*}
\mbox{
  \centerline{
\hspace*{0.0cm}\subfigure{\includegraphics{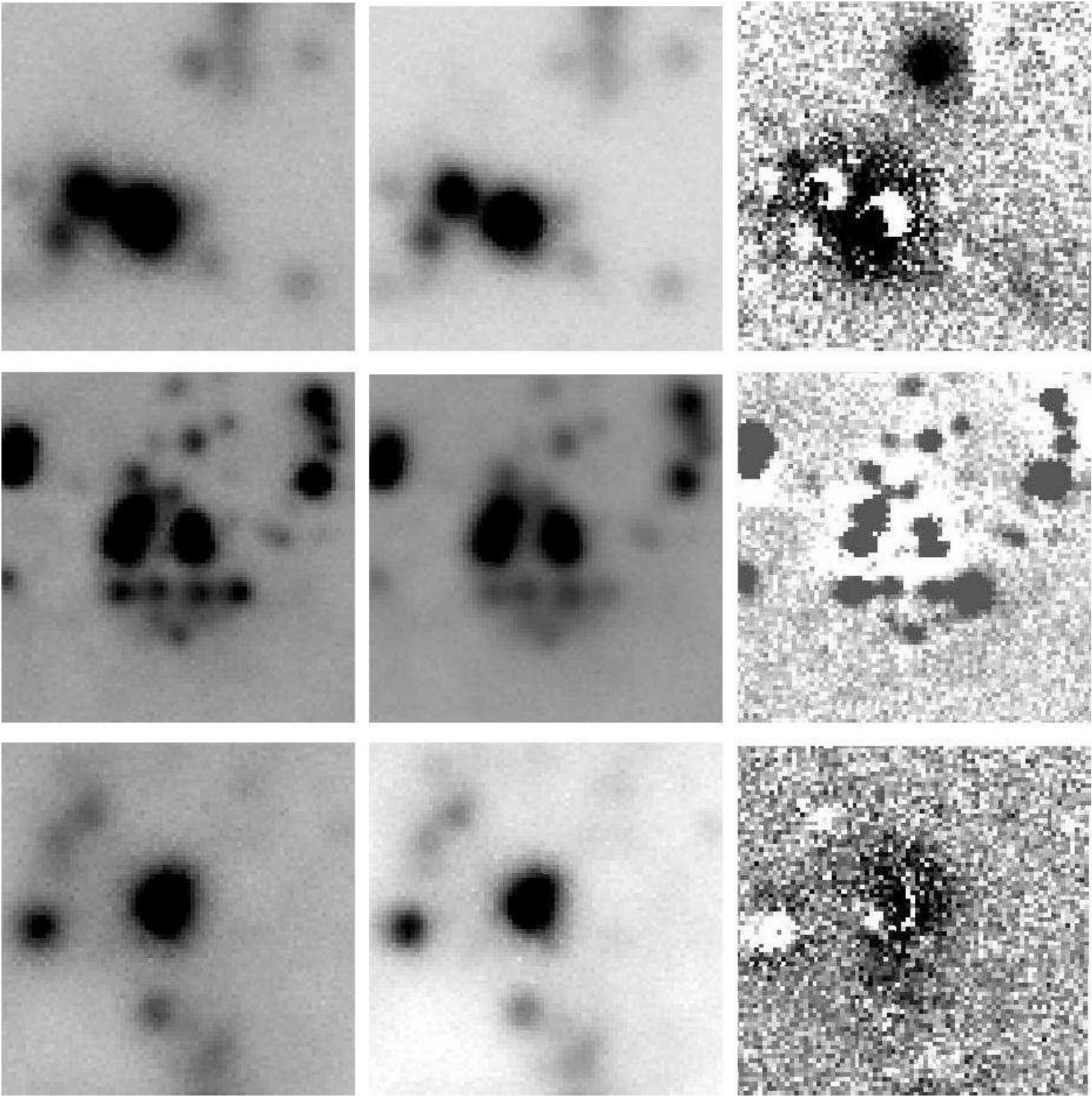}}}}
\caption{HeII emission-line (left
panels), continuum (middle panels) and continuum-subtracted HeII (right
  panels) images for our HII region sample. Top row: MA 1 - Middle row: BCLMP49 - Bottom row: BCLMP40.  Angular size of each
  image is 20$\arcsec$ x 20$\arcsec$ ($\sim$ 80 x 80 pc$^{2}$ at our assumed distance
  of 840 kpc). The spatial resolution and orientation are the same as in
  Figure~\ref{m33} (see Section 2.1 for more details about the narrow-band images).}
\label{heii}
\end{figure*}

\addtocounter{figure}{-1}
\begin{figure*}
\mbox{
 \centerline{
\hspace*{0.0cm}\subfigure{\includegraphics{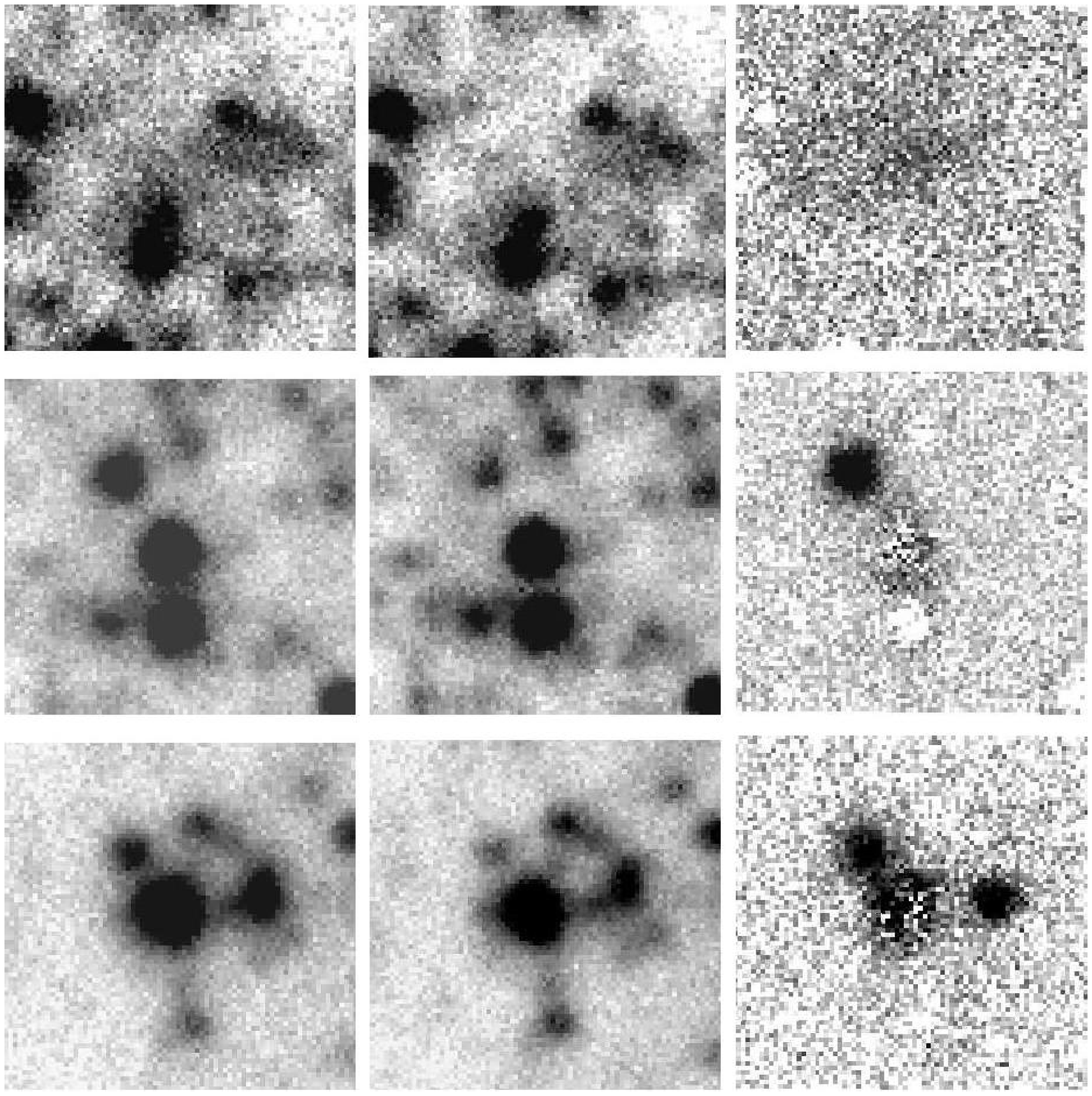}}}}
\caption{{\it Cont.}Top row: HBW673 - Middle row: BCLMP720 - Bottom row: BCLMP715}
\end{figure*}

\addtocounter{figure}{-1}
\begin{figure*}
\mbox{
 \centerline{
\hspace*{0.0cm}\subfigure{\includegraphics{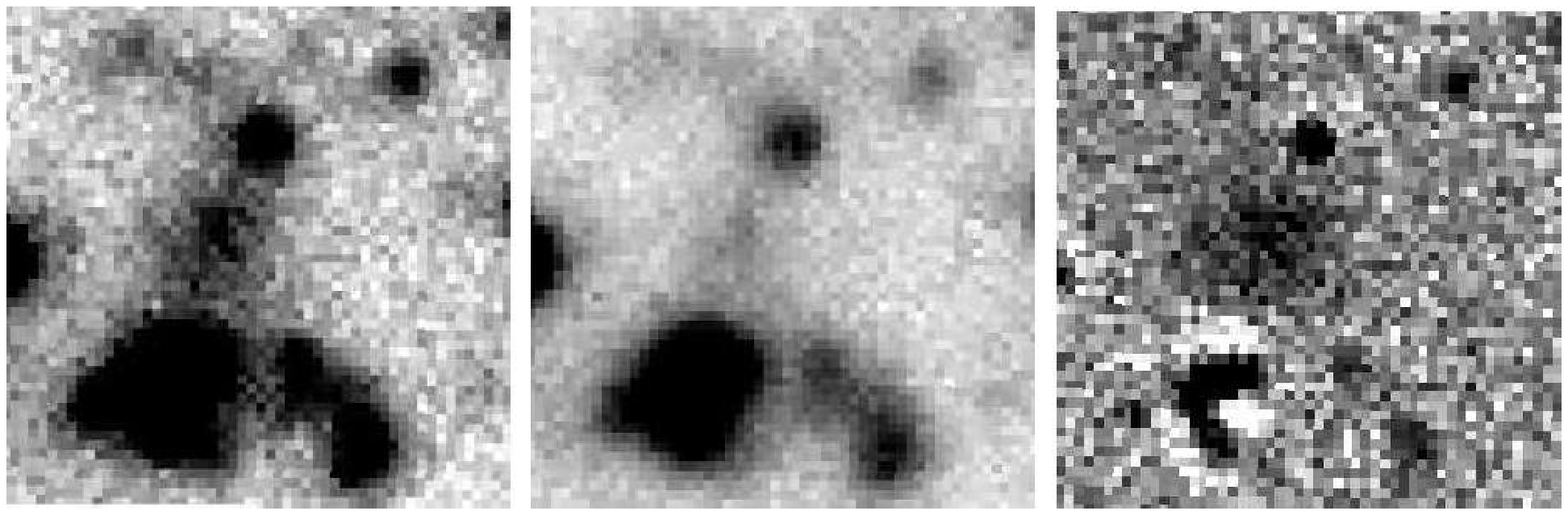}}}}
\caption{{\it Cont.}BCLMP651}
\end{figure*}

\newpage 
 
\begin{figure*}%[!ht] 
 \mbox{ 
  \centerline{ 
\hspace*{0.2cm}\subfigure{\label{}\includegraphics{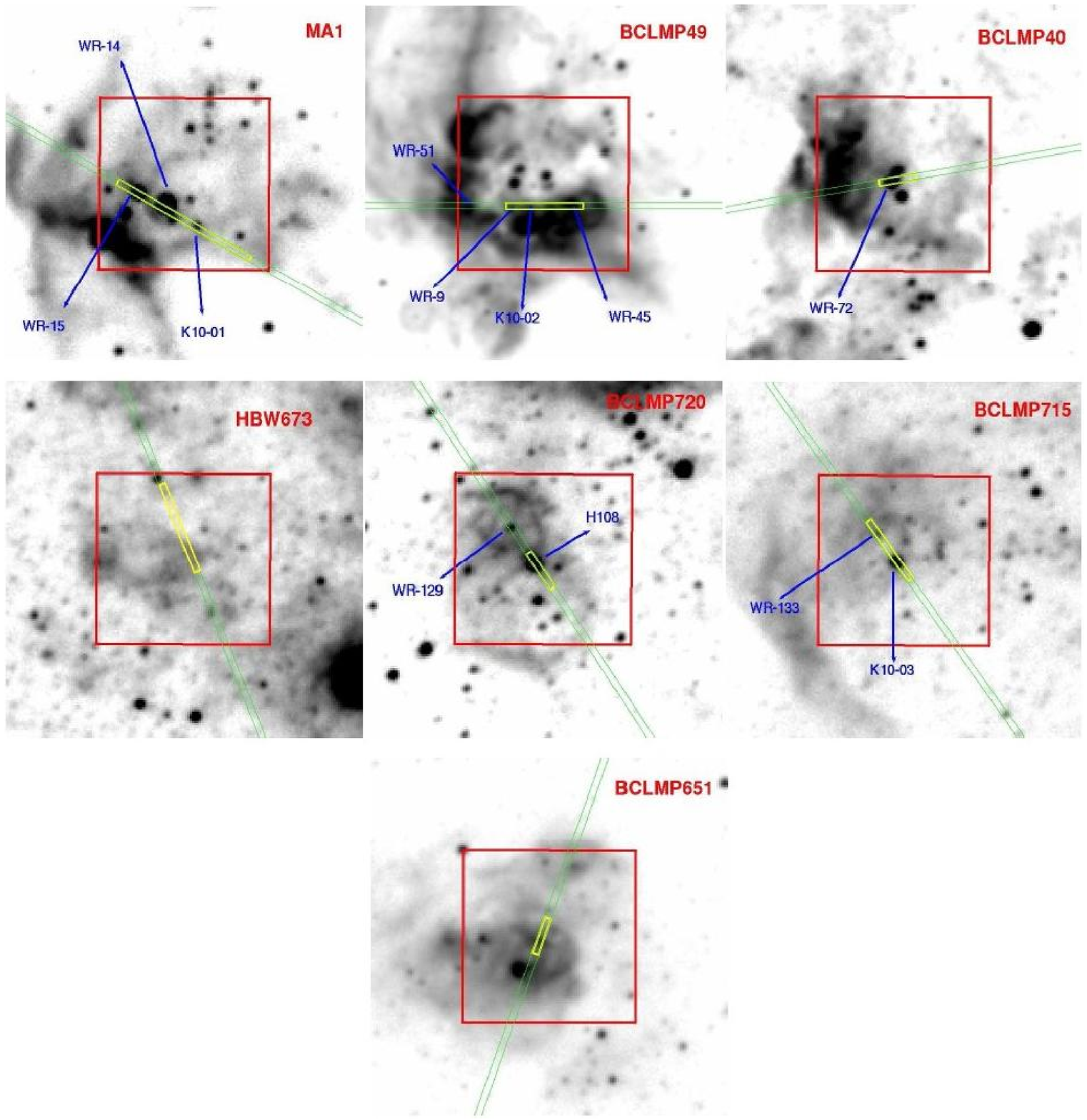}}}} 
%\hspace*{0.0cm}\subfigure{\label{}\includegraphics[width=6.5cm]{/work1/carol/USA/SPECTRA/GMOSN/paper/hiiz29.ha.ps}} 
%\hspace*{0.0cm}\subfigure{\label{}\includegraphics[width=6.5cm]{hiianon.fc.ps}} 
%\hspace*{0.0cm}\subfigure{\label{}\includegraphics[width=6.5cm]{hii49.fc.ps}} 
%\hspace*{0.0cm}\subfigure{\label{}\includegraphics[width=6.5cm]{hii40.fc.ps}} 
%   }} 
% \mbox{ 
%  \centerline{ 
%\hspace*{0.0cm}\subfigure{\label{}\includegraphics[width=6.5cm]{hiiz29.fc.ps}} 
%\hspace*{0.0cm}\subfigure{\label{}\includegraphics[width=6.5cm]{hii720.fc.ps}} 
%\hspace*{0.0cm}\subfigure{\label{}\includegraphics[width=6.5cm]{hii715.fc.new.new.ps}} 
%   }} 
%\mbox{ 
%  \centerline{ 
%\hspace*{0.2cm}\subfigure{\label{}\includegraphics[width=6.5cm]{hii651.fc.ps}} 
%}} 
\caption{Slit position (green/light grey box) for all HII regions overplotted on the 
  H$\alpha$ image from the NOAO Survey of Resolved Stellar Content of Local Group Galaxies
\mbox{(PI: P. Massey)}. Angular size is 40$\arcsec$ x 40$\arcsec$ ($\sim$ 160
x 160 pc$^{2}$ at our assumed distance of 840 kpc). Yellow/light box 
  marks where the nebular HeII emission is located within the slit.  Blue
  arrows indicate the position and name of the associated hot stars. Red/dark grey box
  shows the size of the HeII images (20$\arcsec$ x 20$\arcsec$) as seen in
  Figure~\ref{heii}. The spatial resolution and orientation are the same as in
  Figure~\ref{m33}. The H$\alpha$ image is shown in logarithmic scale to better enhance all the morphological features of the HII regions.} 
\label{finding_charts} 
\end{figure*} 
 
\newpage 
\vspace*{-2.0cm}
\begin{figure*} 
\begin{center}
 \mbox{ 
  \centerline{
%\vspace*{-1.0cm}
\hspace*{0.0cm}\subfigure{\label{}\includegraphics[bb=18 433 592 718,width=7.5cm,clip]{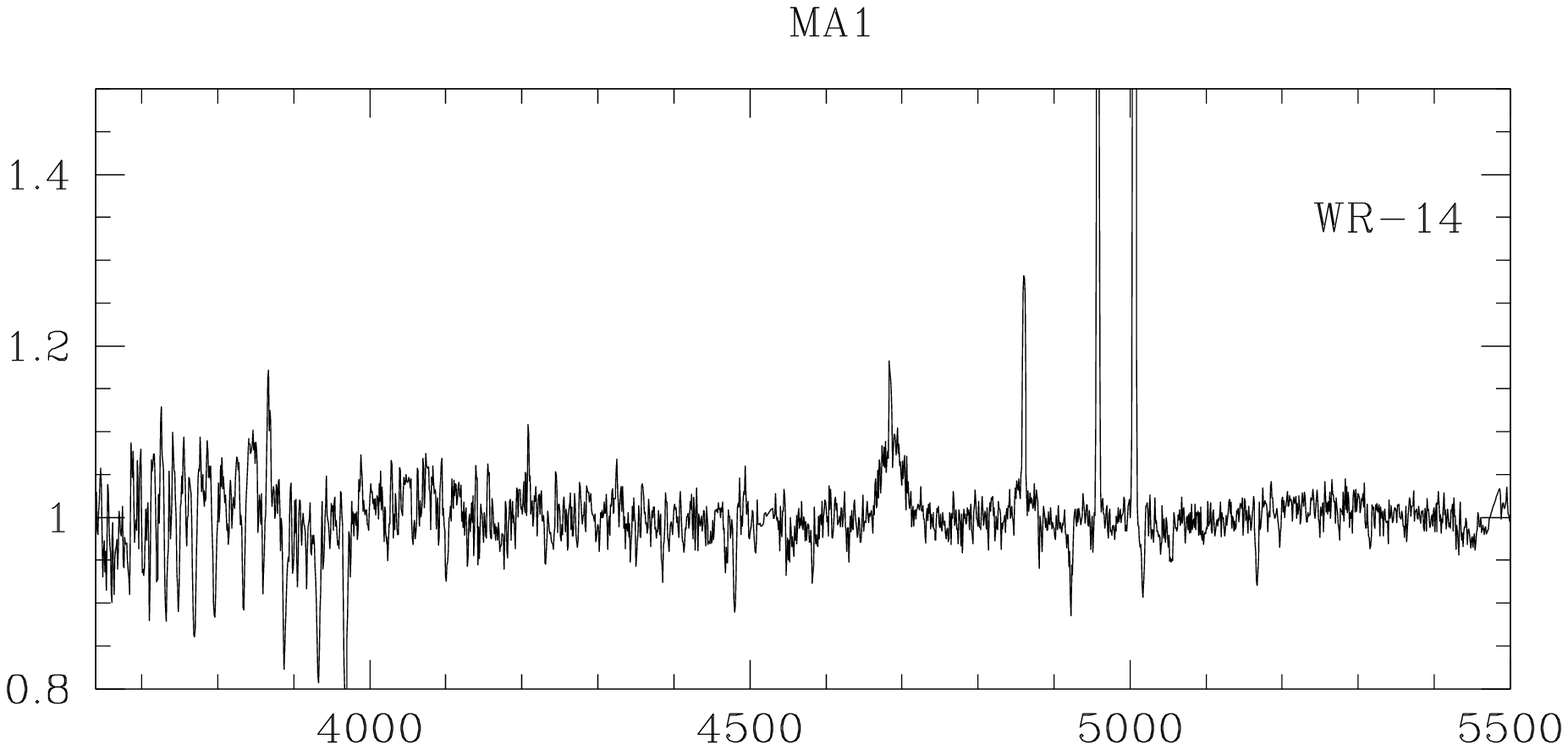}} 
%\vspace*{-6.0cm} 
\hspace*{0.0cm}\subfigure{\label{}\includegraphics[bb=18 433 592 718,width=7.5cm,clip]{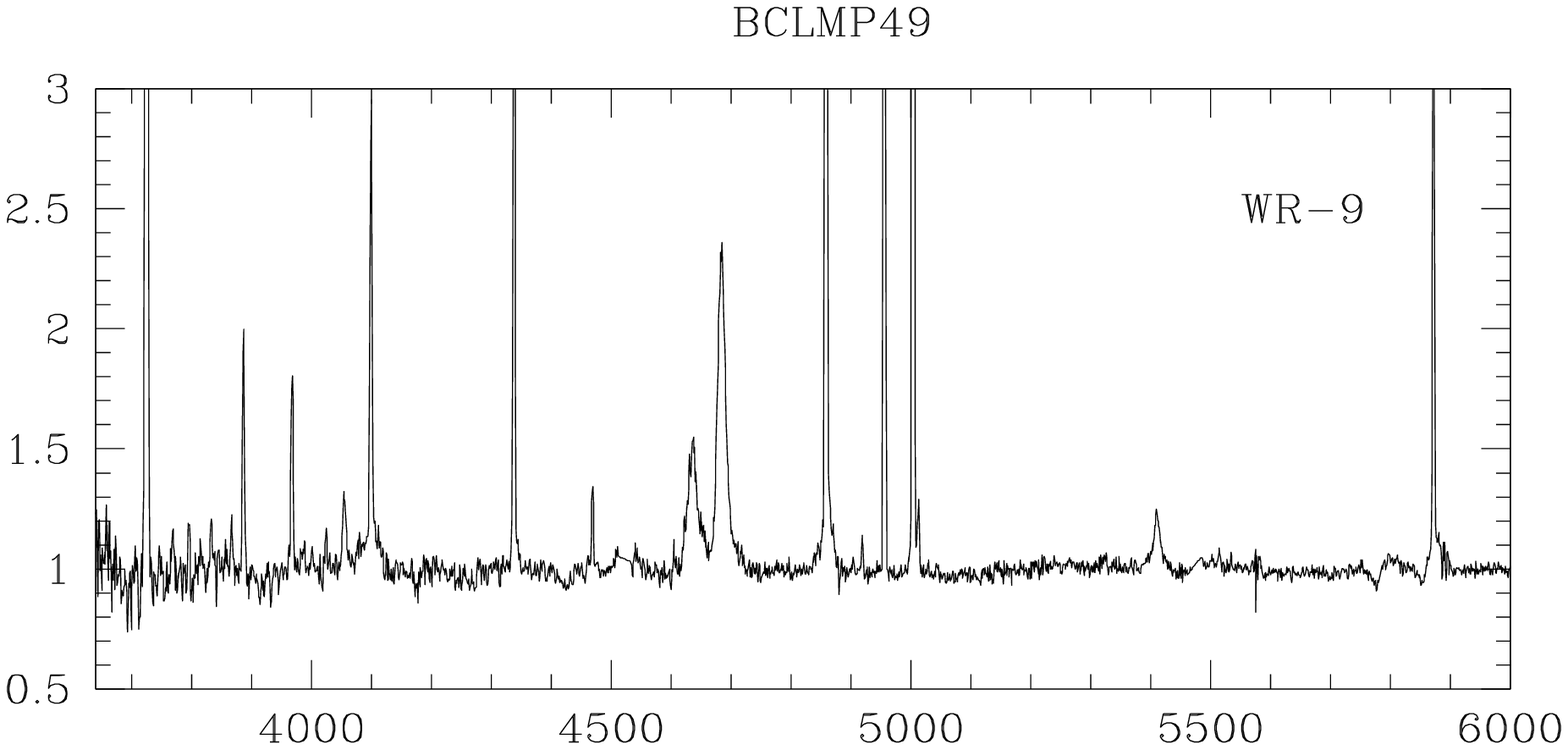}}
}}
%\vspace*{-1.0cm}
\mbox{ 
\centerline{ 
 \hspace*{0.0cm}\subfigure{\label{}\includegraphics[bb=18 433 592 718,width=7.5cm,clip]{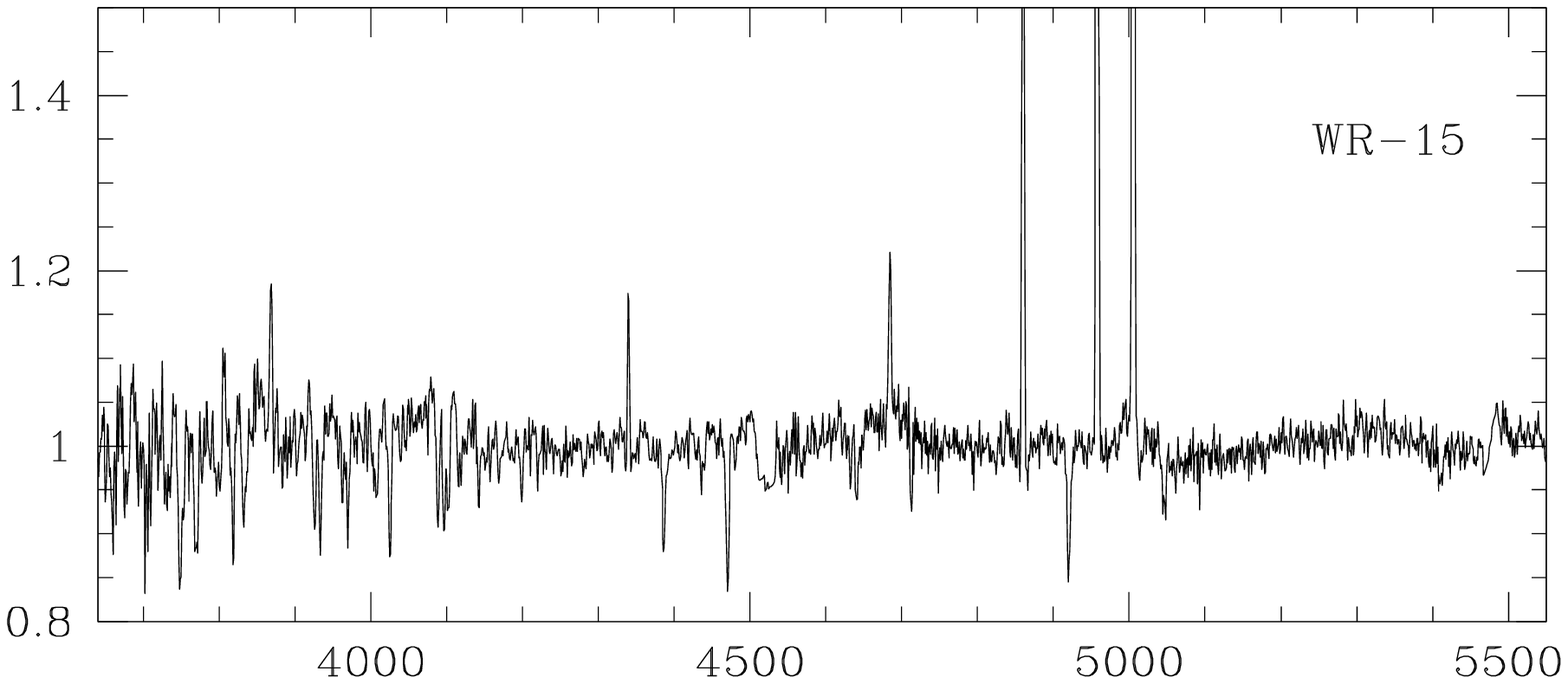}} 
%\vspace*{-0.8cm} 
\hspace*{0.0cm}\subfigure{\label{}\includegraphics[bb=18 433 592 718,width=7.5cm,clip]{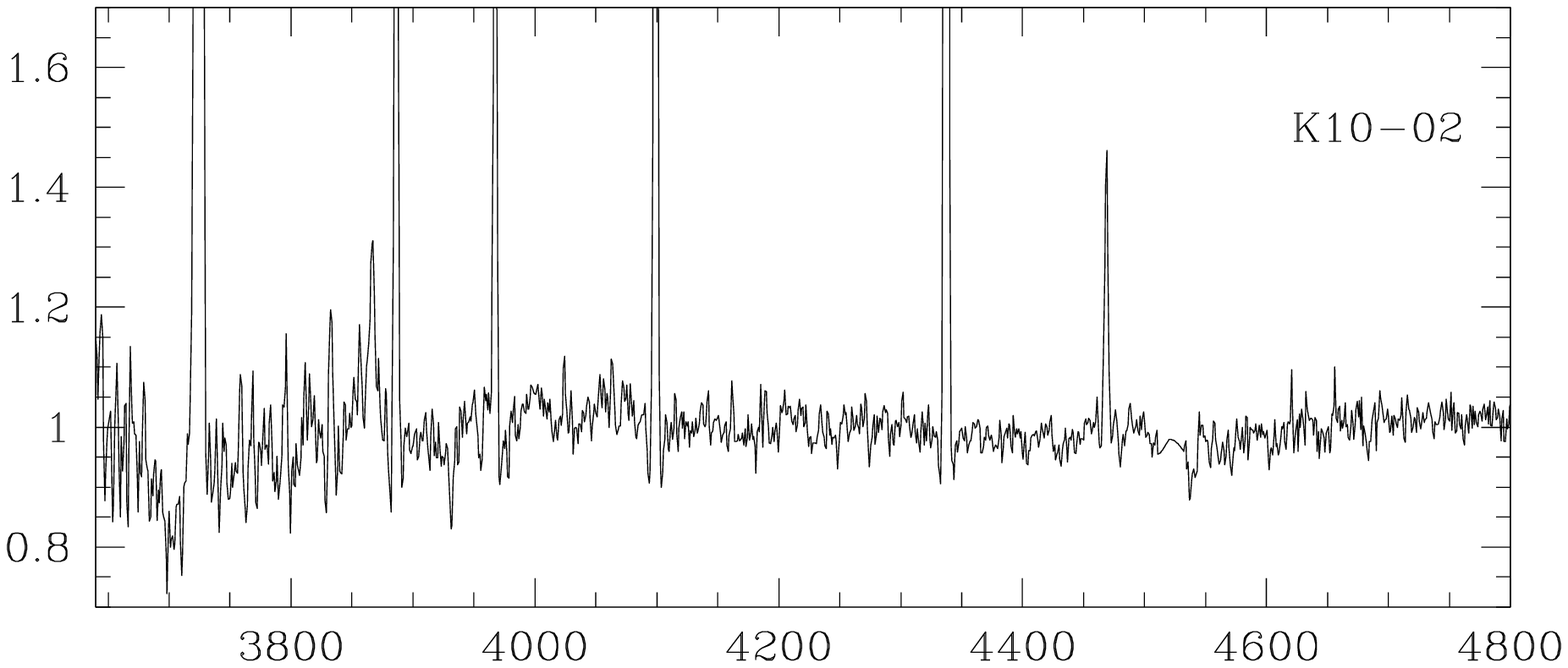}} 
}} 
\mbox{ 
\centerline{ 
\hspace*{0.0cm}\subfigure{\label{}\includegraphics[bb=18 433 592 718,width=7.5cm,clip]{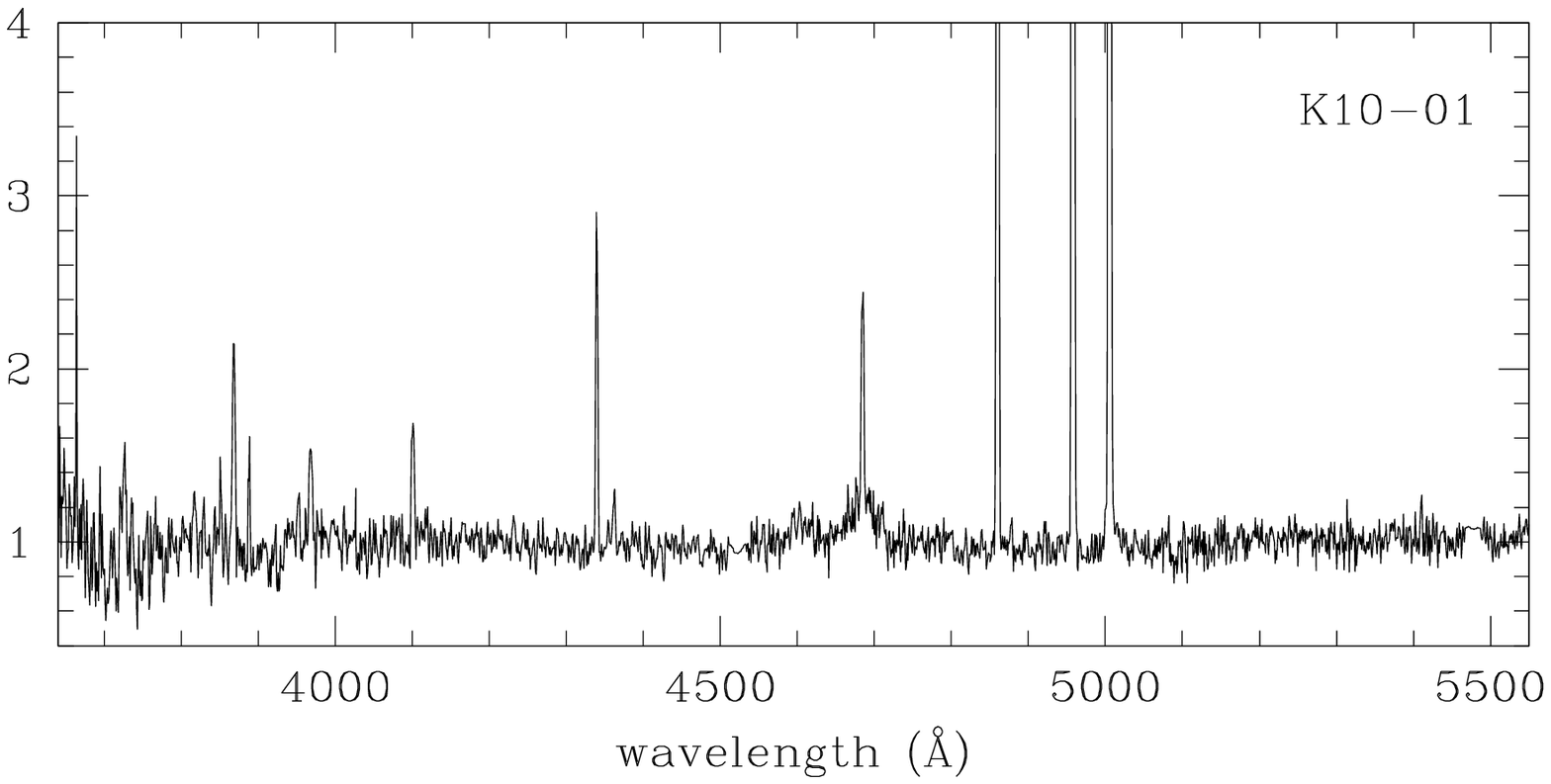}} 
%\vspace*{-0.2cm} 
\hspace*{0.0cm}\subfigure{\label{}\includegraphics[bb=18 433 592 718,width=7.5cm,clip]{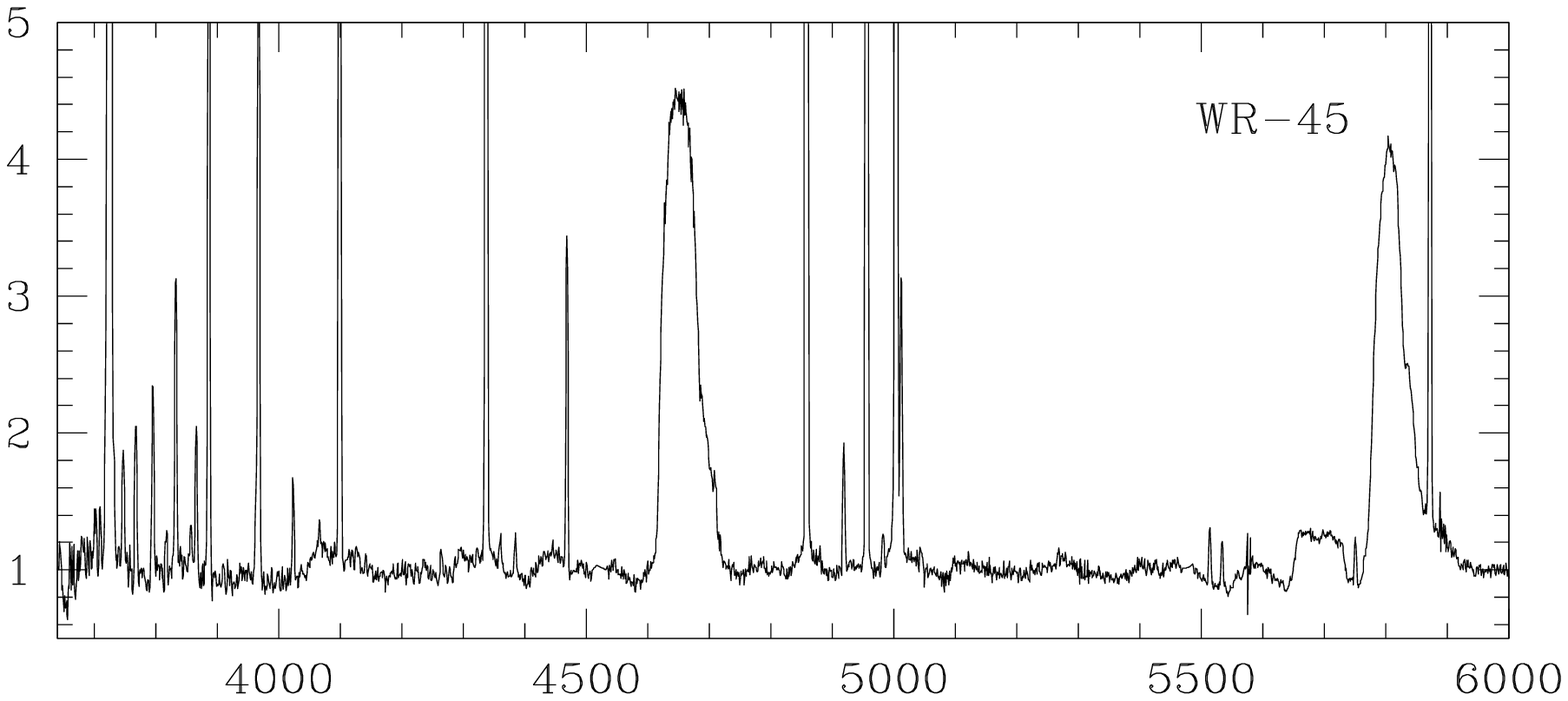}} 
}} 
\mbox{ 
\centerline{ 
\hspace*{0.0cm}\subfigure{\label{}\includegraphics[bb=18 433 592 718,width=7.5cm,clip]{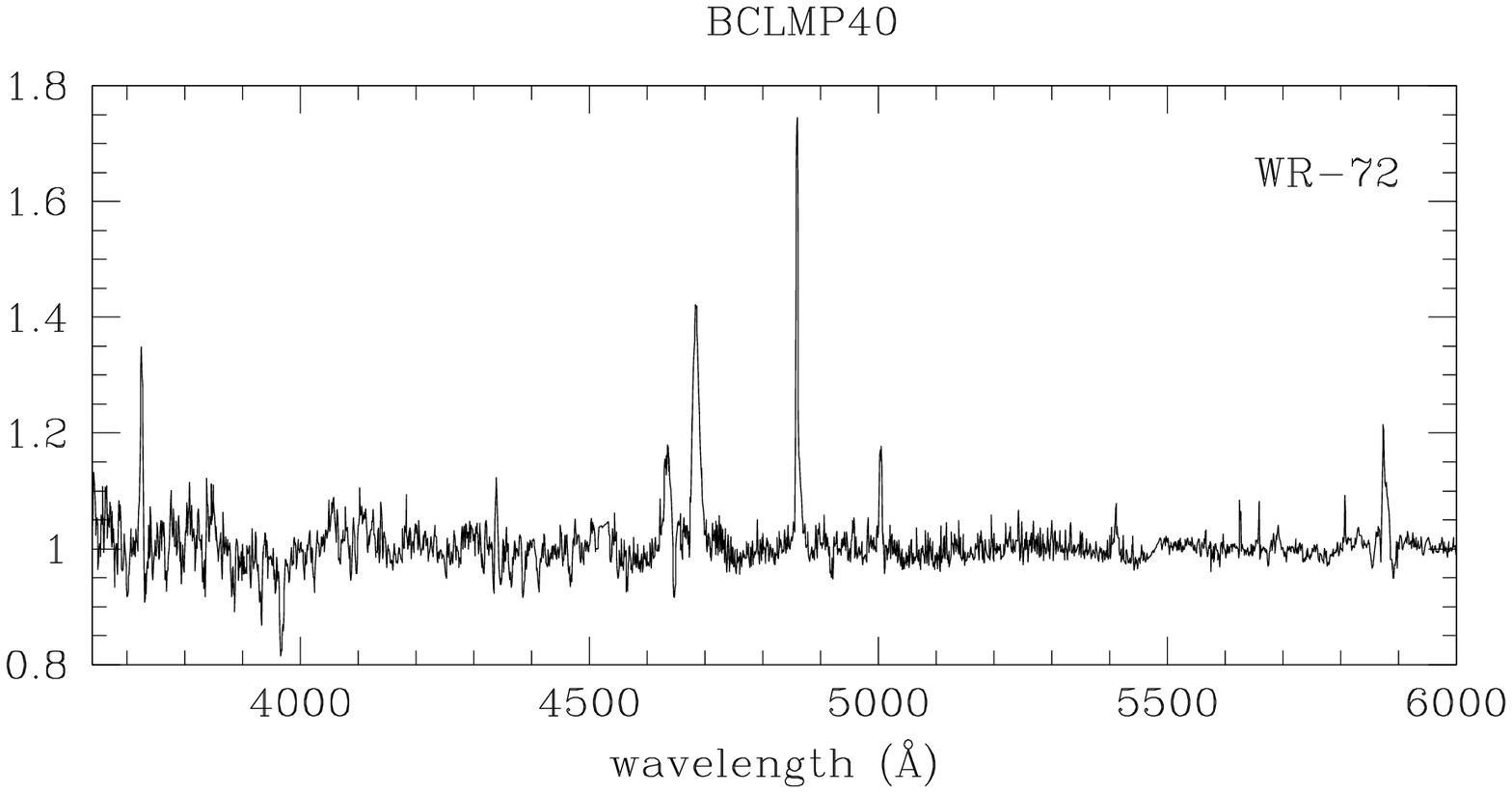}} 
%\vspace*{-0.2cm} 
\hspace*{0.0cm}\subfigure{\label{}\includegraphics[bb=18 433 592 718,width=7.5cm,clip]{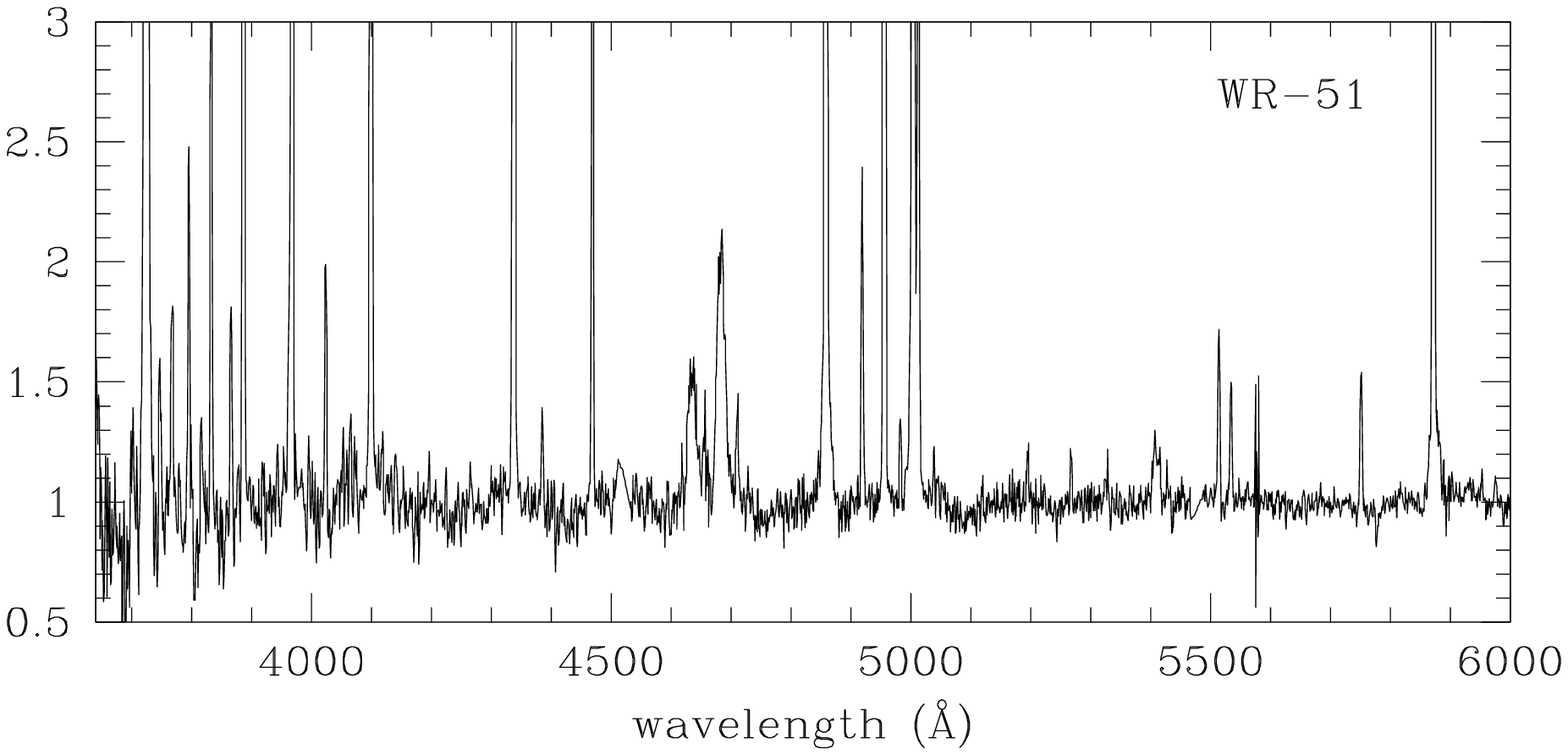}} 
}} 
\mbox{ 
\centerline{ 
\hspace*{0.0cm}\subfigure{\label{}\includegraphics[bb=18 433 592 718,width=7.5cm,clip]{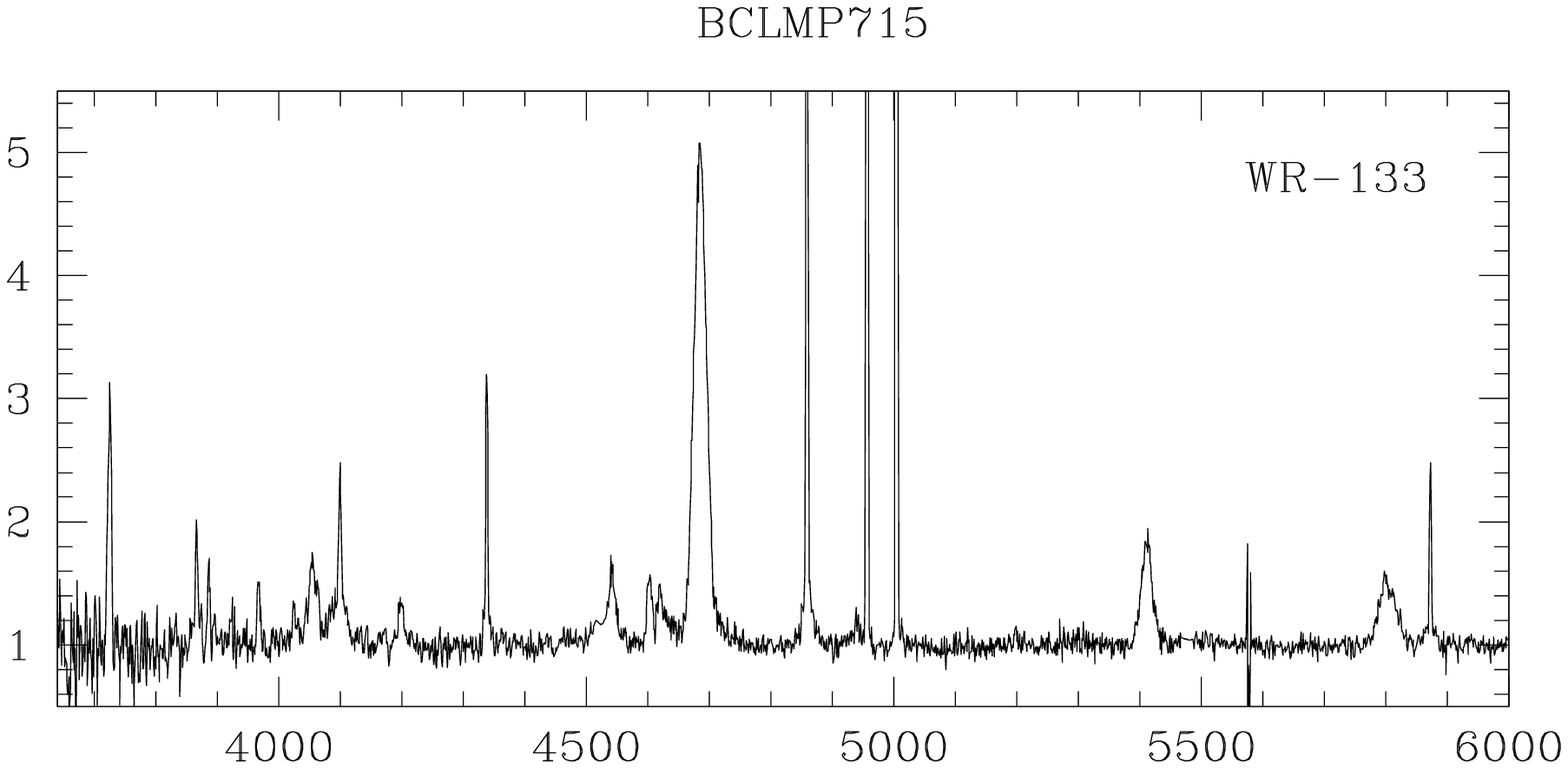}} 
%\vspace*{-0.8cm} 
\hspace*{0.0cm}\subfigure{\label{}\includegraphics[bb=18 433 592 718,width=7.5cm,clip]{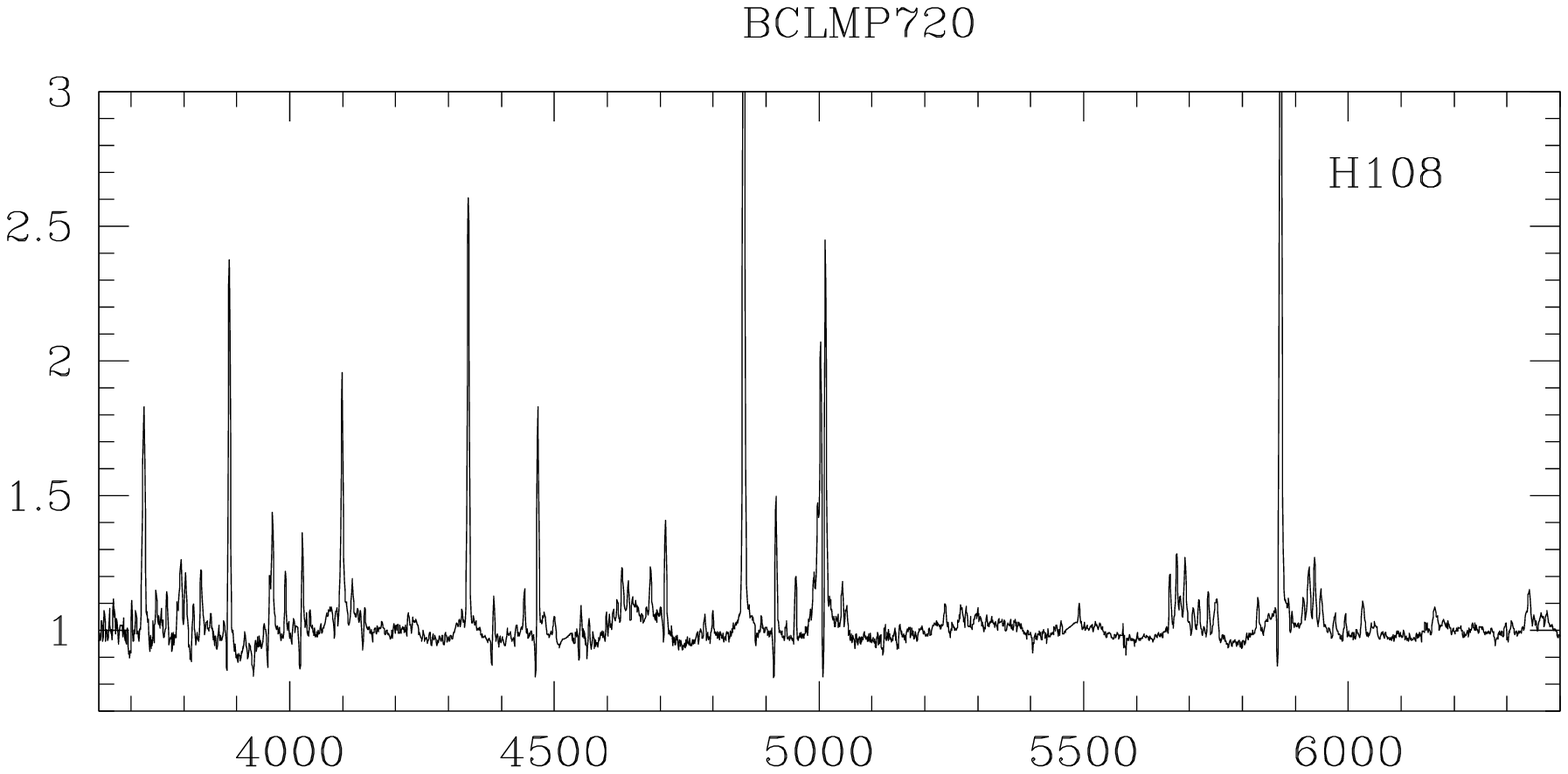}} 
}} 
\mbox{ 
\centerline{ 
\hspace*{0.0cm}\subfigure{\label{}\includegraphics[bb=18 433 592 718,width=7.5cm,clip]{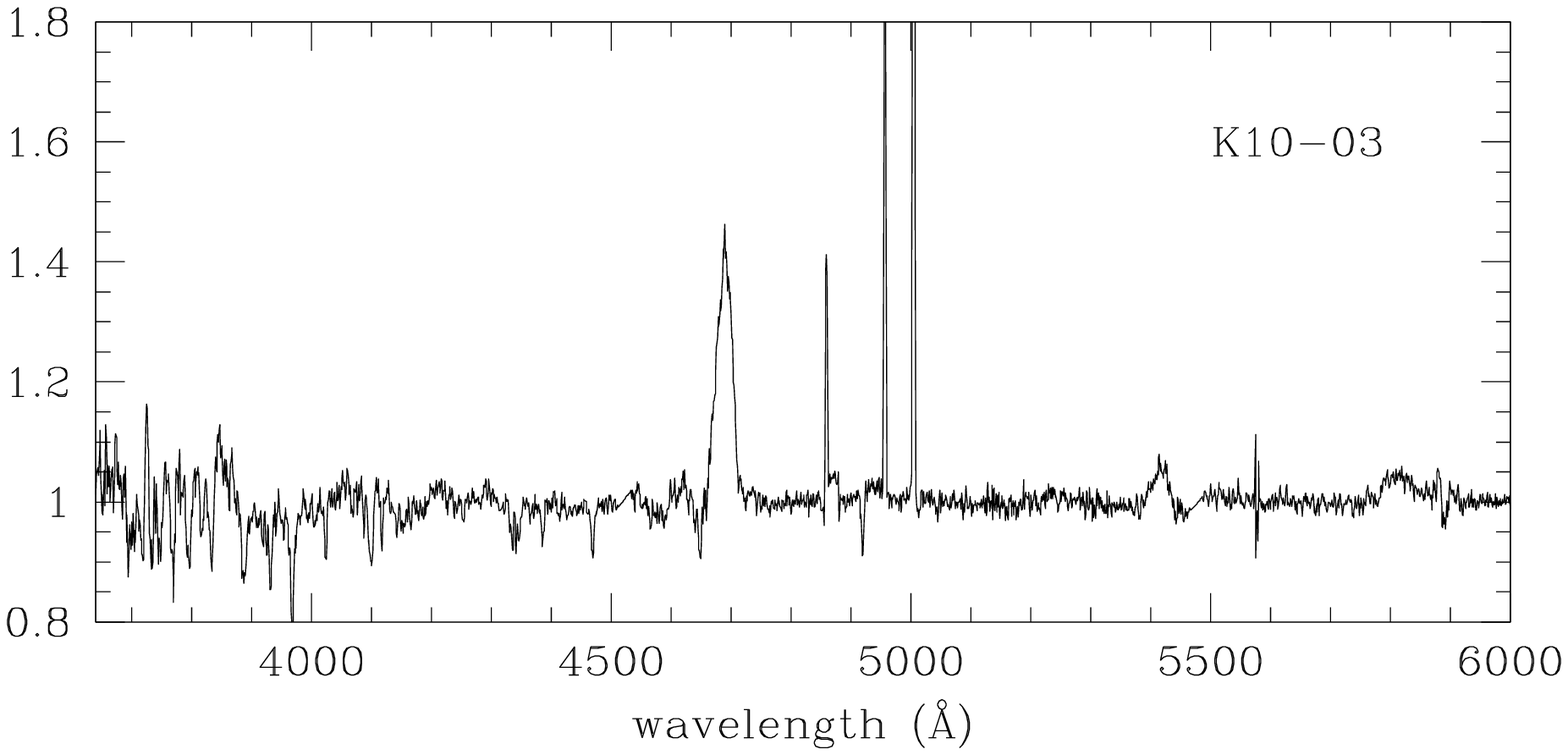}} 
\hspace*{0.0cm}\subfigure{\label{}\includegraphics[bb=18 433 592 718,width=7.5cm,clip]{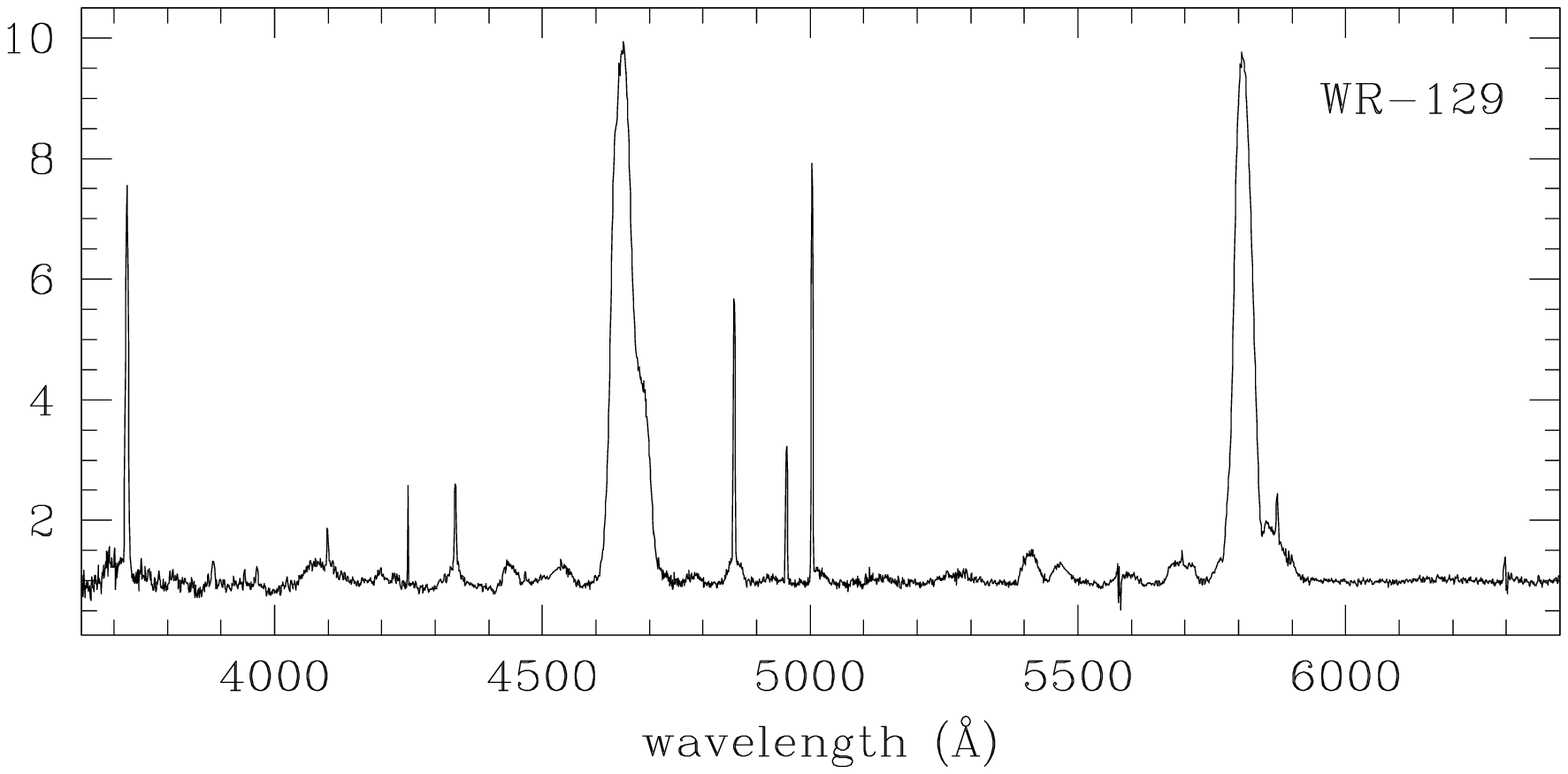}} 
}} 
\end{center} 
\vspace*{-0.5cm} 
\caption{Rectified blue spectra for the ionizing stars listed in Table~\ref{tab4}.} 
\label{star_spectra} 
\end{figure*}

\newpage 
\begin{figure*}%[!ht] 
 \mbox{ 
  \centerline{ 
\hspace*{0.5cm}\subfigure{\label{}\includegraphics[bb=18 428 592 718,width=9cm]{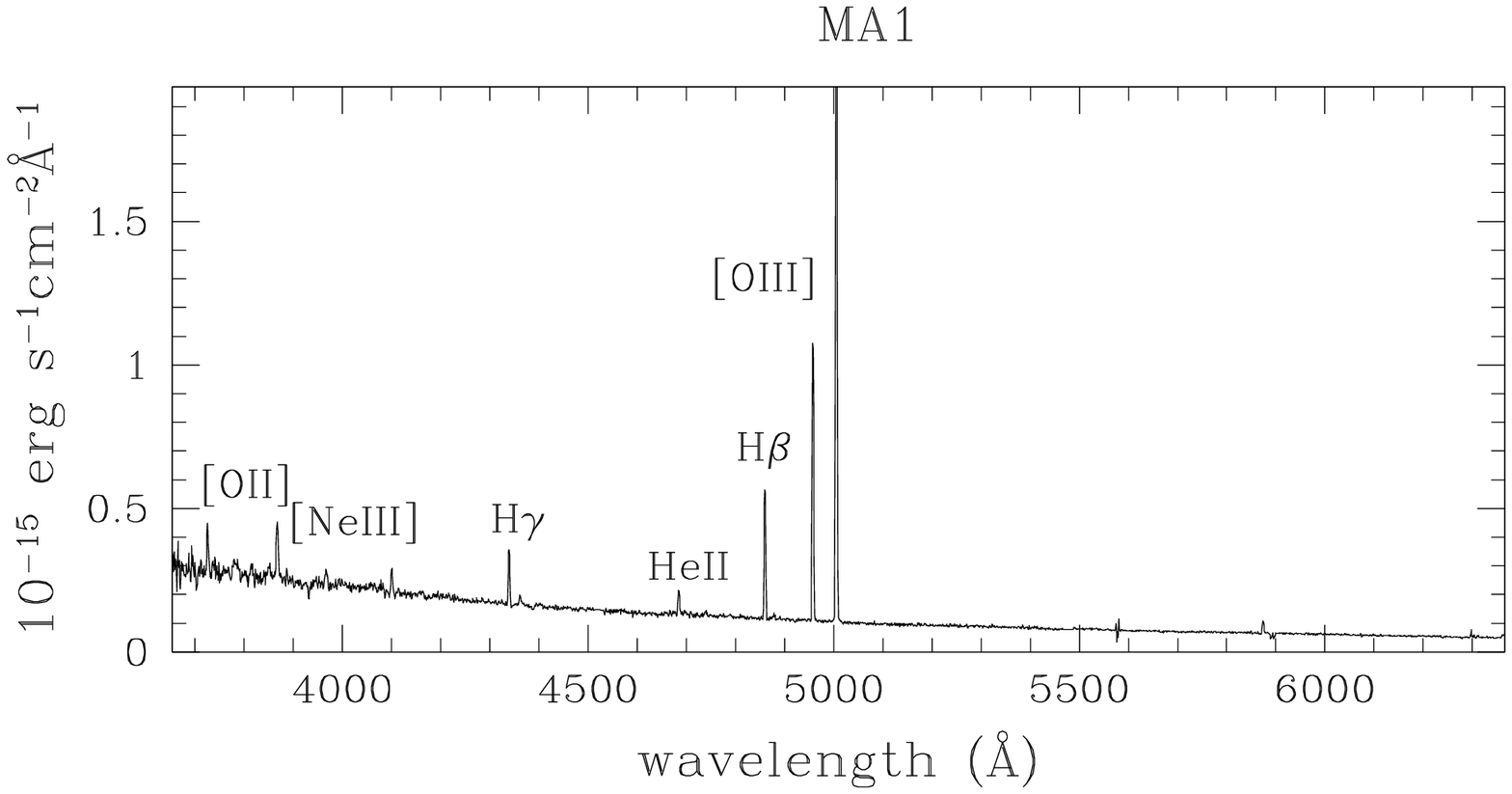}} 
\hspace*{0.0cm}\subfigure{\label{}\includegraphics[bb=18 428 592 718,width=9cm]{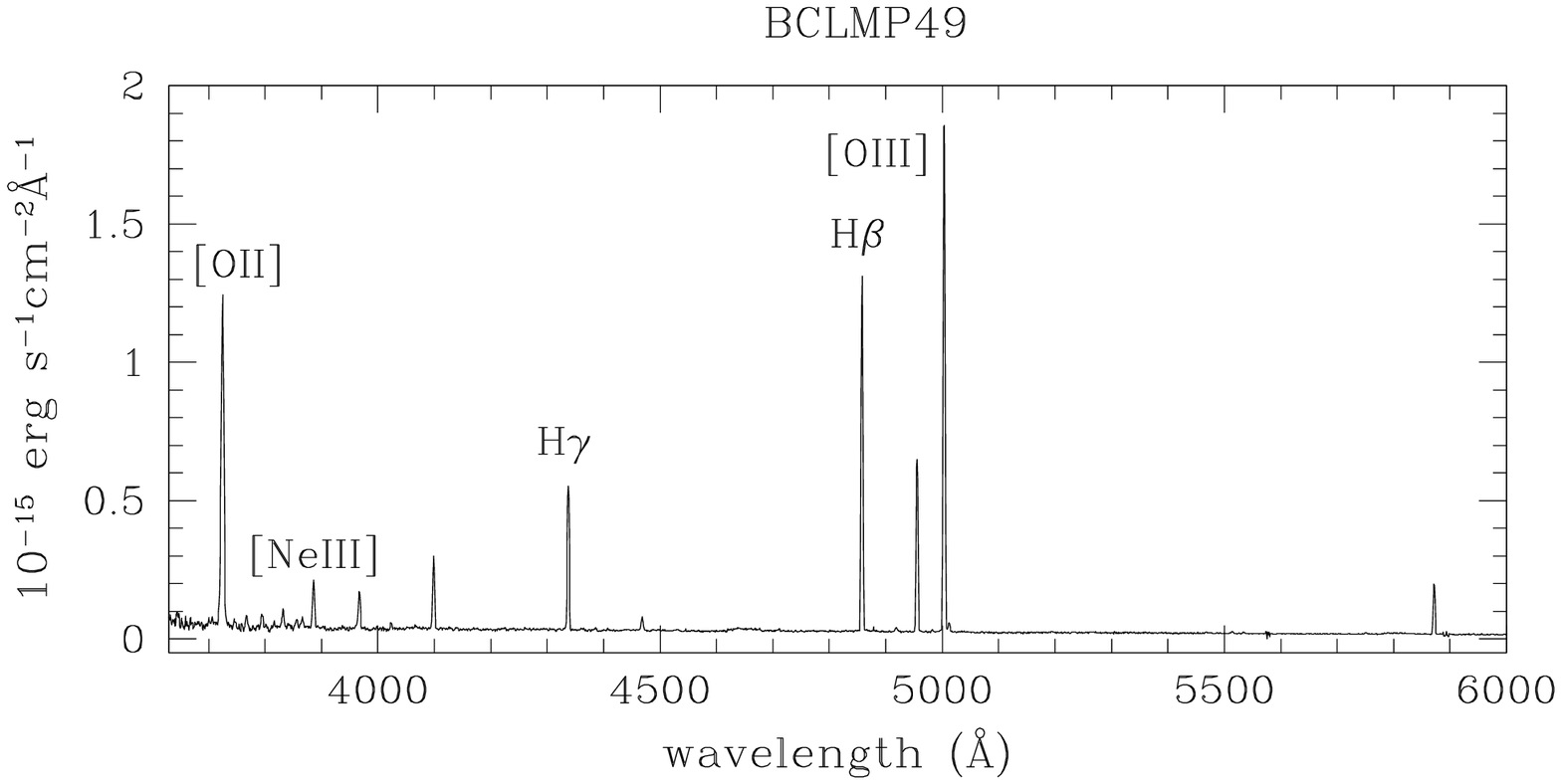}} 
   }} 
 \mbox{ 
  \centerline{ 
\hspace*{0.5cm}\subfigure{\label{}\includegraphics[bb=18 428 592 718,width=9cm]{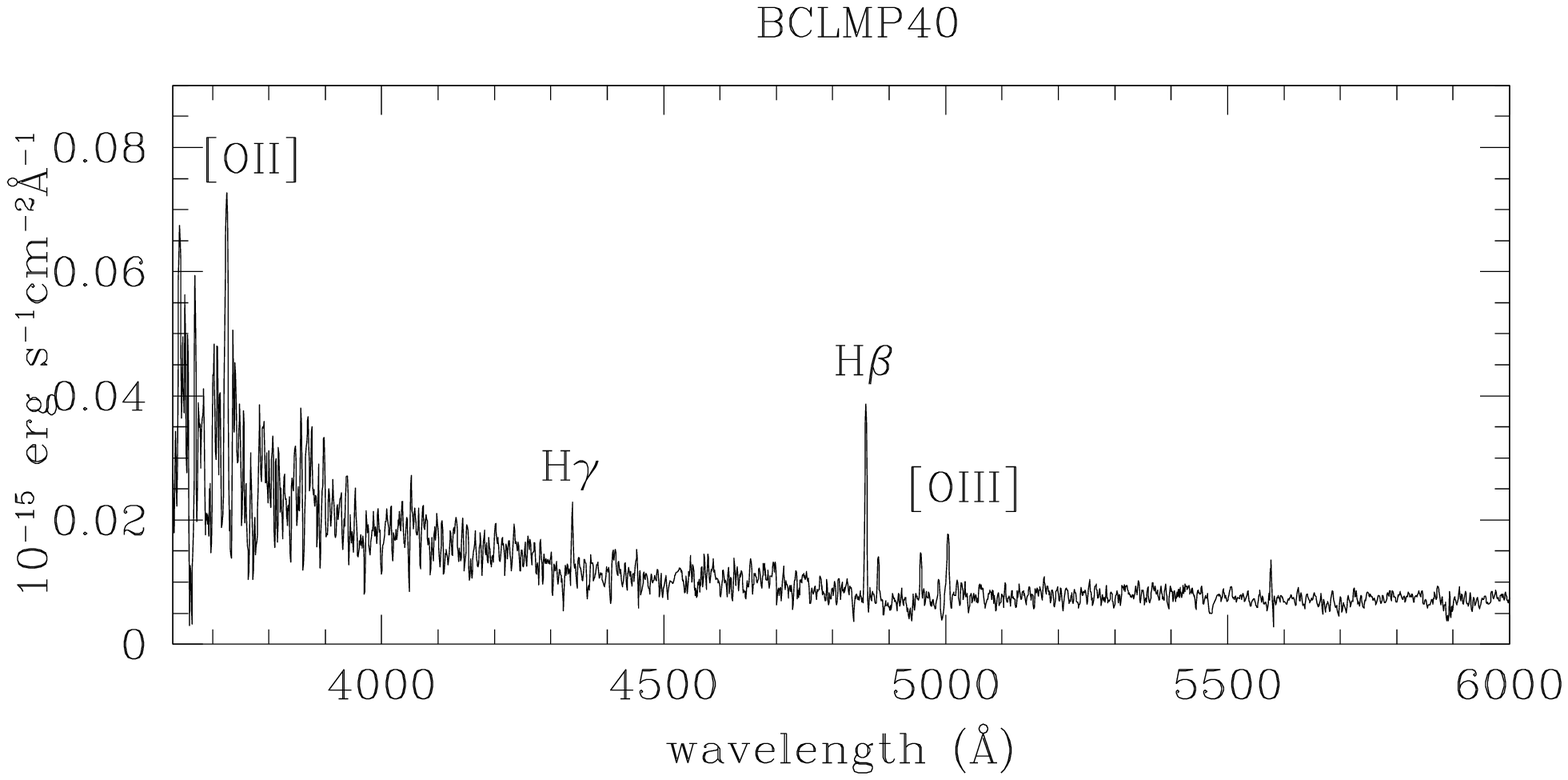}} 
\hspace*{0.0cm}\subfigure{\label{}\includegraphics[bb=18 428 592 718,width=9cm]{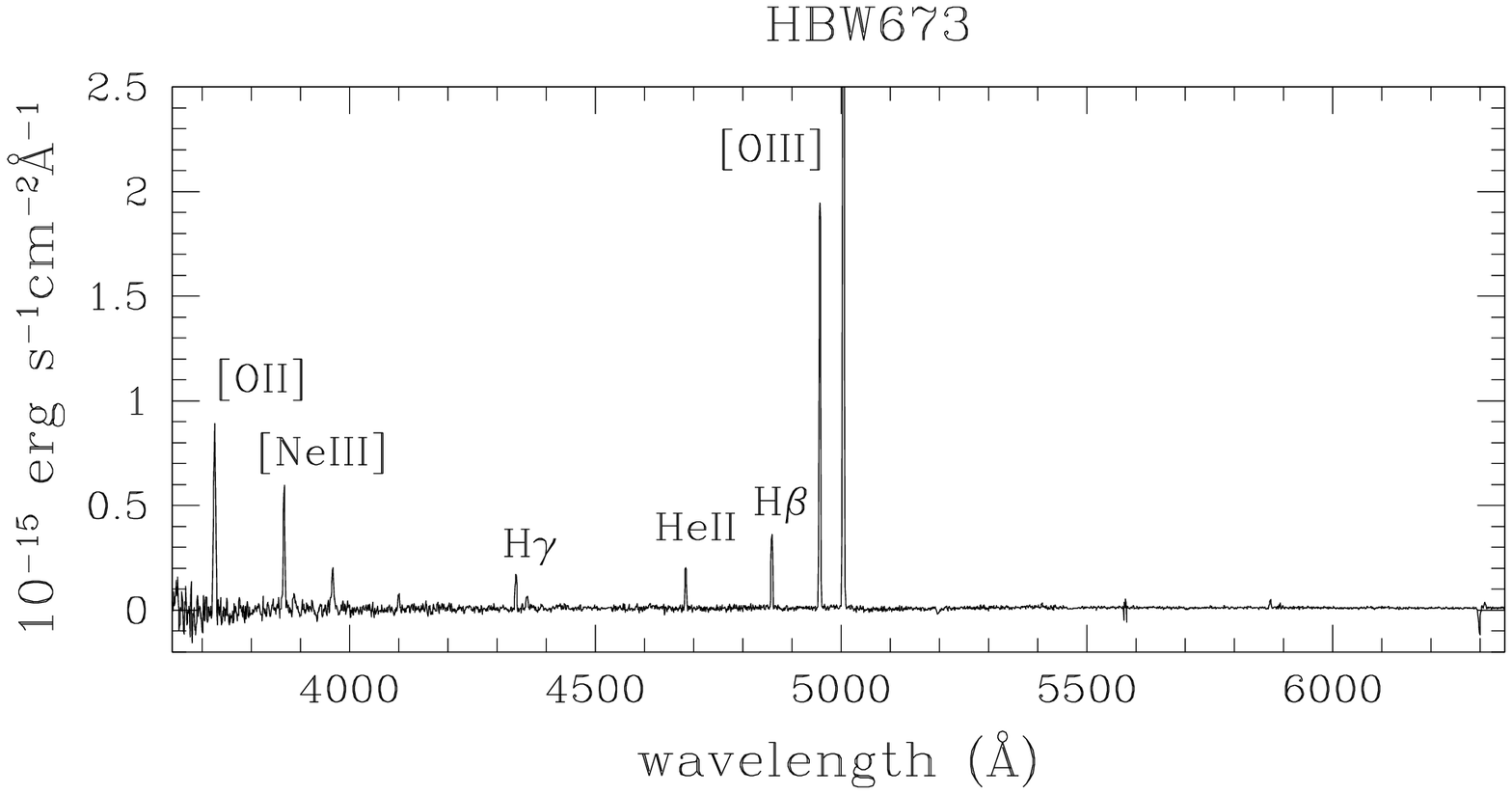}} 
   }} 
 \mbox{ 
  \centerline{ 
\hspace*{0.5cm}\subfigure{\label{}\includegraphics[bb=18 428 592 718,width=9cm]{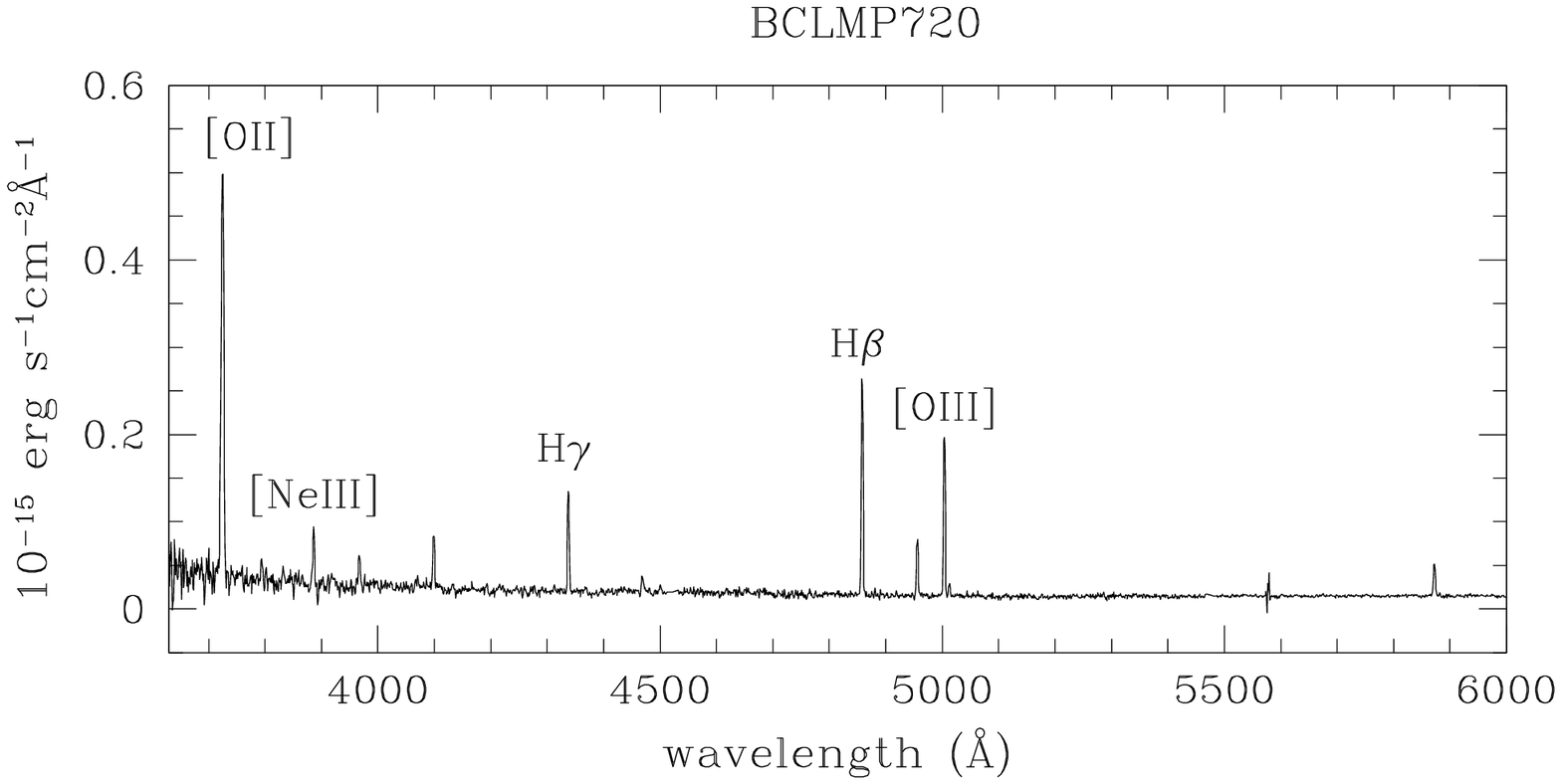}} 
\hspace*{0.0cm}\subfigure{\label{}\includegraphics[bb=18 428 592 718,width=9cm]{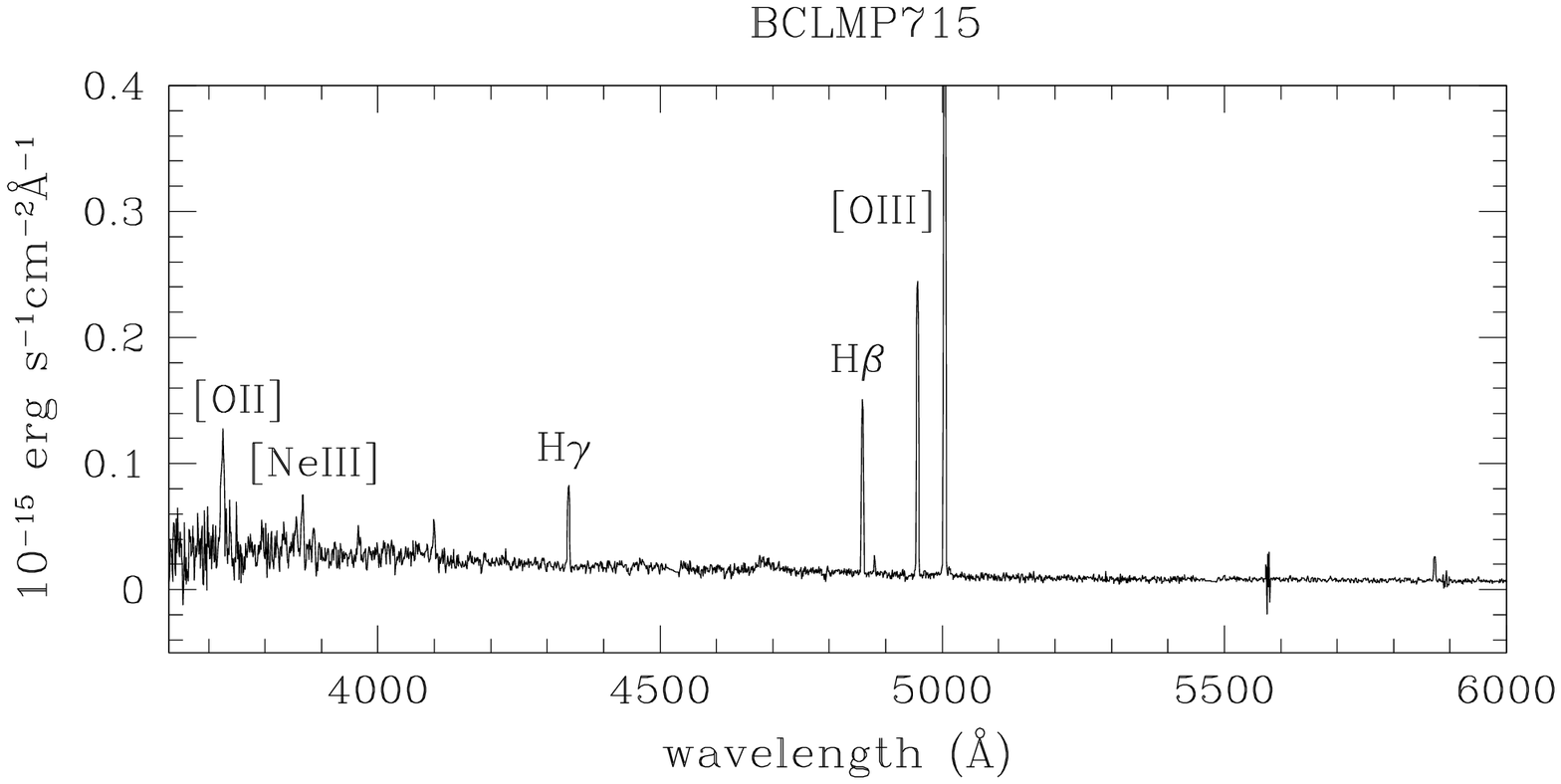}} 
   }} 
\mbox{ 
  \centerline{ 
\hspace*{0.5cm}\subfigure{\label{}\includegraphics[bb=18 428 592 718,width=9cm]{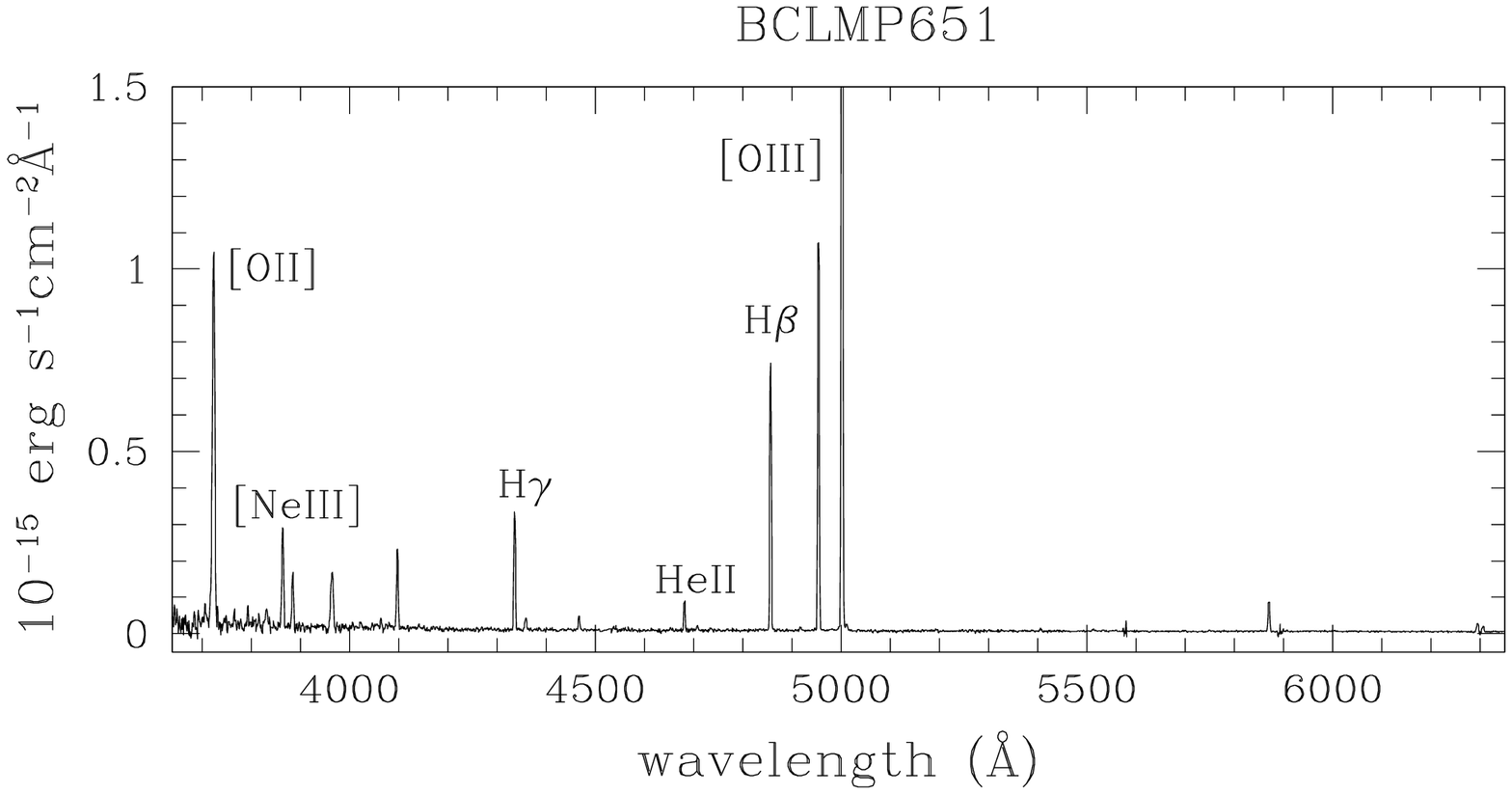}} 
}} 
\caption{Blue spectra in units of 10$^{-15}$ erg 
s$^{-1}$ cm$^{-2}$ \AA$^{-1}$, corresponding to the nebular emission from the zone 
  marked by the yellow box as shown in Figure~\ref{finding_charts} for our HII 
  regions. Selected emission lines are labelled.} 
\label{neb_blue_spectra} 
\end{figure*} 
 
\newpage 
\begin{figure*}%[!ht] 
 \mbox{ 
  \centerline{ 
\hspace*{0.5cm}\subfigure{\label{}\includegraphics[bb=18 428 592 718,width=9cm]{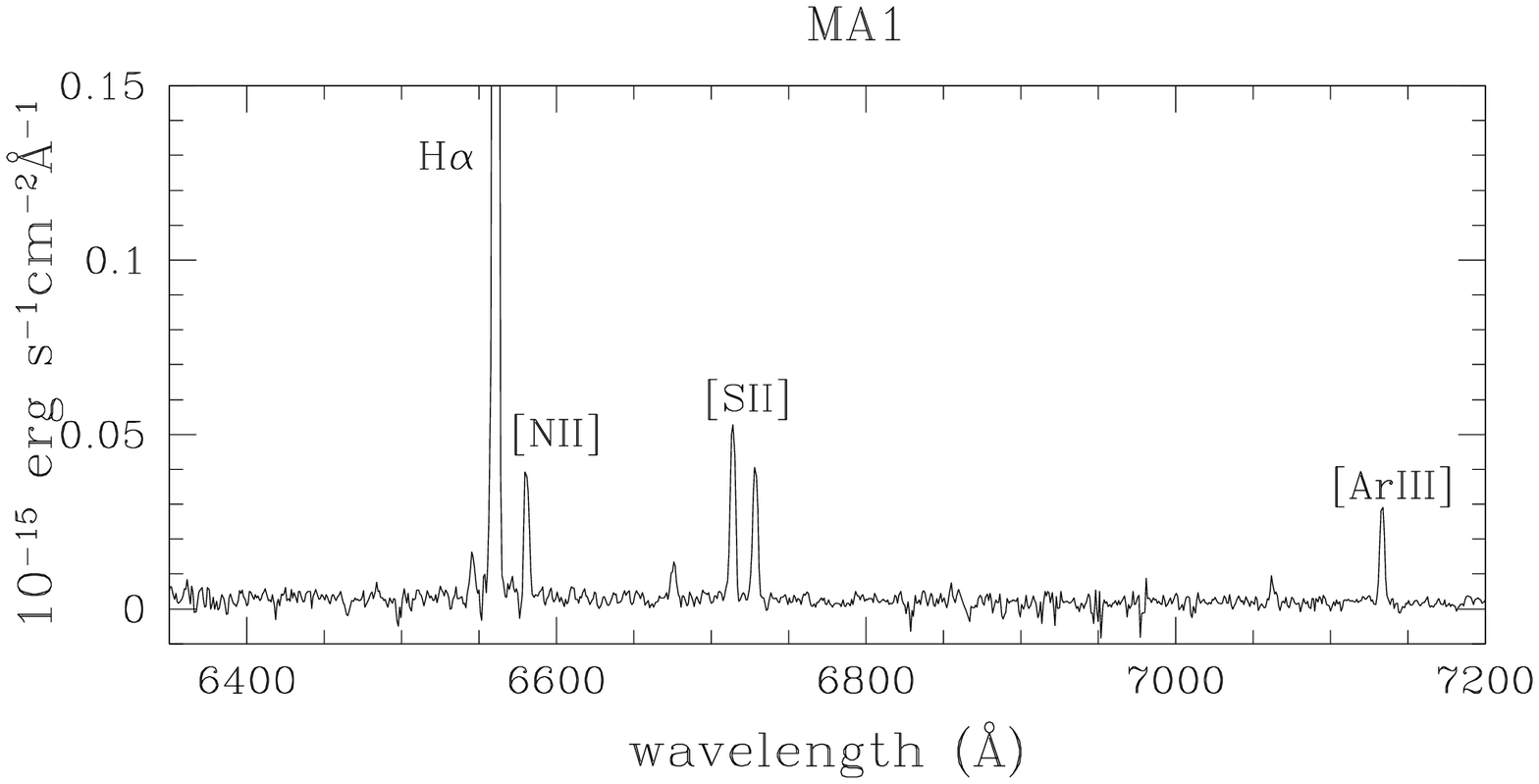}} 
\hspace*{0.0cm}\subfigure{\label{}\includegraphics[bb=18 428 592 718,width=9cm]{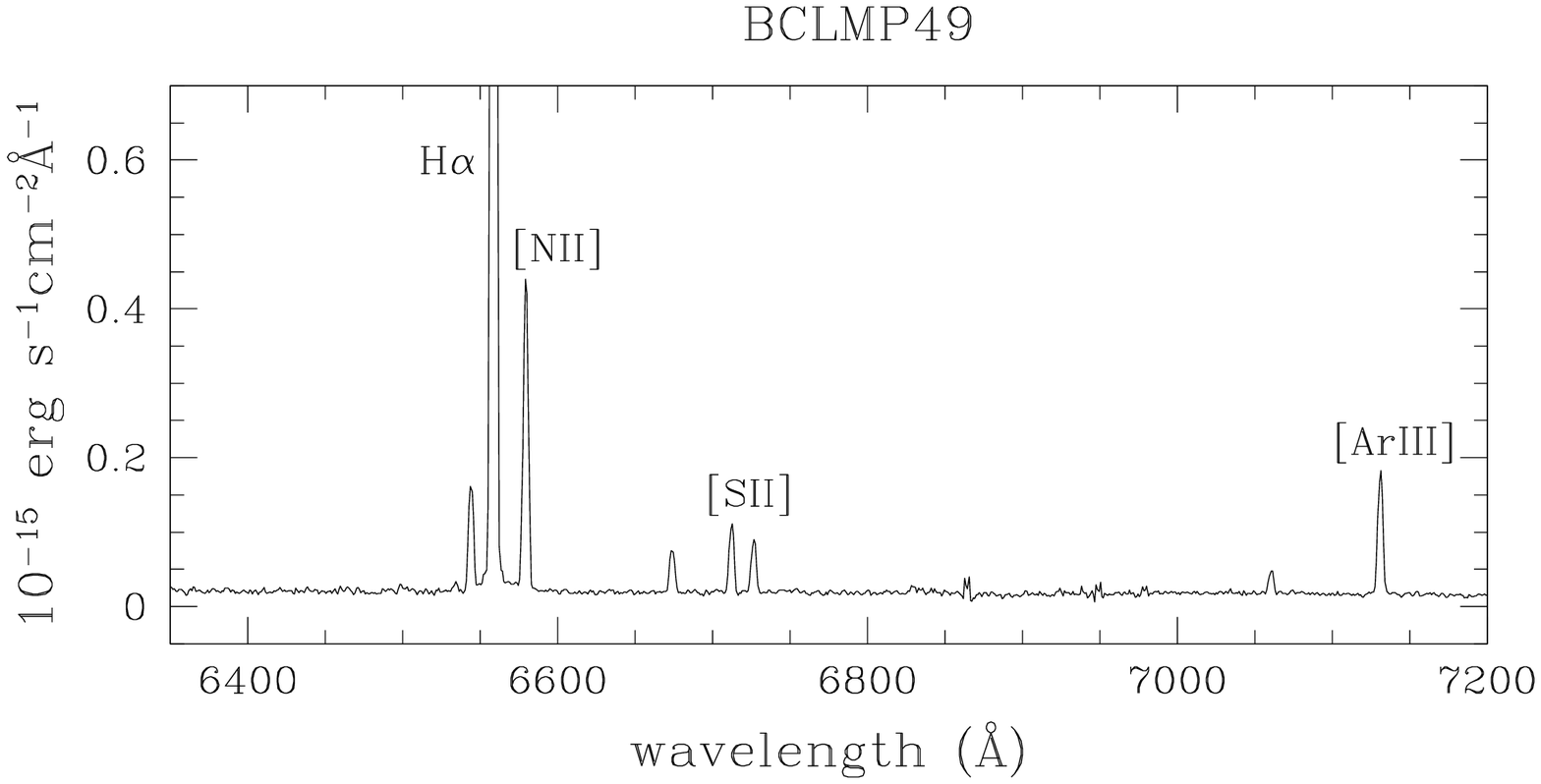}} 
   }} 
 \mbox{ 
  \centerline{ 
\hspace*{0.5cm}\subfigure{\label{}\includegraphics[bb=18 428 592 718,width=9cm]{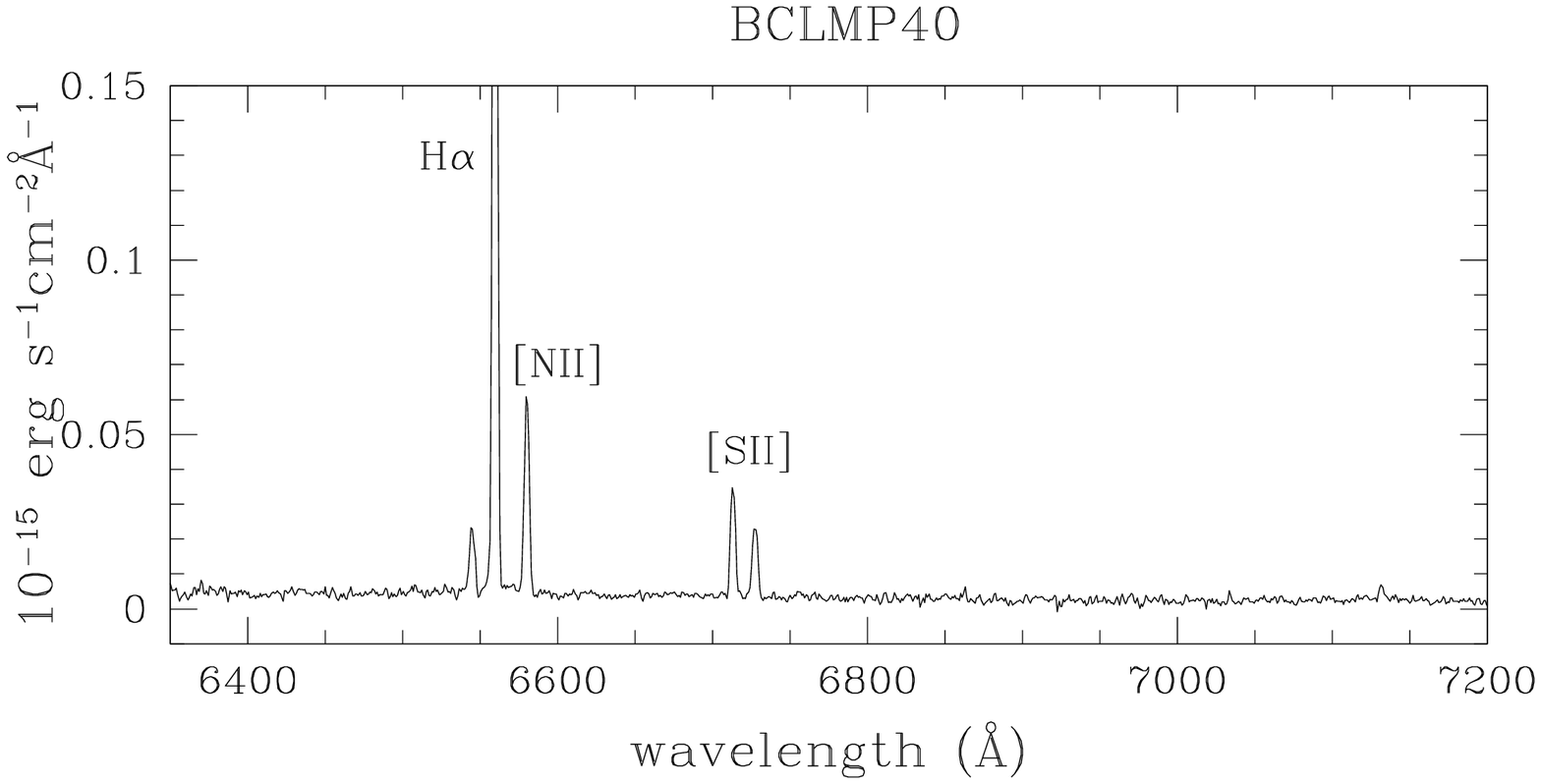}} 
\hspace*{0.0cm}\subfigure{\label{}\includegraphics[bb=18 428 592 718,width=9cm]{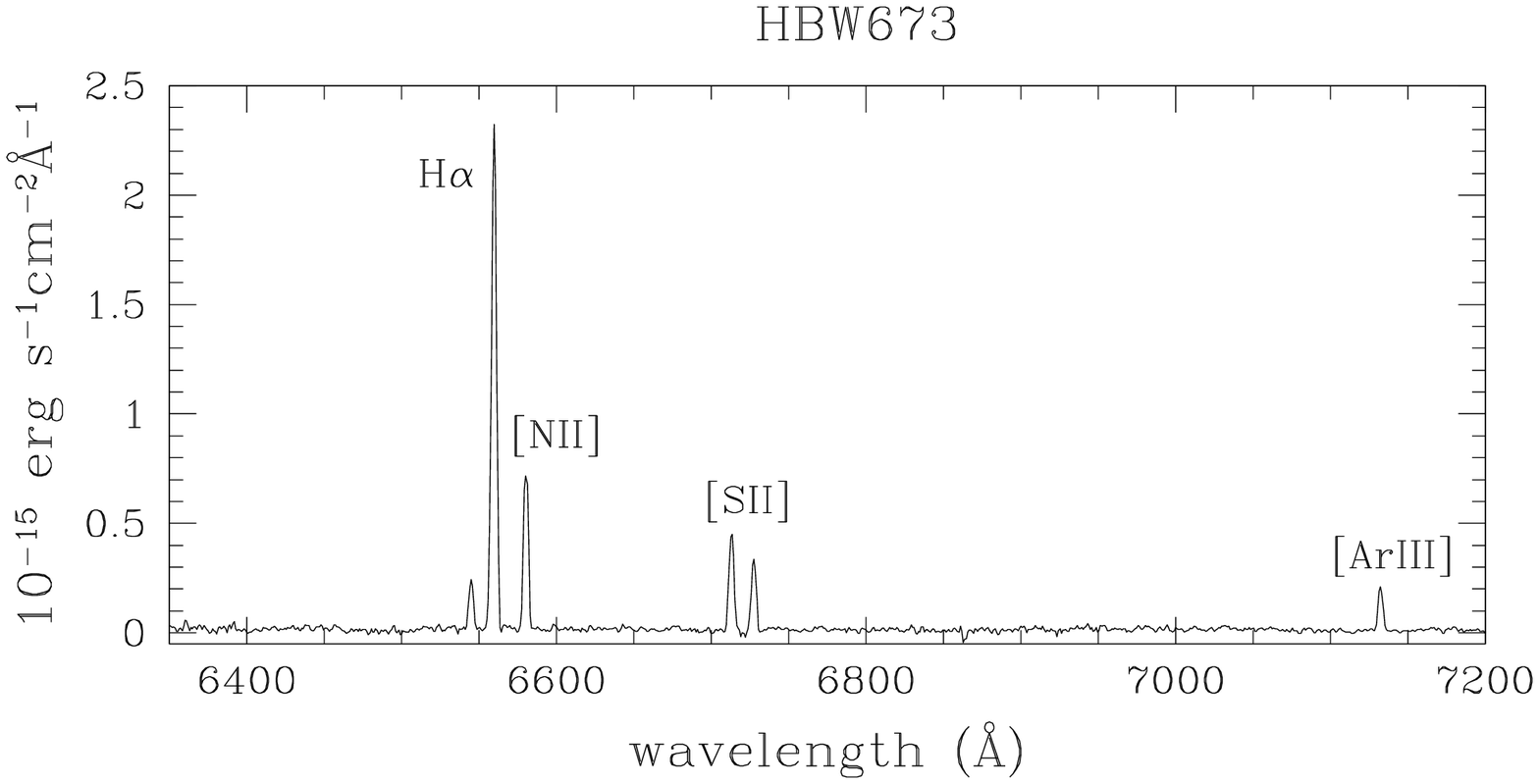}} 
   }} 
 \mbox{ 
  \centerline{ 
\hspace*{0.5cm}\subfigure{\label{}\includegraphics[bb=18 428 592 718,width=9cm]{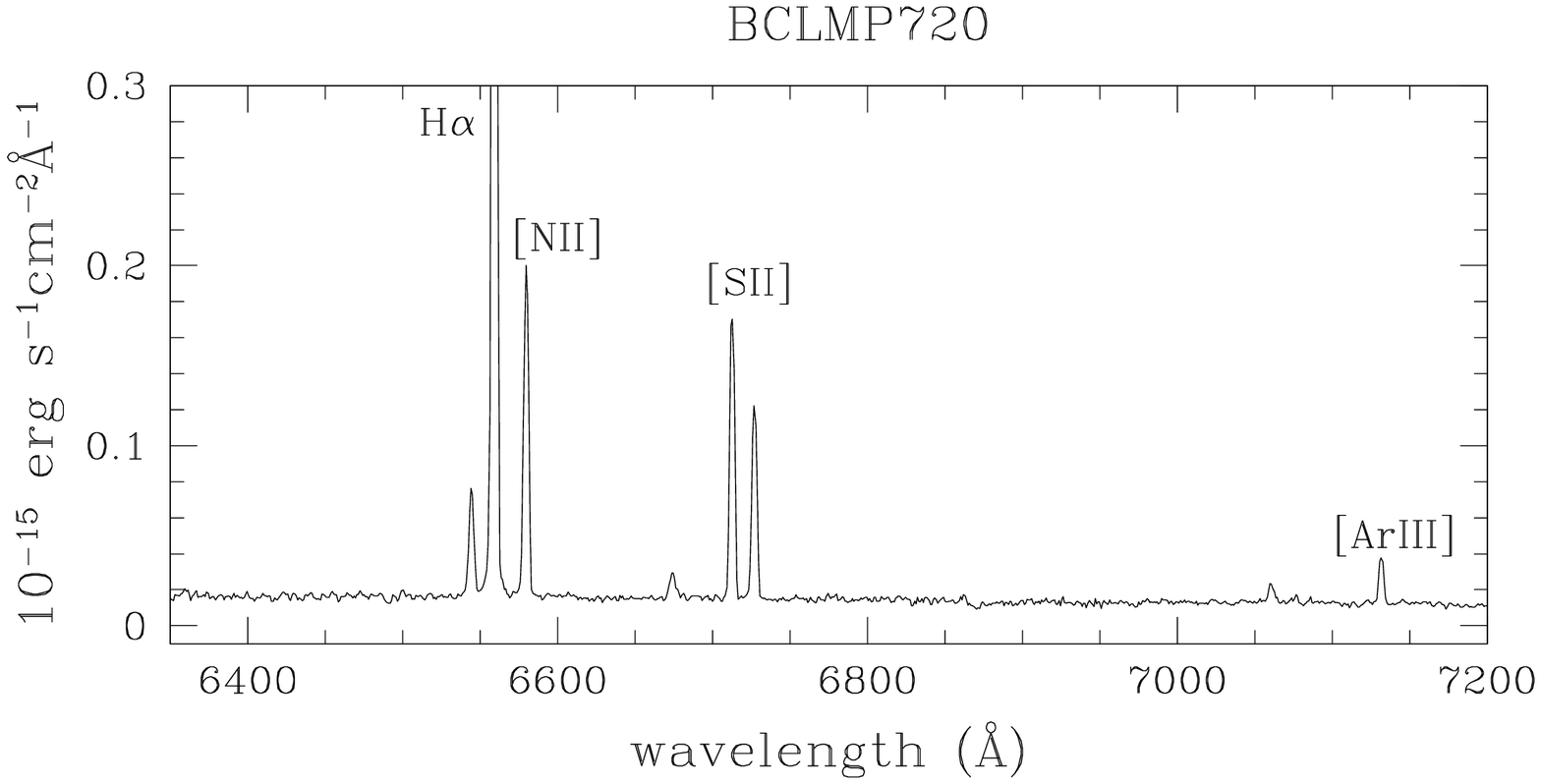}} 
\hspace*{0.0cm}\subfigure{\label{}\includegraphics[bb=18 428 592 718,width=9cm]{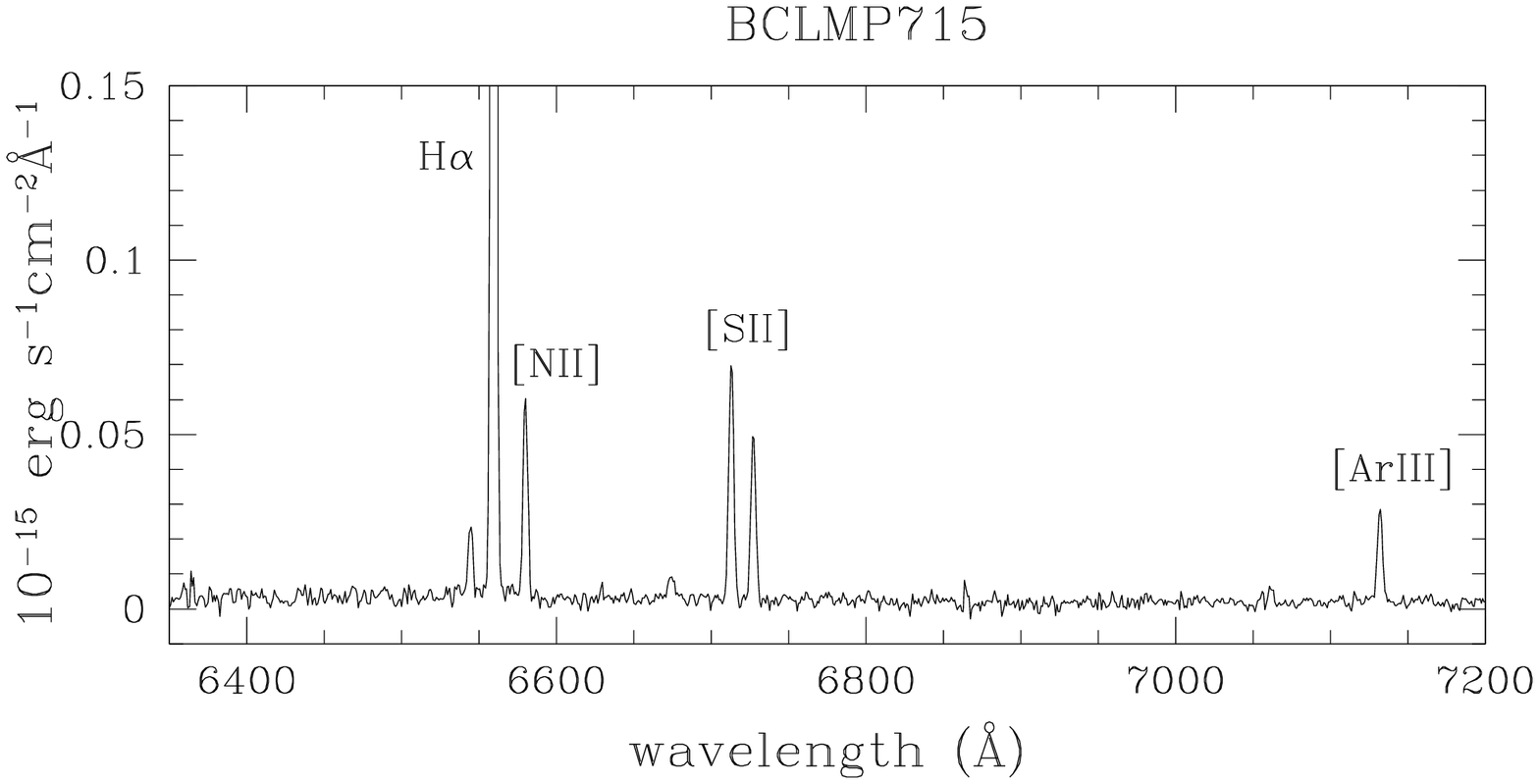}} 
   }} 
\mbox{ 
  \centerline{ 
\hspace*{0.5cm}\subfigure{\label{}\includegraphics[bb=18 428 592 718,width=9cm]{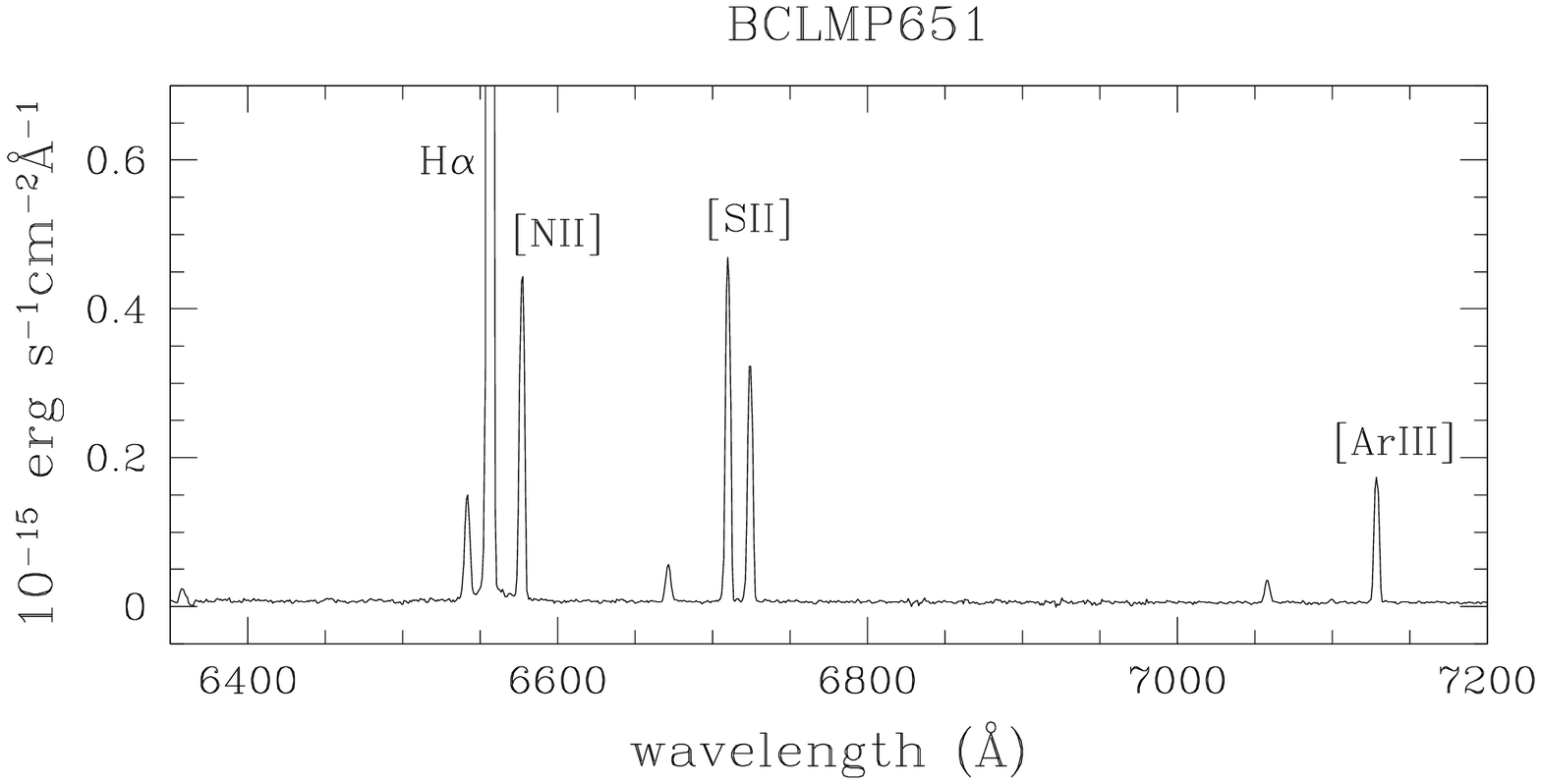}} 
}} 
\caption{Red spectra, in units of 10$^{-15}$ erg 
s$^{-1}$ cm$^{-2}$ \AA$^{-1}$, corresponding to 
  the nebular emission from the zone marked by the yellow box in
  Figure~\ref{finding_charts} for our HII regions. Selected emission lines are
  labelled.} 
\label{neb_red_spectra} 
\end{figure*} 
 
\newpage
 %Red box shows the size of the HeII images (20$\arcsec$ x 20$\arcsec$) as seen in Figure. 
 
\begin{figure*} 
\vspace*{2cm} 
\centering 
\mbox{ 
%\centerline{ 
%\hspace*{1.0cm}\subfigure{\label{}\includegraphics[width=20cm,angle=90.0]{hiianon.heii.new.ps}} 
%\hspace*{1.0cm}\subfigure{\label{}\includegraphics[width=20cm,angle=90.0]{hii651.heii.ps}} 
%\hspace*{1.0cm}\subfigure{\label{}\includegraphics[width=20cm,angle=90.0]{hiiz29.heii.ps}} 
%
%\hspace*{1.0cm}
\subfigure{\label{}\includegraphics[width=20cm,angle=90.0]{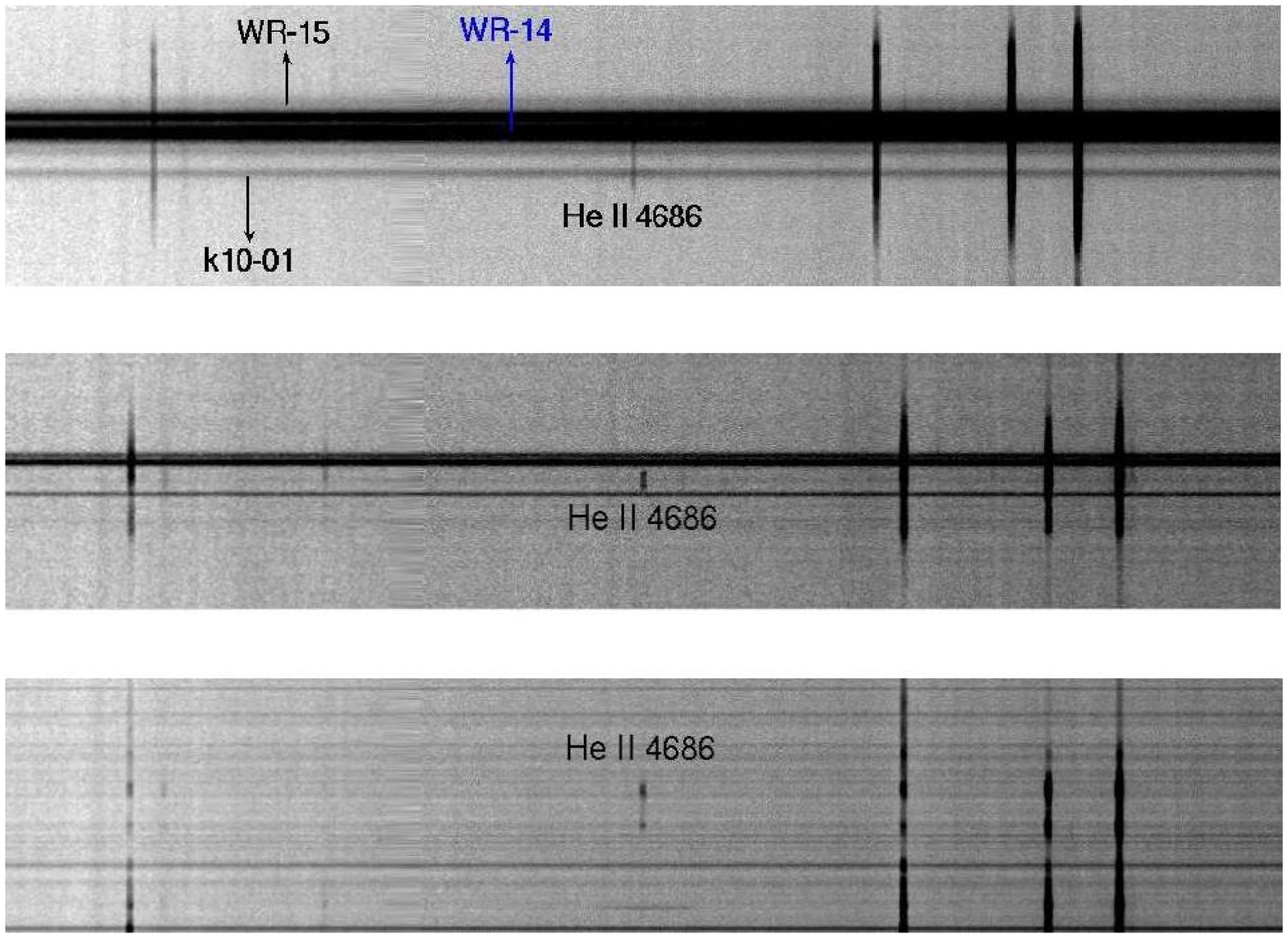}}
\hspace*{1.0cm}\rotcaption[angle=-90.0]{Portion ($\sim$ 4250 - 5100~\AA) of 2D longslit spectral image of MA 1 (top), BCLMP651 (middle) and HBW673 (bottom), showing the spatially extended nature of the nebular HeII $\lambda$ 4686 emission. The width of the image (vertical dimension) is approximately 220$\arcsec$ on the sky. }\label{2dimage}}
\end{figure*}

\begin{figure*}%[!ht] 
 \mbox{ 
  \centerline{ 
\hspace*{0.5cm}\subfigure{\label{}\includegraphics[bb=18 428 592 718,width=9cm]{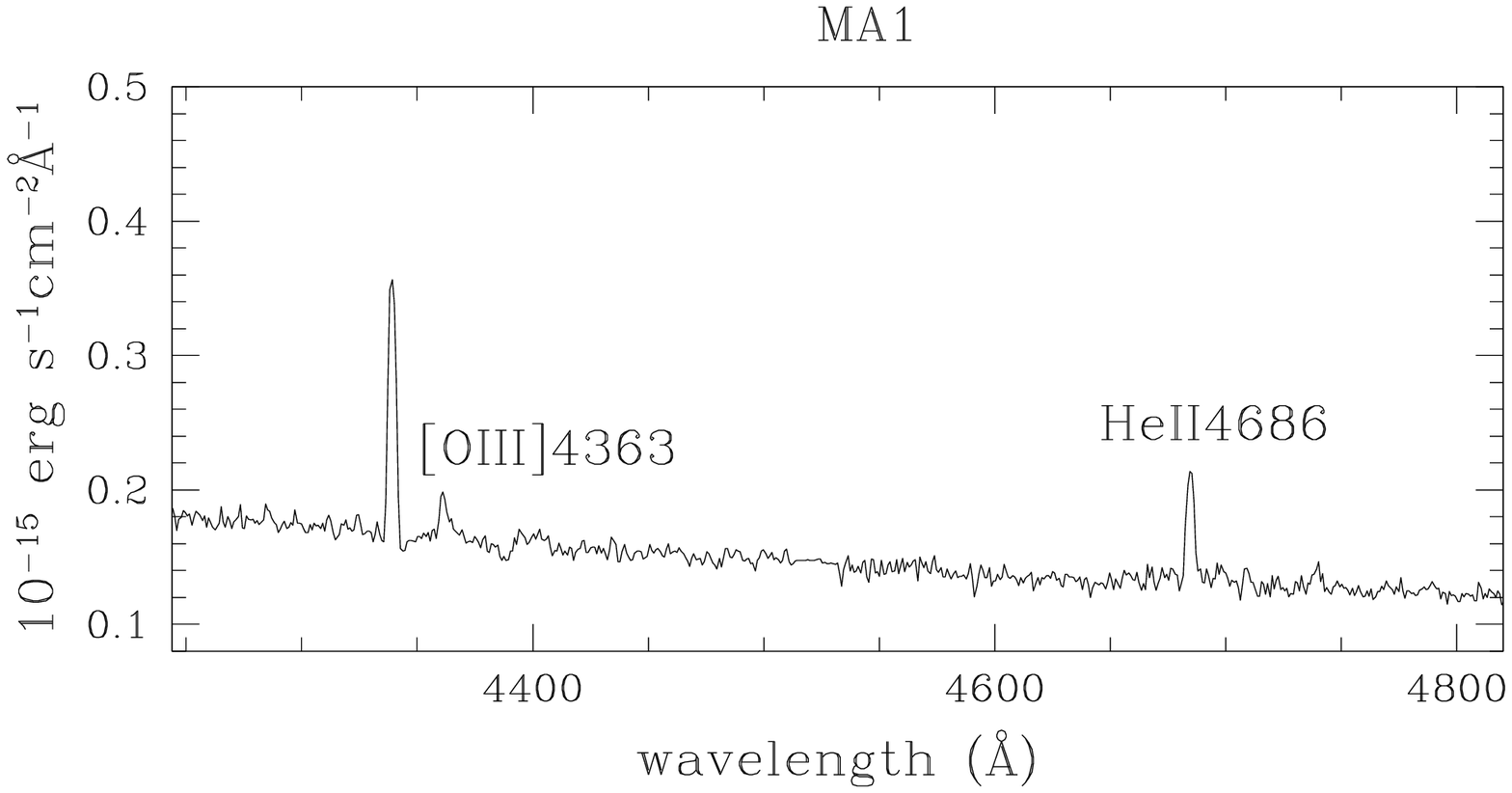}} 
   }} 
 \mbox{ 
  \centerline{ 
\hspace*{0.0cm}\subfigure{\label{}\includegraphics[bb=18 428 592 718,width=9cm]{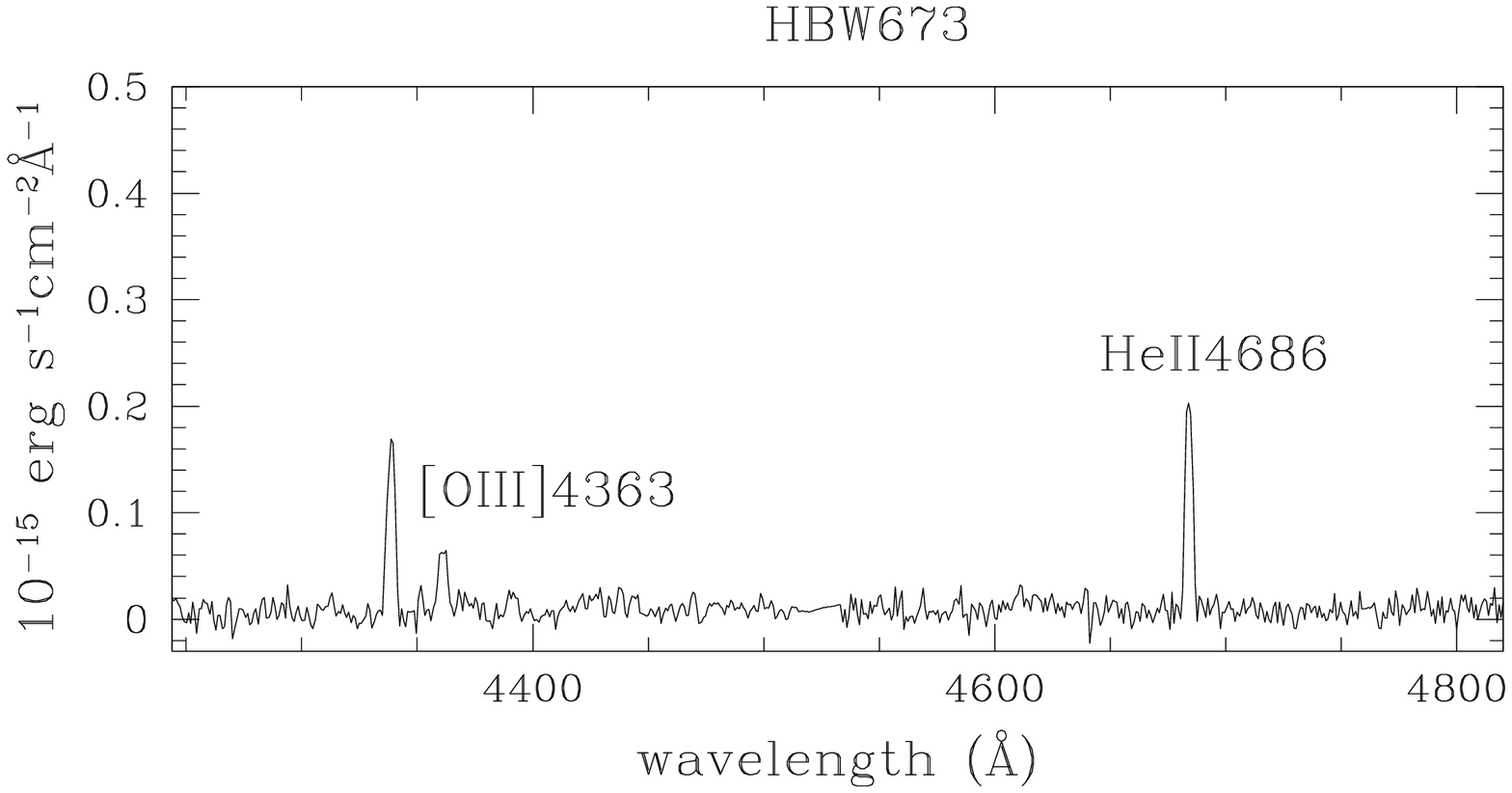}} 
}} 
\mbox{ 
  \centerline{ 
\hspace*{0.0cm}\subfigure{\label{}\includegraphics[bb=18 428 592 718,width=9cm]{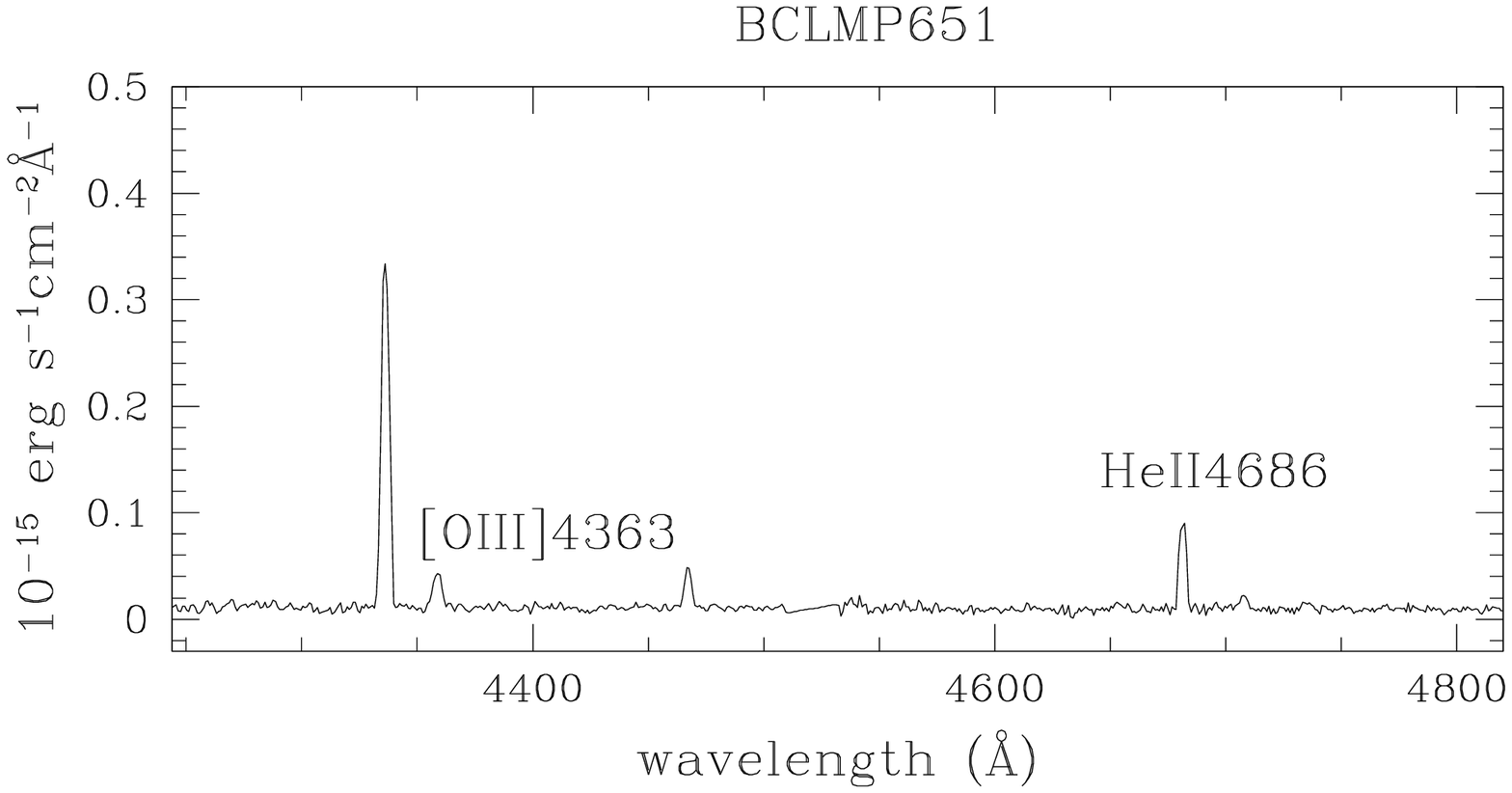}} 
}} 
\caption{MA 1,HBW673,BCLMP651: zoom of the wavelength range $\sim$ 4350-4800 \AA~ 
  showing the faint, temperature sensitive emission line [OIII]$\lambda$4363 
  and the nebular HeII$\lambda$4686 line} 
\label{zoom} 
\end{figure*} 
 
\clearpage 
 
\begin{figure}[ht] 
\mbox{ 
\centerline{ 
\hspace*{0.5cm}\subfigure{\label{}\includegraphics[bb=18 258 592 594,width=9cm,clip]{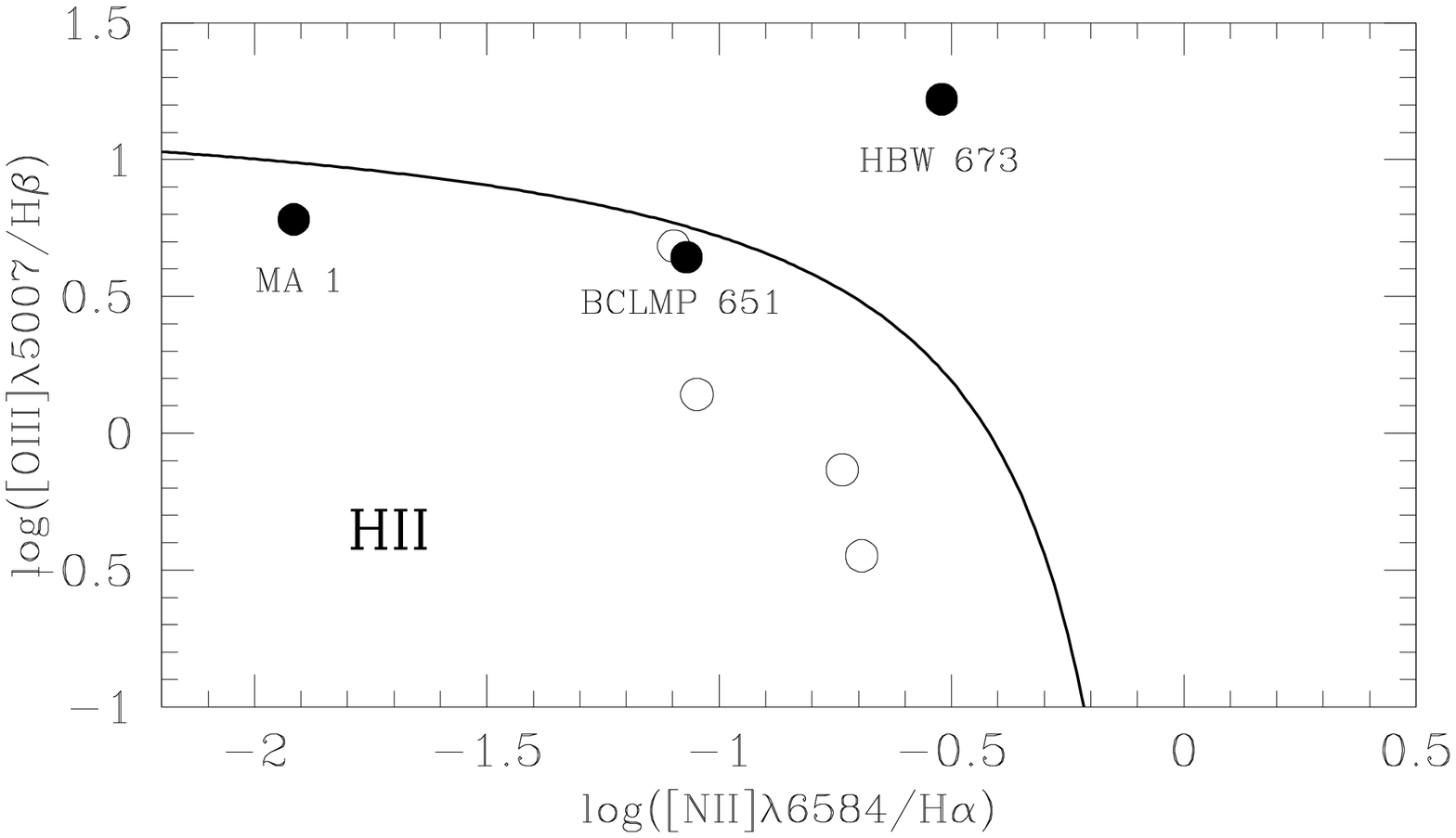}} 
}} 
\caption{BPT diagnostic diagram for our sample of seven candidates for 
HeII-emitting HII regions. Full and empty circles represent the three 
confirmed HeII nebulae (MA 1, BCLMP 651, HBW 673) and the four remaining 
objects, respectively. The solid curve, from \cite{k03}, 
indicates the empirical division between objects photoionized by stellar
radiation (below the solid curve) and regions that 
are not dominated by stellar photoionization (above the solid curve).} 
\label{plots_2}  
\end{figure} 
 
%\clearpage 
 
%\begin{figure} 
%\mbox{ 
%\centerline{ 
%\hspace*{0.5cm}\subfigure{\includegraphics[width=15cm,clip]{/work1/carol/USA/SPECTRA/GMOSN/paper/figures/figure_paul.ps}} 
%}} 
%\caption{Comparison between the GMOS spectroscopy of MC8 (M33-WR14), dereddened 
%by E(B-V)=0.30 mag and synthetic spectra for the WN component (red, $\log L/L_{\odot}$ = 5.9), 
%OBA cluster from starburst99 (green) and composite thereof (blue).} 
%\label{figure_paul}  
%\end{figure}

\begin{figure}
 \resizebox{\hsize}{!}
{\includegraphics[width=15cm,clip]{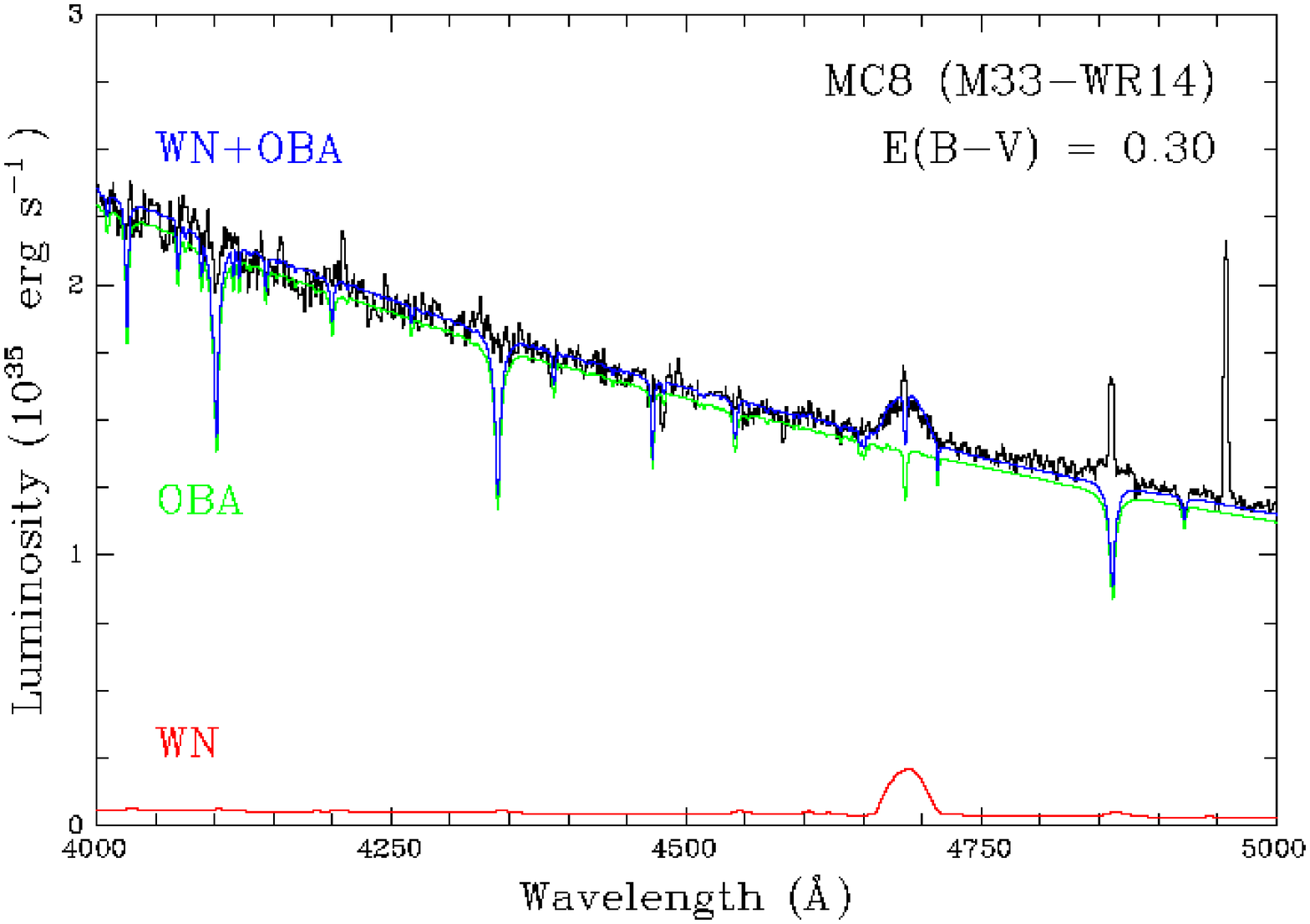}}  
\caption{Comparison between the GMOS spectroscopy of MC8 (M33-WR14), dereddened 
by E(B-V)=0.30 mag and synthetic spectra for the WN component (red, $\log L/L_{\odot}$ = 5.9), 
OBA cluster from starburst99 (green) and composite thereof (blue).} 
\label{figure_paul}  
\end{figure}

\end{document}